\documentclass[apjl,numberedappendix,appendixfloats,twocolumn]{openjournal}

\usepackage{xcolor}                 
\usepackage{textgreek}
\usepackage[utf8]{inputenc}
\usepackage[english]{babel}

\usepackage{colortbl}               
\definecolor{linkcolor}{rgb}{0.0,0.3,0.5}

\usepackage{tensind}
\tensordelimiter{?}
\DeclareGraphicsExtensions{.bmp,.png,.jpg,.pdf}
\usepackage{verbatim}
\usepackage[normalem]{ulem}
\usepackage{soul}

\usepackage{threeparttablex} 

\usepackage{multirow}
\usepackage{graphicx}               
\usepackage{booktabs}
\usepackage{array}
\usepackage{amsfonts,amsopn,amssymb}
\usepackage{flafter}        
\usepackage[section]{placeins}  
\usepackage{float}
\usepackage{longtable,booktabs}   
\usepackage{hyperref}

\usepackage{orcidlink}

\hypersetup{
    unicode,
    colorlinks=true,
    linkcolor=linkcolor,
    citecolor=linkcolor,
    filecolor=linkcolor,
    urlcolor=linkcolor,
}
\graphicspath{{./Plots/}}

\newcommand{\kms}{km~s$^{-1}$}			

\begin{document}

\title{Diameters and Temperatures VII: High-angular resolution measurements of Solar-type stars with the CHARA Array}

\author{Tabetha S. Boyajian$^{1}$\orcidlink{0000-0001-9879-9313}}
\email{boyajian@lsu.edu}
\author{Ashley A. Elliott$^{1}$\orcidlink{0009-0008-3869-5510}}
\author{Emelly D. Tiburcio$^{1}$\orcidlink{0000-0002-3612-5628}}
\author{Kaspar von Braun$^{2}$}
\author{Gail Schaefer$^{3}$}
\author{Timothy R. White$^{4}$}
\author{Daniel Huber$^{5}$}
\author{Michael Ireland$^{6}$}
\author{Jeremy Jones$^{7}$}
\author{Tyler Ellis$^{1}$}
\author{Eliot Hiegel$^{8}$}

\affiliation{
$^{1}$Department of Physics and Astronomy, Louisiana State University, Baton Rouge, LA 70802, USA\\
$^{2}$Lowell Observatory, 1400 West Mars Hill Rd., Flagstaff, AZ 86001, USA\\
$^{3}$The CHARA Array of Georgia State University, Mount Wilson Observatory, Mount Wilson, CA 91023, USA\\
$^{4}$Sydney Informatics Hub, Core Research Facilities, University of Sydney, NSW 2006, Australia\\
$^{5}$Institute for Astronomy, University of Hawaii, Honolulu, HI 96822, USA\\
$^{6}$Astralis-AITC, Research School of Astronomy and Astrophysics, Australian National University, Mount Stromlo Observatory, Cotter Road, Canberra, ACT 2611, Australia\\
$^{7}$Center for High Angular Resolution Astronomy and Department of Physics and Astronomy, Georgia State University, 25 Park Place, Suite 605, Atlanta, GA 30303, USA\\
$^{8}$Department of Earth and Environmental Sciences, University of Rochester, Rochester, NY 14620, USA
}


\begin{abstract}

    We present interferometric measurements of angular diameters for 27 nearby solar-type stars obtained with the Precision Astronomical Visible Observations (PAVO) beam combiner at the CHARA Array. The sample spans a broad range of metallicities, includes several known exoplanet hosts, and covers evolutionary stages from the zero-age main sequence to mildly evolved subgiants. Uniform-disk and limb-darkened angular diameters were measured for each target and combined with bolometric fluxes and Gaia parallaxes to determine precise, model-independent stellar radii, effective temperatures, and luminosities. We achieve typical uncertainties of $\sim1$\% in radius and $\sim1.5$\% in effective temperature.  Comparisons with multiple stellar evolutionary model grids yield mass and age estimates and enable assessment of grid-to-grid systematics, highlighting sensitivities to abundances, evolutionary stage, and proximity to grid boundaries.  Our results provide empirical benchmarks for testing stellar evolutionary theory, refining surface brightness-color relations, and improving the characterization of exoplanet host stars.        
    
\end{abstract}




\maketitle

	
\section{Introduction}\label{s:intro}
	
Interferometric measurements of stellar angular diameters have played a central role in advancing our understanding of stellar physics for over a century, beginning with the pioneering work of \citet{mic21} with the 20-foot Michelson Stellar Interferometer and several decades later by \citet{han74} with the Narrabri Stellar Intensity Interferometer. By directly resolving stellar disks, long-baseline optical/infrared interferometry yields angular sizes with sub-milliarcsecond precision, enabling fundamental determinations of absolute stellar radii when combined with accurate parallaxes. When bolometric fluxes are also available, these measurements provide model-independent estimates of effective temperature via the Stefan–Boltzmann law, bypassing many of the systematic uncertainties inherent in indirect methods such as photometric color-temperature relations or spectral modeling.

\begin{table*}[!t]
\centering
\scriptsize
\begin{ThreePartTable}
\begin{TableNotes}[flushleft]
\item[]\textit{Notes:} Spectral type and magnitudes are from \textsc{SIMBAD} \citep{simbad}. Parallaxes, $\pi$, are zero point corrected Gaia values \citep{GAIA, Lindegren21}. Metallicity, [Fe/H], values are the weighted mean from \citet{soubiran22}. Bolometric fluxes, $F_{\rm bol}$ , are determined in this work from PHOENIX SED fits to the available broadband photometry using RADPy; see \S\ref{s:properties}. Number of known planets, $N_{planets}$, is from the NASA Exoplanet Archive (as of July 2025; \citealt{nexsci}).
\end{TableNotes}

\begin{longtable}{lcccccccc}
\caption{General properties of the science sample.}\label{tab:sample_tab}\\
\toprule
\textbf{Star} & 
\textbf{Spectral} &
\boldmath{$V$} &
\boldmath{$K$} & 
\boldmath{$\pi\pm\sigma$} &
\textbf{[Fe/H]} &
\boldmath{$F_{\rm bol}\pm\sigma\times10^{-8}$ } &
\boldmath{$N_{planets}$} & 
\textbf{Note} \\
\textbf{Name} &
\textbf{Type} &
\textbf{(mag)}      &
\textbf{(mag)} &
\textbf{(mas)}  &
\textbf{(dex)}  &
\boldmath{$\rm(erg/s/cm^{2})$}  &
  &
  \\
\midrule
\endfirsthead

\caption[]{General properties of the science sample (continued).}\\
\toprule
\textbf{Star} & 
\textbf{Spectral} &
\boldmath{$V$} &
\boldmath{$K$} & 
\boldmath{$\pi\pm\sigma$} &
\textbf{[Fe/H]} &
\boldmath{$F_{\rm bol}\times10^{-8}$ } &
\boldmath{$N_{planets}$} & 
\textbf{Note} \\
\textbf{Name} &
\textbf{Type} &
\textbf{(mag)}      &
\textbf{(mag)} &
\textbf{(mas)}  &
\textbf{(dex)}  &
\boldmath{$\rm(erg/s/cm^{2})$}  &
  &
  \\
\midrule
\endhead

\bottomrule
\insertTableNotes
\endlastfoot
HD 1461  &  G3VFe0.5  &  6.46  &  4.90  &  42.695 $\pm$ 0.030  &  $0.19$  &  6.89 $\pm$0.19  &  2  &    \\ 
HD 8574  &  F8  &  7.12  &  5.78  &  22.269 $\pm$ 0.022  &  $0.01$  &  3.67 $\pm$0.10  &  1  &    \\ 
HD 8673  &  F7V  &  6.34  &  5.17  &  26.165 $\pm$ 0.037  &  $0.15$  &  7.24 $\pm$0.18  &  1  &    \\ 
HD 10697  &  G3Va  &  6.28  &  4.60  &  30.115 $\pm$ 0.040  &  $0.14$  &  8.55 $\pm$0.24  &  1  &    \\ 
HD 16141  &  G2V  &  6.78  &  5.27  &  26.474 $\pm$ 0.027  &  $0.14$  &  4.93 $\pm$0.15  &  1  &    \\ 
HD 20367  &  F8V  &  6.40  &  5.04  &  38.337 $\pm$ 0.032  &  $0.12$  &  7.07 $\pm$0.18  &  1  &    \\ 
HD 22879  &  G0VmF2  &  6.67  &  5.18  &  38.292 $\pm$ 0.031  &  $-0.84$\phs  &  5.77 $\pm$0.15  &  0  &  metal poor  \\ 
HD 30562  &  G2IV  &  5.77  &  4.31  &  38.222 $\pm$ 0.040  &  $0.22$  &  13.02 $\pm$0.44\phn  &  1  &    \\ 
HD 42807  &  G5V  &  6.43  &  4.85  &  53.731 $\pm$ 0.032  &  $0.02$  &  7.12 $\pm$0.19  &  0  &    \\ 
HD 58551  &  F6V  &  6.54  &  5.25  &  32.642 $\pm$ 0.082  &  $-0.53$\phs  &  6.47 $\pm$0.18  &  0  &  metal poor  \\ 
HD 76932  &  G2VFe-1.8CH-1  &  5.86  &  4.36  &  46.661 $\pm$ 0.042  &  $-0.88$\phs  &  13.06 $\pm$0.35\phn  &  0  &  metal poor  \\ 
HD 108510  &  G1V\_Fe-0.7  &  6.74  &  5.35  &  34.375 $\pm$ 0.047  &  $0.02$  &  5.19 $\pm$0.13  &  0  &  SB1 - low mass ratio  \\ 
HD 127334  &  G5VCH0.3  &  6.36  &  4.74  &  42.725 $\pm$ 0.017  &  $0.18$  &  7.68 $\pm$0.20  &  0  &    \\ 
HD 144579  &  G8V  &  6.67  &  4.76  &  69.616 $\pm$ 0.014  &  $-0.68$\phs  &  6.51 $\pm$0.17  &  0  &  metal poor  \\ 
HD 154345  &  G9  &  6.76  &  5.00  &  54.714 $\pm$ 0.018  &  $-0.10$\phs  &  5.53 $\pm$0.14  &  1  &    \\ 
HD 164922  &  G9V  &  7.01  &  5.11  &  45.468 $\pm$ 0.017  &  $0.18$  &  4.62 $\pm$0.12  &  4  &    \\ 
HD 181096  &  F6IV:  &  6.02  &  6.47  &  23.793 $\pm$ 0.022  &  $-0.25$\phs  &  10.03 $\pm$0.27\phn  &  0  &    \\ 
HD 182488  &  K0V  &  6.37  &  4.49  &  64.043 $\pm$ 0.016  &  $0.19$  &  8.63 $\pm$0.32  &  0  &  unconfirmed planet host  \\ 
HD 184499  &  G0V  &  6.64  &  5.08  &  31.388 $\pm$ 0.017  &  $-0.57$\phs  &  6.07 $\pm$0.16  &  0  &  metal poor  \\ 
HD 185269  &  G2V  &  6.67  &  5.26  &  19.213 $\pm$ 0.018  &  $0.14$  &  5.67 $\pm$0.17  &  1  &    \\ 
HD 190228  &  G5IV  &  7.30  &  5.35  &  15.869 $\pm$ 0.016  &  $-0.25$\phs  &  3.59 $\pm$0.09  &  1  &    \\ 
HD 193664  &  G3V  &  5.91  &  4.45  &  57.182 $\pm$ 0.021  &  $-0.10$\phs  &  11.63 $\pm$0.39\phn  &  0  &    \\ 
HD 195019  &  G1V  &  6.87  &  5.26  &  26.614 $\pm$ 0.022  &  $0.05$  &  4.82 $\pm$0.13  &  1  &    \\ 
HD 196885  &  F8V  &  6.39  &  5.07  &  29.371 $\pm$ 0.027  &  $0.23$  &  7.03 $\pm$0.18  &  1  &    \\ 
HD 202568  &  G0  &  6.60  &  4.40  &  14.973 $\pm$ 0.017  &  $-0.19$\phs  &  7.56 $\pm$0.22  &  0  &    \\ 
HD 210277  &  G8V  &  8.57  &  4.80  &  46.814 $\pm$ 0.028  &  $0.21$  &  6.53 $\pm$0.19  &  1  &    \\ 
HD 222155  &  G0  &  7.11  &  5.51  &  19.773 $\pm$ 0.016  &  $-0.16$\phs  &  3.92 $\pm$0.10  &  1  &    \\ 

\end{longtable}
\end{ThreePartTable}
\end{table*}

Over the past two decades, the commissioning of several long-baseline interferometers such as The CHARA Array, the Very Large Telescope Interferometer (VLTI), and the Navy Precision Optical Interferometer (NPOI) has enabled a dramatic increase in the number and precision of directly measured stellar diameters. Large, homogeneous surveys have extended high-accuracy measurements beyond the brightest giants to encompass hundreds of nearby main-sequence and subgiant stars across a range of spectral types (e.g., \citealt{vanbelle09, boy12b, White2018, ligi16, rabus19, karo22a,karo22b, baines25}; see also Figure~\ref{fig:pavo_hrd}). These efforts have provided stringent empirical constraints on stellar effective temperature scales (e.g., \citealt{code76,boy13a, mann15}), surface brightness–color relations (e.g., \citealt{ker04, Adams2018}), and mass–radius relations (e.g., \citealt{demory09, mann15}). The combination of interferometric angular diameters with high-quality parallaxes from Gaia \citep{GAIA} and bolometric flux measurements has yielded effective temperatures with typical uncertainties of $<1$\%, rivaling or surpassing those achievable through spectroscopy alone.

Such precise, model-independent stellar parameters have proven especially valuable for characterizing exoplanet host stars, where uncertainties in stellar radius propagate directly into planet radius and density estimates (e.g., \citealt{vonbraun12, ligi12, Ellis2021, caba22}). Interferometry has been used to benchmark planetary systems across a wide range of architectures from hot Jupiters (e.g., \citealt{boy15}) to small, potentially rocky planets (e.g., \citealt{kane17}), and those on wide visual orbits (e.g., \citealt{baines12,elliott24}).  In addition, stellar angular diameter work has great synergy with microlensing planet surveys, where they provide the foundation for surface brightness-color relations that are required to determine sizes of the microlensing source stars in order to derive host star and planet masses in microlensing events (e.g., \citealt{gould92, an02, bennett16}).  Interferometry has also been used to validate stellar radii inferred from asteroseismology (e.g. \citealt{huber12, johnson14, white15}), providing independent checks to validate the range and capabilities of such techniques. More broadly, empirically determined stellar radii serve as essential anchors for calibrating stellar evolutionary models, particularly for metal-poor or $\alpha$-enhanced stars where theoretical predictions diverge (e.g., \citealt{boy08, creevey12, karo18}). Continued interferometric surveys targeting nearby solar-type stars are therefore critical not only for improving stellar models, but also for refining the physical properties of the growing exoplanet populations from current and upcoming missions.

In this work, we present high-angular-resolution measurements of 27 nearby solar-type stars.  Our sample spans a range of metallicities, includes several known exoplanet hosts, and covers evolutionary stages from the zero-age main sequence to mildly evolved subgiant stars. We derive uniform-disk and limb-darkened angular diameters for each target, and combine these with bolometric fluxes and Gaia parallaxes to determine precise radii, effective temperatures, and luminosities. We then compare these empirical measurements to predictions from multiple stellar evolutionary model grids to infer stellar masses and ages, and to assess model-to-model systematics.

The following paper is organized as follows.  In Section~\ref{s:observations} we describe the sample selection, observations, and data calibration procedures.  Section~\ref{s:properties} details our determination of stellar fundamental properties from the interferometric measurements.  In section~\ref{s:models} we present comparisons with stellar evolutionary models and discuss implications for both stellar astrophysics and exoplanet characterization. We summarize our conclusions in Section~\ref{s:conclusion}.


\section{Interferometric Observations}\label{s:observations}

Our science sample consists of 27 solar-type stars -- primarily late-F and G type -- either on or near the main sequence.  They are moderately bright (range in magnitudes from $V\sim5.8$ to 8.6~mag), and reside nearby (distances spanning $d\sim14$ to 67~pc).  Basic properties for this sample are presented in Table~\ref{tab:sample_tab}.

\begin{table*}[!t]
\centering
\begin{ThreePartTable}
\begin{TableNotes}[flushleft]
\item[]\textit{Notes:} Uncertainties on values are $1\sigma$.  
$\mu_R$ is the linear limb-darkening coefficient in $R$-band.  
$R_{\star}$, $T_{\rm eff}$, and $L_{\star}$ are derived using measured angular diameters, zero-point corrected parallaxes, and bolometric fluxes; see \S\ref{s:properties} for details.  
\end{TableNotes}

\begin{longtable}{lcccccc}
\caption{Fundamental properties of the science sample.}\label{tab:prop_tab}\\
\toprule
\textbf{Star} &
\boldmath{$\theta_{\rm UD}\pm\sigma$} &
\boldmath{$\mu_R$} &
\boldmath{$\theta_{\rm LD}\pm\sigma$} &
\boldmath{$R_{\star}\pm\sigma$} &
\boldmath{$T_{\rm eff}\pm\sigma$} &
\boldmath{$L_{\star}\pm\sigma$} \\
\textbf{Name} &
\textbf{(mas)} &
      &
\textbf{(mas)} &
\textbf{(R$_{\odot}$)} &
\textbf{(K)} &
\textbf{(L$_{\odot}$)} \\
\midrule
\endfirsthead

\caption[]{Fundamental properties of the science sample (continued).}\\
\toprule
\textbf{Star} &
\boldmath{$\theta_{\rm UD}\pm\sigma$} &
\boldmath{$\mu_R$} &
\boldmath{$\theta_{\rm LD}\pm\sigma$} &
\boldmath{$R_{\star}\pm\sigma$} &
\boldmath{$T_{\rm eff}\pm\sigma$} &
\boldmath{$L_{\star}\pm\sigma$} \\
\textbf{Name} &
\textbf{(mas)} &
      &
\textbf{(mas)} &
\textbf{(R$_{\odot}$)} &
\textbf{(K)} &
\textbf{(L$_{\odot}$)} \\
\midrule
\endhead

\bottomrule
\insertTableNotes
\endlastfoot
HD 1461  &  0.433$\pm$0.006  & 0.544  &  0.456$\pm$0.006  &  1.151$\pm$0.015  &  5616$\pm$53 &  1.18$\pm$0.03  \\ 
HD 8574  &  0.307$\pm$0.010  & 0.520  &  0.322$\pm$0.010  &  1.557$\pm$0.048  &  5713$\pm$96 &  2.31$\pm$0.06  \\ 
HD 8673  &  0.347$\pm$0.003  & 0.454  &  0.361$\pm$0.003  &  1.486$\pm$0.014  &  6390$\pm$51 &  3.30$\pm$0.08  \\ 
HD 10697  &  0.486$\pm$0.007  & 0.542  &  0.512$\pm$0.007  &  1.832$\pm$0.024  &  5593$\pm$54 &  2.94$\pm$0.08  \\ 
HD 16141  &  0.360$\pm$0.003  & 0.532  &  0.378$\pm$0.004  &  1.537$\pm$0.014  &  5675$\pm$50 &  2.20$\pm$0.07  \\ 
HD 20367  &  0.395$\pm$0.004  & 0.500  &  0.413$\pm$0.005  &  1.160$\pm$0.013  &  5940$\pm$51 &  1.50$\pm$0.04  \\ 
HD 22879  &  0.383$\pm$0.002  & 0.466  &  0.399$\pm$0.003  &  1.122$\pm$0.008  &  5743$\pm$42 &  1.23$\pm$0.03  \\ 
HD 30562  &  0.539$\pm$0.005  & 0.508  &  0.565$\pm$0.006  &  1.591$\pm$0.016  &  5918$\pm$58 &  2.78$\pm$0.09  \\ 
HD 42807  &  0.425$\pm$0.003  & 0.519  &  0.446$\pm$0.004  &  0.893$\pm$0.007  &  5728$\pm$44 &  0.77$\pm$0.02  \\ 
HD 58551  &  0.338$\pm$0.002  & 0.439  &  0.351$\pm$0.002  &  1.158$\pm$0.009  &  6303$\pm$49 &  1.90$\pm$0.05  \\ 
HD 76932  &  0.543$\pm$0.005  & 0.453  &  0.567$\pm$0.005  &  1.309$\pm$0.011  &  5910$\pm$47 &  1.87$\pm$0.05  \\ 
HD 108510  &  0.339$\pm$0.002  & 0.495  &  0.354$\pm$0.003  &  1.109$\pm$0.008  &  5941$\pm$44 &  1.37$\pm$0.04  \\ 
HD 127334  &  0.466$\pm$0.006  & 0.549  &  0.491$\pm$0.007  &  1.237$\pm$0.017  &  5564$\pm$52 &  1.31$\pm$0.03  \\ 
HD 144579  &  0.497$\pm$0.004  & 0.549  &  0.523$\pm$0.004  &  0.809$\pm$0.006  &  5170$\pm$39 &  0.42$\pm$0.01  \\ 
HD 154345  &  0.407$\pm$0.005  & 0.543  &  0.427$\pm$0.005  &  0.841$\pm$0.009  &  5492$\pm$47 &  0.58$\pm$0.02  \\ 
HD 164922  &  0.393$\pm$0.005  & 0.584  &  0.416$\pm$0.005  &  0.986$\pm$0.012  &  5318$\pm$47 &  0.70$\pm$0.02  \\ 
HD 181096  &  0.432$\pm$0.002  & 0.453  &  0.450$\pm$0.003  &  2.037$\pm$0.013  &  6211$\pm$47 &  5.53$\pm$0.15  \\ 
HD 182488  &  0.517$\pm$0.008  & 0.569  &  0.545$\pm$0.009  &  0.916$\pm$0.015  &  5436$\pm$67 &  0.66$\pm$0.02  \\ 
HD 184499  &  0.390$\pm$0.004  & 0.479  &  0.407$\pm$0.004  &  1.396$\pm$0.015  &  5763$\pm$48 &  1.92$\pm$0.05  \\ 
HD 185269  &  0.359$\pm$0.003  & 0.504  &  0.376$\pm$0.003  &  2.106$\pm$0.018  &  5893$\pm$51 &  4.79$\pm$0.15  \\ 
HD 190228  &  0.380$\pm$0.007  & 0.586  &  0.401$\pm$0.008  &  2.720$\pm$0.051  &  5089$\pm$58 &  4.45$\pm$0.12  \\ 
HD 193664  &  0.519$\pm$0.005  & 0.495  &  0.542$\pm$0.006  &  1.021$\pm$0.011  &  5872$\pm$58 &  1.11$\pm$0.04  \\ 
HD 195019  &  0.360$\pm$0.005  & 0.529  &  0.377$\pm$0.006  &  1.525$\pm$0.022  &  5651$\pm$56 &  2.12$\pm$0.06  \\ 
HD 196885  &  0.376$\pm$0.006  & 0.491  &  0.393$\pm$0.006  &  1.440$\pm$0.024  &  6082$\pm$64 &  2.54$\pm$0.07  \\ 
HD 202568  &  0.608$\pm$0.006  & 0.627  &  0.646$\pm$0.006  &  4.647$\pm$0.044  &  4830$\pm$42 &  10.53$\pm$0.31\phn  \\ 
HD 210277  &  0.457$\pm$0.006  & 0.557  &  0.481$\pm$0.006  &  1.107$\pm$0.015  &  5396$\pm$53 &  0.93$\pm$0.03  \\ 
HD 222155  &  0.334$\pm$0.006  & 0.527  &  0.350$\pm$0.007  &  1.904$\pm$0.037  &  5570$\pm$65 &  3.13$\pm$0.08  \\ 

\end{longtable}
\end{ThreePartTable}
\end{table*}

We observed the sample over the 2010-2018 observing seasons using The CHARA Array, a long-baseline optical/infrared interferometer located at Mount Wilson Observatory \citep{ttb05}.  The six telescopes available at the CHARA Array are arranged in a Y configuration with separations between telescopes (baseline) ranging from $\sim 35 - 330$~meters. Each of the Array's three arms: South (S), East (E), and West (W) has two telescopes, denoted as a '1' or '2', '2' being located more central in the Array.  We primarily observed in configurations using the longest baselines ($>278$~meters) in order to maximize the interferometers resolution. All data were taken with the Precision Astronomical Visible Observations (PAVO) beam combiner \citep{Ire08}, which measures interference fringes over a $630-950$~nm dispersed bandwidth with 2 telescopes.  An observation log is presented in Appendix Table~\ref{tab:obs_log}.

Interferometric data are calibrated to remove instrumental and atmospheric effects in the data. The ideal approach to minimize any calibration or systematic errors requires a star to be observed with two unique baselines, on two different nights, and with two calibrators \citep[e.g.,][]{boy12b}. We select calibrator stars to be unresolved point sources with similar brightness and close in proximity to the science object. For this initial calibrator selection, we use the \textsc{SearchCal} tool developed by the Jean-Marie Mariotti Center\footnote{Available at http://www.jmmc.fr/searchcal}. In order to further remove bad calibrators, we did a literature search to screen for multiplicity (any companions must have separation $> 2 - 3$~arcsec) and rapid rotation (projected rotational velocities must be $v \sin i < 100$~\kms).  Table~\ref{tab:obs_log} lists the calibrator star(s) used for each science target, and Table~\ref{tab:cals} lists the calibrator angular diameter estimates that we use to calibrate our data. Table~\ref{tab:bad_cals} lists bad calibrator stars identified in our data set.

The stellar angular diameter for each science star is determined using \texttt{RADPy}\footnote{https://github.com/spaceashley/radpy} \citep[][]{radpy}, which fits the measured calibrated visibilities to the uniform-disk ($\theta_{UD}$) and limb-darkened ($\theta_{LD}$) functions as defined in \citet{han74}.  Specifically for these CHARA PAVO data, \texttt{RADPy} uses the ``flux conserving'' linear limb-darkening coefficients in $R$-band $\mu_R$ from \citet{cla11}, after an iteration based on $T_{\rm eff}$ values resulting from the uniform disk solution (see Section~\ref{s:properties}). 

Uncertainties on $\theta_{UD}$ and $\theta_{LD}$ are computed via a Monte Carlo with bracket bootstrapping: in each iteration, visibilities within a observation ``bracket'' are resampled from their reported errors and re-fit, and the final diameter is taken as the mean of the accepted trials with the $1\sigma$ uncertainty given by the mean absolute deviation of that distribution. For $\theta_{LD}$, the limb-darkening coefficient is drawn from a normal prior with $\sigma=0.02$ to propagate its uncertainty. Because the resampling is performed at the calibration–bracket level, the final quoted $1\sigma$ errors empirically include contributions from both random noise and bracket-to-bracket calibration variability.  Table~\ref{tab:prop_tab} shows the final results from \texttt{RADPy}, and the data and solutions for each star are presented in  Figures~\ref{fig:diameter_plots1}, \ref{fig:diameter_plots2}, \ref{fig:diameter_plots3}, \ref{fig:diameter_plots4}, and \ref{fig:diameter_plots5}.

We note that some observations, particularly at low spatial frequencies, show slightly elevated calibrated visibilities in the reddest wavelength channels. This behavior has previously been seen in PAVO data and is likely associated with wavelength-dependent calibration effects, potentially including reduced flux at the bandpass edges and variable atmospheric/telluric transmission. These low-spatial-frequency measurements have little leverage on the fitted angular diameters, which are primarily constrained by the more highly resolved data.


\section{Stellar Properties}\label{s:properties}

\begin{figure*}[!t]
    \centering
    \includegraphics[width=0.95\textwidth]
    {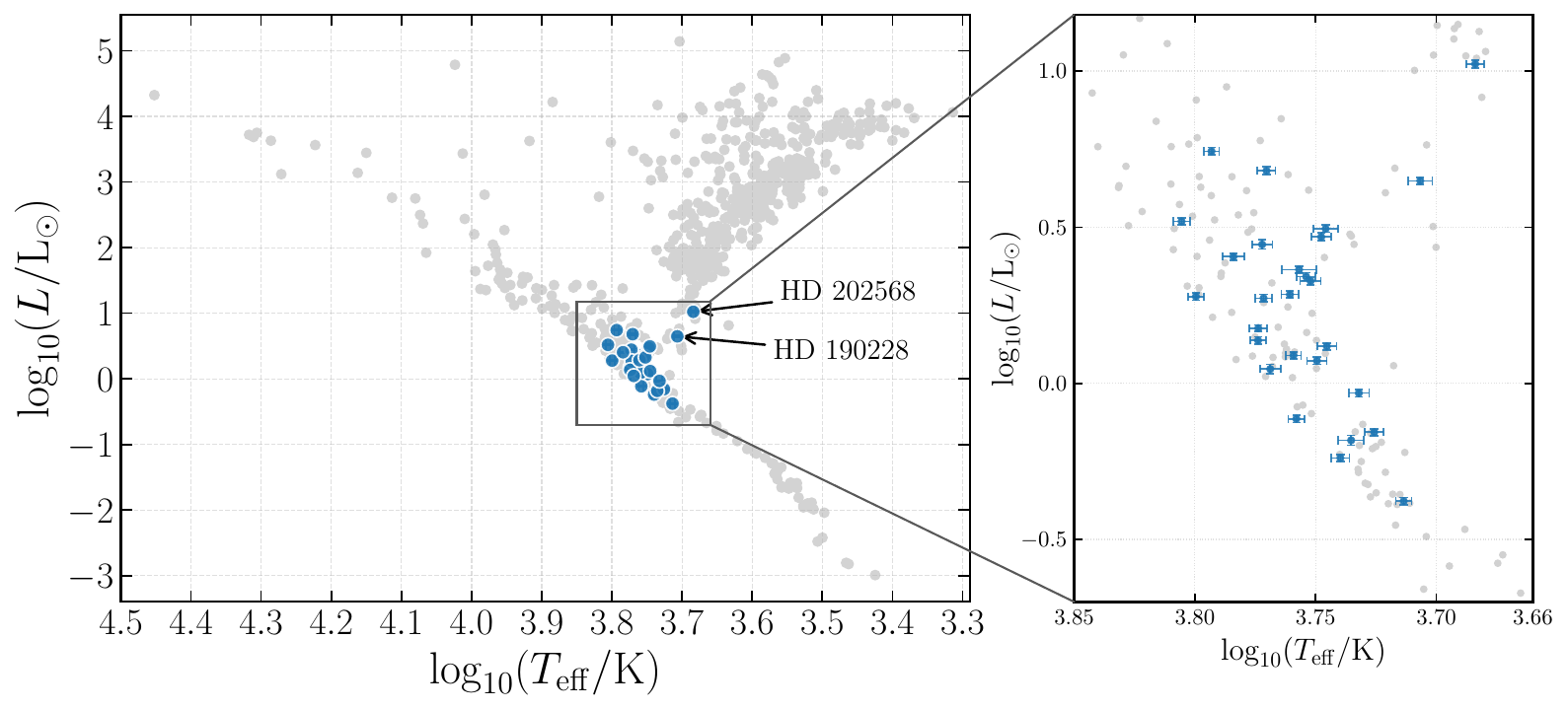}
    \caption{Empirical H–R diagram showing our sample (blue circles) together with stars having precise angular-diameter measurements (gray circles; data source: \url{https://www.chara.gsu.edu/science-highlights/stellar-diameters}). Arrows identify the two post-main-sequence subgiants, HD 190228 and HD 202568, discussed in \S4. The right panel enlarges the boxed region of the H–R diagram; error bars on the blue points show the 1$\sigma$ uncertainties in the measured temperature and luminosity.}     \label{fig:pavo_hrd}
\end{figure*}

We determine the stellar bolometric flux $F_{\rm bol}$ using the SED-fitting module in \texttt{RADPy}, which employs the \texttt{SEDFit} framework \citep{sedfit} to fit stellar-atmosphere models to broadband photometry. For each target, we fitted PHOENIX model spectra to the Johnson, Tycho, 2MASS, TESS, and Gaia photometry where available. Interstellar extinction was fixed at $A_V=0$ for all stars due to their close proximity. The resulting best-fitting model spectral energy distributions were integrated over wavelength to determine $F_{\rm bol}$, which can be found in Table~\ref{tab:sample_tab}. Uncertainties were estimated through Monte Carlo resampling of the input photometry and repeated SED fitting; the reported uncertainty is the standard deviation of the resulting $F_{\rm bol}$ distribution with an additional 2\% uncertainty added in quadrature to account for unknown systematic offsets and zero-points in photometric systems \citep{boh14}. 

We compute the absolute stellar luminosities, $L$, from these measurements of bolometric flux and the zero-point corrected Gaia parallax \citep{GAIA, Lindegren21} (also in Table~\ref{tab:sample_tab}). Our luminosity values are expressed in solar units, using the solar luminosity value of $L_{\odot} = 3.8270 \times 10^{33}$~erg/s \citep{kopp11,mam12}.  We use the corrected Gaia parallax and our measured angular diameter to calculate the linear radius $R$. Effective temperatures $T_{\rm eff}$ are derived through the Stefan-Boltzmann equation, rearranged in terms of observables: 

\begin{eqnarray}
T_{\rm eff} (\rm{K}) = 2341 (F_{\rm bol}/\theta_{\rm LD}^2)^{0.25}
\end{eqnarray}

\noindent where $F_{\rm bol}$ is the bolometric flux expressed in units of $10^{-8}$~erg/s/cm$^{2}$ and $\theta_{\rm LD}$ is the limb-darkened angular diameter expressed in units of milliarcseconds. These stellar properties are listed in Table~\ref{tab:prop_tab}, and Figure~\ref{fig:pavo_hrd} presents our sample of stars plotted on an H-R diagram.

\section{Stellar Models and Discussion}\label{s:models}

Our sample comprises 27 solar-type dwarfs and mild subgiants with effective temperatures spanning 4830 to 6390~K (HD 202568 and HD 8673). This confined parameter space is ideally suited for comparative tests of stellar evolutionary models, where our model-independent radii and effective temperatures are measured with typical uncertainties  $\sim$1\% and 1.5\%, respectively, providing tight anchors on the H-R diagram. In this homogeneous regime, an analysis of variations in inferred mass and age across multiple stellar evolutionary models can be interpreted as model behavior due to input physics rather than data heterogeneity, enabling controlled checks of cross-model consistency and clear identification of where the grids agree, diverge, or are weakly discriminating.  

We use stellar evolutionary models to determine stellar masses and ages for the CHARA sample using the \texttt{Kiauhoku} interpolation toolkit \citep{claytor20}, as implemented in \citet{tayar22}, leveraging our precise, empirically determined stellar properties obtained in this work (\S\ref{s:properties}). For each star, we input measured effective temperatures, luminosities, and metallicities into \texttt{Kiauhoku}, where the temperatures and luminosities are derived from our interferometric analysis (Table~\ref{tab:prop_tab}), and the input metallicities ([Fe/H]; Table~\ref{tab:sample_tab}) are the weighted averages drawn from \citet{soubiran22}. \texttt{Kiauhoku} includes four independent evolutionary model grids: YREC, MIST, Dartmouth, and Garstec. The \texttt{Kiauhoku} routine performs a deterministic interpolation for a given star to identify the closest matching point in each stellar evolutionary grid and return the corresponding model mass and age.    

Table~\ref{tab:model_results} shows these results for each star in our sample: four sets of (Mass,Age) estimates, the mean ($\overline{\mathrm{Mass}}$,$\overline{\mathrm{Age}}$) and standard deviation ($\sigma$Mass,$\sigma$Age) across the grids (where the standard deviation $\sigma$ represents model-dependent systematics rather than formal measurement errors) and the fractional grid to grid offsets. For a quantity $X$ (mass or age), we define the fractional offset as $(X_{\rm max}-X_{\rm min})/{X_{\rm max}}\times 100$\%, where $X_{\rm max}$ and $X_{\rm min}$ are the maximum and minimum values returned by the evolutionary grids.
Such an analysis of the grid-to-grid scatter provides insight into the reliability of different regions of stellar parameter space.  We identify several consistent patterns that emerged across the CHARA sample to note:

\begin{table*}
\centering
\scriptsize
\begin{threeparttable}
\caption{\texttt{Kiauhoku} results for the stars in our sample.}
\label{tab:model_results}

\begin{tabular}{lcc|cc|cc|cc|cccccc}
\toprule
 & 
\multicolumn{2}{c}{\textbf{YREC}}  & 
\multicolumn{2}{c}{\textbf{MIST}} &  
\multicolumn{2}{c}{\textbf{Dartmouth}} &  
\multicolumn{2}{c}{\textbf{Garstec}} &  
\multicolumn{4}{c}{ }  &
\multicolumn{1}{c}{\textbf{Fractional}}  &
\multicolumn{1}{c}{\textbf{Fractional}}           \\
\textbf{Star} & 
\multicolumn{1}{c}{\textbf{Mass}} & 
\multicolumn{1}{c}{\textbf{Age}} & 
\multicolumn{1}{c}{\textbf{Mass}} & 
\multicolumn{1}{c}{\textbf{Age}} & 
\multicolumn{1}{c}{\textbf{Mass}} & 
\multicolumn{1}{c}{\textbf{Age}} & 
\multicolumn{1}{c}{\textbf{Mass}} & 
\multicolumn{1}{c}{\textbf{Age}} & 
\multicolumn{1}{c}{\boldmath{$\overline{\rm{Mass}}$}} & 
\multicolumn{1}{c}{\boldmath{$\sigma$}\textbf{Mass}} & 
\multicolumn{1}{c}{\boldmath{$\overline{\rm{Age}}$} } & 
\multicolumn{1}{c}{\boldmath{$\sigma$}\textbf{Age}} & 
\multicolumn{1}{c}{\textbf{Offset}} & 
\multicolumn{1}{c}{\textbf{Offset}} 
       \\
       
\textbf{Name} & 
\multicolumn{1}{c}{\boldmath{(M$_{\odot}$)}} & 
\multicolumn{1}{c}{\textbf{(Gyr)}} & 
\multicolumn{1}{c}{\boldmath{(M$_{\odot}$)}} & 
\multicolumn{1}{c}{\textbf{(Gyr)}} & 
\multicolumn{1}{c}{\boldmath{(M$_{\odot}$)}} & 
\multicolumn{1}{c}{\textbf{(Gyr)}} & 
\multicolumn{1}{c}{\boldmath{(M$_{\odot}$)}} & 
\multicolumn{1}{c}{\textbf{(Gyr)}} & 
\multicolumn{1}{c}{\boldmath{(M$_{\odot}$)}} & 
\multicolumn{1}{c}{\boldmath{(M$_{\odot}$)}} & 
\multicolumn{1}{c}{\textbf{(Gyr)}} & 
\multicolumn{1}{c}{\textbf{(Gyr)}} & 
\multicolumn{1}{c}{\textbf{Mass}} & 
\multicolumn{1}{c}{\textbf{Age}}   \\

\hline
\midrule

HD 1461 & 0.943 & 13.056 & 0.975 & \phn9.329 & 0.986 & \phn8.754 & 0.978 & \phn9.659 & 0.971 & 0.019 & 10.200 & 1.941 & \phn4.3\% & 32.9\% \\
HD 8574 & 1.016 & 10.553 & 1.009 & \phn9.578 & 1.036 & \phn9.263 & 1.031 & \phn9.528 & 1.023 & 0.012 & \phn9.731 & 0.565 & \phn2.6\% & 12.2\% \\
HD 8673 & 1.301 & \phn2.636 & 1.297 & \phn2.317 & 1.307 & \phn2.292 & 1.315 & \phn2.183 & 1.305 & 0.008 & \phn2.357 & 0.195 & \phn1.3\% & 17.2\% \\
HD 10697 & 1.108 & \phn8.176 & 1.091 & \phn8.067 & 1.112 & \phn7.546 & 1.112 & \phn7.797 & 1.106 & 0.010 & \phn7.896 & 0.283 & \phn1.9\% & \phn7.7\% \\
HD 16141 & 1.028 & 10.728 & 1.033 & \phn9.407 & 1.049 & \phn8.906 & 1.045 & \phn9.480 & 1.039 & 0.010 & \phn9.630 & 0.775 & \phn2.0\% & 17.0\% \\
HD 20367 & 1.059 & \phn5.982 & 1.076 & \phn4.297 & 1.090 & \phn4.108 & 1.089 & \phn4.189 & 1.079 & 0.015 & \phn4.644 & 0.895 & \phn2.8\% & 31.3\% \\
HD 22879 & 0.673 & 29.217 & 0.668 & 25.251 & 0.700 & 22.547 & 0.704 & 21.609 & 0.686 & 0.018 & 24.656 & 3.410 & \phn5.0\% & 26.0\% \\
HD 30562 & 1.230 & \phn5.139 & 1.228 & \phn4.476 & 1.163 & \phn5.619 & 1.242 & \phn4.515 & 1.216 & 0.035 & \phn4.937 & 0.547 & \phn6.3\% & 20.3\% \\
HD 42807 & 1.012 & \phn0.263 & 1.012 & \phn0.041 & 1.029 & \phn0.039 & 1.026 & \phn0.044 & 1.020 & 0.009 & \phn0.097 & 0.111 & \phn1.7\% & 85.0\% \\
HD 58551 & 0.951 & \phn6.746 & 0.924 & \phn7.204 & 0.942 & \phn6.819 & 0.976 & \phn5.347 & 0.948 & 0.022 & \phn6.529 & 0.814 & \phn5.4\% & 25.8\% \\
HD 76932 & 0.745 & 19.650 & 0.735 & 17.945 & 0.755 & 17.397 & 0.769 & 16.007 & 0.751 & 0.014 & 17.749 & 1.507 & \phn4.4\% & 18.5\% \\
HD 108510 & 1.025 & \phn6.371 & 1.035 & \phn4.722 & 1.058 & \phn4.343 & 1.063 & \phn4.050 & 1.045 & 0.018 & \phn4.872 & 1.037 & \phn3.5\% & 36.4\% \\
HD 127334 & 0.927 & 15.421 & 0.955 & 11.688 & 0.966 & 11.038 & 0.958 & 12.118 & 0.951 & 0.017 & 12.566 & 1.954 & \phn4.0\% & 28.4\% \\
HD 144579 & 0.608 & 34.671 & 0.600 & 31.240 & 0.600 & 36.220 & 0.600 & 36.044 & 0.602 & 0.004 & 34.544 & 2.309 & \phn1.5\% & 13.7\% \\
HD 154345 & 0.889 & \phn4.132 & 0.882 & \phn3.592 & 0.904 & \phn2.580 & 0.918 & \phn1.198 & 0.898 & 0.016 & \phn2.875 & 1.290 & \phn3.9\% & 71.0\% \\
HD 164922 & 0.827 & 21.306 & 0.865 & 13.757 & 0.877 & 12.723 & 0.868 & 14.269 & 0.859 & 0.022 & 15.514 & 3.915 & \phn5.7\% & 40.3\% \\
HD 181096 & 1.192 & \phn4.787 & 1.166 & \phn4.792 & 1.186 & \phn4.795 & 1.204 & \phn4.465 & 1.187 & 0.016 & \phn4.710 & 0.163 & \phn3.1\% & \phn6.9\% \\
HD 182488 & 0.903 & 10.047 & 0.948 & \phn4.105 & 0.943 & \phn4.753 & 0.952 & \phn4.170 & 0.937 & 0.023 & \phn5.768 & 2.867 & \phn5.2\% & 59.1\% \\
HD 184499 & 0.804 & 16.894 & 0.792 & 15.748 & 0.809 & 15.259 & 0.821 & 14.813 & 0.806 & 0.012 & 15.678 & 0.896 & \phn3.5\% & 12.3\% \\
HD 185269 & 1.268 & \phn4.940 & 1.254 & \phn4.752 & 1.273 & \phn4.538 & 1.277 & \phn4.619 & 1.268 & 0.010 & \phn4.712 & 0.175 & \phn1.8\% & \phn8.1\% \\
HD 190228 & 1.063 & \phn8.056 & 1.059 & \phn7.354 & 1.064 & \phn7.714 & 1.050 & \phn8.233 & 1.059 & 0.006 & \phn7.839 & 0.389 & \phn1.3\% & 10.7\% \\
HD 193664 & 0.964 & \phn6.618 & 0.963 & \phn5.608 & 0.985 & \phn5.078 & 0.994 & \phn4.611 & 0.977 & 0.015 & \phn5.479 & 0.862 & \phn3.1\% & 30.3\% \\
HD 195019 & 0.999 & 11.394 & 0.994 & 10.400 & 1.018 & 10.003 & 1.013 & 10.305 & 1.006 & 0.011 & 10.525 & 0.603 & \phn2.3\% & 12.2\% \\
HD 196885 & 1.218 & \phn4.329 & 1.223 & \phn3.529 & 1.230 & \phn3.468 & 1.233 & \phn3.573 & 1.226 & 0.007 & \phn3.724 & 0.405 & \phn1.2\% & 19.9\% \\
HD 202568 & 0.915 & 15.019 & 0.993 & 10.009 & 0.939 & 13.156 & 0.859 & 18.256 & 0.926 & 0.056 & 14.110 & 3.452 & 13.5\% & 45.2\% \\
HD 210277 & 0.859 & 20.571 & 0.896 & 14.361 & 0.905 & 13.387 & 0.893 & 15.150 & 0.888 & 0.020 & 15.867 & 3.218 & \phn5.2\% & 34.9\% \\
HD 222155 & 1.034 & \phn9.206 & 1.007 & \phn9.096 & 1.032 & \phn8.893 & 1.035 & \phn8.862 & 1.027 & 0.013 & \phn9.014 & 0.164 & \phn2.7\% & \phn3.7\% \\

\bottomrule
\end{tabular}

\vspace{4pt}
\begin{tablenotes}[flushleft]
\footnotesize
\item[]\textit{Notes:} Results for HD~22879, HD~58551, HD~76932, HD~144579, and HD~184499 are likely unreliable; some grids for HD~42807 and HD~154345 did not converge.  Fractional offsets are defined as $(X_{\rm max}-X_{\rm min})/X_{\rm max}$ and are expressed as percentages. See \S\ref{s:models} for details.
\end{tablenotes}

\end{threeparttable}
\end{table*}

\begin{itemize}
    \item \textbf{Model-to-model mass agreement:} We find that inferred stellar masses agree to within $\sim6$\% across all four grids for the majority of stars. This level of agreement reflects a convergence in evolutionary predictions for main-sequence and mildly evolved subgiant stars, particularly in the range of solar-type stars, and supports the use of 1-D models calibrated on solar analogs to estimate stellar masses with moderate confidence. These mass constraints in turn enhance the precision of derived surface gravities, a critical parameter for exoplanetary transit modeling. In contrast, inferred stellar ages exhibit a much broader dispersion, and should be interpreted as plausible ranges, rather than precise estimates.

    \item \textbf{Metal-poor stars:}  Four of the five metal-poor stars in our sample (HD 22879, HD 76932, HD 144579, and HD 184499) yield ages that exceed the age of the Universe in \textit{all four} grids. These extreme values underscore the challenges evolutionary models face at even moderately low metallicity, where differences in boundary conditions, convective efficiency, and diffusion treatments become especially consequential.  Moreover, increased $\alpha$-element enhancement, which typically tracks with reduction of metallicity, is also ignored in all model grids used in this analysis. To that end, we note that this set of four metal-poor stars has moderate enrichment ([$\alpha$/Fe] ranging from 0.29 to 0.33; \citealt{soubiran05}). On the other hand, the sole metal-poor star in our sample with model ages $<14$~Gyr, HD 58551, has an $\alpha$-element enhancement roughly half of the others ([$\alpha$/Fe]$=0.16$; \citealt{soubiran05}), further supporting the shortcomings in model outcomes for metal-poor, $\alpha$-element enhanced stars.

    \item \textbf{Low-mass stars:} Stellar evolutionary models are quite insensitive to age for stars less than $\sim0.8$~M$_{\odot}$ as such stars evolve very slowly on the main-sequence. In our analysis, we do not propagate measurement uncertainties when interpolating stellar parameters onto the model grids to estimate errors on the model-derived values (our aim in this work is to investigate grid-to-grid differences). If we did, the age uncertainties for stars in this mass range would be of order the inferred ages themselves (i.e., fractional uncertainties of $\sim100\%$ or greater), in some cases leaving the ages effectively unconstrained over the age of the Universe \citep[e.g.,][]{angus19}. This reflects the slow evolution of low-mass main-sequence stars and the closely spaced, nearly overlapping isochrones in this region of the H-R diagram. Also to note, the model grids used here do not extend below $0.6$~M$_{\odot}$.  HD 144579 lies close to this lower boundary, and should therefore be interpreted with caution, as edge effects and sparse grid coverage can further reduce the reliability of its mass and age estimates.

    \item \textbf{Model convergence with respect to stellar masses:} For the stars in our sample more massive than 0.8~M$_{\odot}$, we find that the fractional grid-to-grid offsets in both mass and age decrease with increasing stellar mass. For example, the typical age fractional offsets drop from $\sim35$\% to $\sim10$\%, and typical mass fractional offsets drop from $\sim4.5$\% to $<2$\% when comparing stars with mean model mass $\sim0.9$~M$_{\odot}$ and 1.3~M$_{\odot}$.  This is illustrated in Figure~\ref{fig:mass_age_offsets}, where we show these offsets plotted against mean model mass for each star, and the results from robust linear fits  plotted as dotted (for age offset) and dashed (for mass offset) lines.  This convergence suggests that, within the temperature-luminosity range probed here, higher-mass main-sequence stars occupy regions of the models that are less sensitive to differences in input physics, likely because they are farther from rapid evolutionary phases where small changes in assumptions within the chosen evolutionary model can yield large differences in inferred properties.

\begin{figure}
    \centering
    \includegraphics[width=1.0\linewidth]{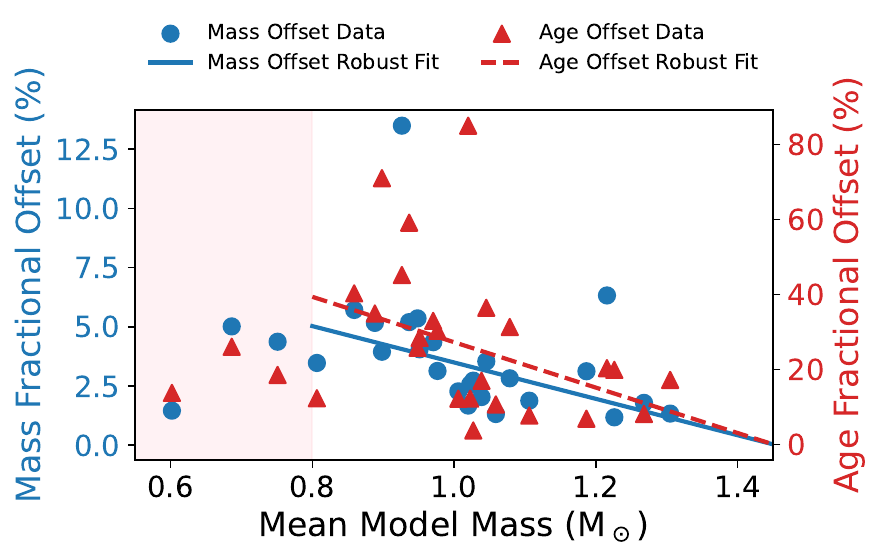}
    \caption{Mass (blue circles, left axis) and age (red triangles, right axis) fractional offsets between stellar model grids as a function of mean model mass (Table~\ref{tab:model_results}). The shaded region marks stars with mean mass $\overline{\rm{Mass}}<0.8$~M$_{\odot}$, where stellar evolutionary models are relatively insensitive to age. Solid and dashed lines show robust linear fits to the mass and age offsets, respectively, computed using only stars with mean mass $\overline{\rm{Mass}}>0.8$~M$_{\odot}$.}
    \label{fig:mass_age_offsets}
\end{figure}

    \item \textbf{Young stars:} We find that some grids were unable to converge for HD 42807 and HD 154345, which are also the stars that have the highest fractional age offset of 85\% and 71\%, respectively (Table~\ref{tab:model_results}). These stars fall near the zero-age main sequence (ZAMS), and consequentially reach the minimum available equivalent evolutionary point in the model grids, preventing convergence and a reliable age estimate. While, for instance, a young age for HD 42807 is supported by age estimates from stellar rotation, lithium abundance, and chromospheric activity \citep[e.g.,][]{ramirez12, lorenzo19, stanford20, shej24}, careful attention to edge cases such as these highlight the sensitivity of young star fits to a evolutionary model's choice boundary conditions. A closer look shows that the mass estimates for both stars are fairly consistent across all four grids, with the absolute ages being uniformly young. 
    The extreme fractional offsets arise only because modest absolute differences of a few hundred Myr are amplified when the denominator is small. No single grid stands out as systematically worse, but the near-vertical structure of ZAMS tracks makes all models equally fragile in this regime. Thus, the inflated fractional offsets presented here illustrate the inherent insensitivity of evolutionary tracks to ages at early main-sequence stages, rather than genuine model-to-model disagreement.

    \item \textbf{Subgiant stars:} Two stars in our sample, HD 190228 and HD 202568, are moderately evolved off the main-sequence (marked in Figure~\ref{fig:pavo_hrd}). We note that model results for these two are strikingly different:  HD 202568 shows the highest fractional mass offset of 13.5\%, as well as a large fractional age offset (45.2\%), whereas HD 190228 has nearly the lowest statistics of our entire sample (fractional offset for mass and age are 1.3\% and 10.7\%, respectively).  The differences in model results for these two subgiants arises from the fact that HD 202568 (more luminous of the two) has reached the bottom of the giant branch evolutionary phase, where the model tracks are compact and overlapping, making them hard to distinguish between, whereas HD 190228 solidly resides on the subgiant branch, where the short evolutionary timescales enable precise characterization. 
\end{itemize}

\section{Conclusion}\label{s:conclusion}

In this work, we have presented high-angular-resolution interferometric measurements for 27 nearby solar-type stars using the PAVO beam combiner at the CHARA Array. By combining uniform-disk and limb-darkened angular diameters with bolometric fluxes and Gaia parallaxes, we have derived precise, model-independent stellar radii (typical uncertainties $\sim1$\%) and effective temperatures (typical uncertainties  of 1.5\%). These measurements populate a range of metallicities and evolutionary states, including several exoplanet host stars and a subset of metal-poor, $\alpha$-enhanced stars.

Comparisons with multiple stellar evolutionary model grids reveal generally consistent stellar mass estimates (agreement within $\sim6$\% for most stars), but larger discrepancies in inferred ages, particularly for metal-poor stars and for stars at evolutionary phases where model tracks are tightly clustered. These results highlight both the strengths and current limitations of widely used model grids, emphasizing the need for expanded grids that include $\alpha$-element enhancement and better treatments of boundary conditions. Our empirical measurements provide robust calibration points for stellar evolutionary theory, surface brightness-color relations, and mass-radius-temperature relations. They also contribute directly to the improved characterization of exoplanetary systems, where stellar radius and mass uncertainties propagate to planet properties.

Future work will revisit these comparisons using the updated MIST models of \citet{dotter26}, which incorporate $\alpha$-enhanced compositions self-consistently in both the stellar interiors and synthetic spectra. Applying these models to the metal-poor, $\alpha$-enhanced stars in our sample will test whether explicitly accounting for [$\alpha$/Fe] reduces the unphysically large inferred ages and grid-to-grid discrepancies identified here, while providing a more stringent assessment of model systematics across metallicity and evolutionary state.


\section*{Acknowledgments}
T.S.B. and A.E. acknowledge support by the National Science Foundation under Grant No. AST-2205914.

This work is based upon observations obtained with the Georgia State University Center for High Angular Resolution Astronomy Array at Mount Wilson Observatory.  The CHARA Array is supported by the National Science Foundation under Grant No. AST-2034336 and AST-2407956. Institutional support has been provided from the GSU College of Arts and Sciences, GSU Office of the Vice President for Research and Economic Development, and GSU Office of the Provost.

CHARA telescope time was granted by NOIRLab through the Mid-Scale Innovations Program (MSIP). MSIP is funded by NSF. Access was granted through the NOIRLab community-access programs NOIRlab proposal (NOIRLab PropID: 2018A-0210; PI: T. Ellis).

This research has made use of the SIMBAD database,
operated at CDS, Strasbourg, France.

This research has made use of the VizieR catalogue access tool, CDS,
Strasbourg, France \citep{10.26093/cds/vizier}. The original description 
of the VizieR service was published in \citet{vizier2000}.

This research has made use of the Jean-Marie Mariotti Center \texttt{SearchCal} service, which involves the JSDC and JMDC catalogues \citep{chelli16}.
\footnote{Available at https://www.jmmc.fr/searchcal}. This research has made use of the Jean-Marie Mariotti Center \texttt{Aspro} service \citep{bourges16}.
\footnote{Available at http://www.jmmc.fr/aspro}

	
	


\bibliographystyle{apsrev4-1}

\bibliography{paper}

@ARTICLE{nexsci,
       author = {{Christiansen}, Jessie L. and {McElroy}, Douglas L. and {Harbut}, Marcy and {Ciardi}, David R. and {Crane}, Megan and {Good}, John and {Hardegree-Ullman}, Kevin K. and {Kesseli}, Aurora Y. and {Lund}, Michael B. and {Lynn}, Meca and {Muthiar}, Ananda and {Nilsson}, Ricky and {Oluyide}, Toba and {Papin}, Michael and {Rivera}, Amalia and {Swain}, Melanie and {Susemiehl}, Nicholas D. and {Tam}, Raymond and {van Eyken}, Julian and {Beichman}, Charles},
        title = "{The NASA Exoplanet Archive and Exoplanet Follow-up Observing Program: Data, Tools, and Usage}",
      journal = {arXiv e-prints},
         year = 2025,
        month = jun,
          eid = {arXiv:2506.03299},
        pages = {arXiv:2506.03299},
          doi = {10.48550/arXiv.2506.03299},
archivePrefix = {arXiv},
       eprint = {2506.03299},
 primaryClass = {astro-ph.EP},
       adsurl = {https://ui.adsabs.harvard.edu/abs/2025arXiv250603299C}
}

@ARTICLE{angus19,
       author = {{Angus}, Ruth and {Morton}, Timothy D. and {Foreman-Mackey}, Daniel and {van Saders}, Jennifer and {Curtis}, Jason and {Kane}, Stephen R. and {Bedell}, Megan and {Kiman}, Rocio and {Hogg}, David W. and {Brewer}, John},
        title = "{Toward Precise Stellar Ages: Combining Isochrone Fitting with Empirical Gyrochronology}",
      journal = {\aj},
         year = 2019,
        month = nov,
       volume = {158},
       number = {5},
          eid = {173},
        pages = {173},
          doi = {10.3847/1538-3881/ab3c53},
archivePrefix = {arXiv},
       eprint = {1908.07528},
 primaryClass = {astro-ph.SR},
       adsurl = {https://ui.adsabs.harvard.edu/abs/2019AJ....158..173A}
}

@ARTICLE{simbad,
       author = {{Wenger}, M. and {Ochsenbein}, F. and {Egret}, D. and {Dubois}, P. and {Bonnarel}, F. and {Borde}, S. and {Genova}, F. and {Jasniewicz}, G. and {Lalo{\"e}}, S. and {Lesteven}, S. and {Monier}, R.},
        title = "{The SIMBAD astronomical database. The CDS reference database for astronomical objects}",
      journal = {\aaps},
         year = 2000,
        month = apr,
       volume = {143},
        pages = {9-22},
          doi = {10.1051/aas:2000332},
archivePrefix = {arXiv},
       eprint = {astro-ph/0002110},
 primaryClass = {astro-ph},
       adsurl = {https://ui.adsabs.harvard.edu/abs/2000A&AS..143....9W}
}

@ARTICLE{vanbelle09,
       author = {{van Belle}, Gerard T. and {von Braun}, Kaspar},
        title = "{Directly Determined Linear Radii and Effective Temperatures of Exoplanet Host Stars}",
      journal = {\apj},
         year = 2009,
        month = apr,
       volume = {694},
       number = {2},
        pages = {1085-1098},
          doi = {10.1088/0004-637X/694/2/1085},
archivePrefix = {arXiv},
       eprint = {0901.1206},
 primaryClass = {astro-ph.SR},
       adsurl = {https://ui.adsabs.harvard.edu/abs/2009ApJ...694.1085V}
}

@ARTICLE{ker04,
       author = {{Kervella}, P. and {Th{\'e}venin}, F. and {Di Folco}, E. and {S{\'e}gransan}, D.},
        title = "{The angular sizes of dwarf stars and subgiants. Surface brightness relations calibrated by interferometry}",
      journal = {\aap},
         year = 2004,
        month = oct,
       volume = {426},
        pages = {297-307},
          doi = {10.1051/0004-6361:20035930},
archivePrefix = {arXiv},
       eprint = {astro-ph/0404180},
 primaryClass = {astro-ph},
       adsurl = {https://ui.adsabs.harvard.edu/abs/2004A&A...426..297K}
}

@ARTICLE{creevey12,
       author = {{Creevey}, O.~L. and {Th{\'e}venin}, F. and {Boyajian}, T.~S. and {Kervella}, P. and {Chiavassa}, A. and {Bigot}, L. and {M{\'e}rand}, A. and {Heiter}, U. and {Morel}, P. and {Pichon}, B. and {Mc Alister}, H.~A. and {ten Brummelaar}, T.~A. and {Collet}, R. and {van Belle}, G.~T. and {Coud{\'e} du Foresto}, V. and {Farrington}, C. and {Goldfinger}, P.~J. and {Sturmann}, J. and {Sturmann}, L. and {Turner}, N.},
        title = "{Fundamental properties of the Population II fiducial stars <ASTROBJ>HD 122563</ASTROBJ> and <ASTROBJ>Gmb 1830</ASTROBJ> from CHARA interferometric observations}",
      journal = {\aap},
         year = 2012,
        month = sep,
       volume = {545},
          eid = {A17},
        pages = {A17},
          doi = {10.1051/0004-6361/201219651},
archivePrefix = {arXiv},
       eprint = {1207.5954},
 primaryClass = {astro-ph.SR},
       adsurl = {https://ui.adsabs.harvard.edu/abs/2012A&A...545A..17C}
}

@ARTICLE{boy08,
       author = {{Boyajian}, Tabetha S. and {McAlister}, Harold A. and {Baines}, Ellyn K. and {Gies}, Douglas R. and {Henry}, Todd and {Jao}, Wei-Chun and {O'Brien}, David and {Raghavan}, Deepak and {Touhami}, Yamina and {ten Brummelaar}, Theo A. and {Farrington}, Chris and {Goldfinger}, P.~J. and {Sturmann}, Laszlo and {Sturmann}, Judit and {Turner}, Nils H. and {Ridgway}, Stephen},
        title = "{Angular Diameters of the G Subdwarf {\ensuremath{\mu}} Cassiopeiae A and the K Dwarfs {\ensuremath{\sigma}} Draconis and HR 511 from Interferometric Measurements with the CHARA Array}",
      journal = {\apj},
         year = 2008,
        month = aug,
       volume = {683},
       number = {1},
        pages = {424-432},
          doi = {10.1086/589554},
archivePrefix = {arXiv},
       eprint = {0804.2719},
 primaryClass = {astro-ph},
       adsurl = {https://ui.adsabs.harvard.edu/abs/2008ApJ...683..424B}
}

@ARTICLE{karo18,
       author = {{Karovicova}, I. and {White}, T.~R. and {Nordlander}, T. and {Lind}, K. and {Casagrande}, L. and {Ireland}, M.~J. and {Huber}, D. and {Creevey}, O. and {Mourard}, D. and {Schaefer}, G.~H. and {Gilmore}, G. and {Chiavassa}, A. and {Wittkowski}, M. and {Jofr{\'e}}, P. and {Heiter}, U. and {Th{\'e}venin}, F. and {Asplund}, M.},
        title = "{Accurate effective temperatures of the metal-poor benchmark stars HD 140283, HD 122563, and HD 103095 from CHARA interferometry}",
      journal = {\mnras},
         year = 2018,
        month = mar,
       volume = {475},
       number = {1},
        pages = {L81-L85},
          doi = {10.1093/mnrasl/sly010},
archivePrefix = {arXiv},
       eprint = {1801.03274},
 primaryClass = {astro-ph.SR},
       adsurl = {https://ui.adsabs.harvard.edu/abs/2018MNRAS.475L..81K}
}

@ARTICLE{dotter26,
       author = {{Dotter}, Aaron and {Bauer}, Evan B. and {Park}, Minjung and {Conroy}, Charlie and {Milone}, Antonino P. and {Joyce}, Meridith and {Cantiello}, Matteo},
        title = "{MESA Isochrones and Stellar Tracks (MIST). II. Models with {\ensuremath{\alpha}}-enhanced Chemical Composition}",
      journal = {\apjs},
         year = 2026,
        month = apr,
       volume = {283},
       number = {2},
          eid = {64},
        pages = {64},
          doi = {10.3847/1538-4365/ae48f3},
archivePrefix = {arXiv},
       eprint = {2602.22012},
 primaryClass = {astro-ph.SR},
       adsurl = {https://ui.adsabs.harvard.edu/abs/2026ApJS..283...64D}
}

@ARTICLE{boh14,
       author = {{Bohlin}, Ralph C. and {Gordon}, Karl D. and {Tremblay}, P.-E.},
        title = "{Techniques and Review of Absolute Flux Calibration from the Ultraviolet to the Mid-Infrared}",
      journal = {\pasp},
         year = 2014,
        month = aug,
       volume = {126},
       number = {942},
        pages = {711},
          doi = {10.1086/677655},
archivePrefix = {arXiv},
       eprint = {1406.1707},
 primaryClass = {astro-ph.IM},
       adsurl = {https://ui.adsabs.harvard.edu/abs/2014PASP..126..711B}
}

@software{sedfit,
	author = {{Kounkel}, Marina},
	doi = {10.5281/zenodo.8076500},
	month = jun,
	publisher = {Zenodo},
	title = {SEDFit},
	url = {https://doi.org/10.5281/zenodo.8076500},
	year = 2023}

@ARTICLE{karo22a,
       author = {{Karovicova}, I. and {White}, T.~R. and {Nordlander}, T. and {Casagrande}, L. and {Ireland}, M. and {Huber}, D.},
        title = "{Fundamental stellar parameters of benchmark stars from CHARA interferometry. II. Dwarf stars}",
      journal = {\aap},
         year = 2022,
        month = feb,
       volume = {658},
          eid = {A47},
        pages = {A47},
          doi = {10.1051/0004-6361/202141833},
archivePrefix = {arXiv},
       eprint = {2109.06203},
 primaryClass = {astro-ph.SR},
       adsurl = {https://ui.adsabs.harvard.edu/abs/2022A&A...658A..47K}
}

@ARTICLE{karo22b,
       author = {{Karovicova}, I. and {White}, T.~R. and {Nordlander}, T. and {Casagrande}, L. and {Ireland}, M. and {Huber}, D.},
        title = "{Fundamental stellar parameters of benchmark stars from CHARA interferometry. III. Giant and subgiant stars}",
      journal = {\aap},
         year = 2022,
        month = feb,
       volume = {658},
          eid = {A48},
        pages = {A48},
          doi = {10.1051/0004-6361/202142100},
archivePrefix = {arXiv},
       eprint = {2109.13258},
 primaryClass = {astro-ph.SR},
       adsurl = {https://ui.adsabs.harvard.edu/abs/2022A&A...658A..48K}
}

@ARTICLE{demory09,
       author = {{Demory}, B. -O. and {S{\'e}gransan}, D. and {Forveille}, T. and {Queloz}, D. and {Beuzit}, J. -L. and {Delfosse}, X. and {di Folco}, E. and {Kervella}, P. and {Le Bouquin}, J. -B. and {Perrier}, C. and {Benisty}, M. and {Duvert}, G. and {Hofmann}, K. -H. and {Lopez}, B. and {Petrov}, R.},
        title = "{Mass-radius relation of low and very low-mass stars revisited with the VLTI}",
      journal = {\aap},
         year = 2009,
        month = oct,
       volume = {505},
       number = {1},
        pages = {205-215},
          doi = {10.1051/0004-6361/200911976},
archivePrefix = {arXiv},
       eprint = {0906.0602},
 primaryClass = {astro-ph.SR},
       adsurl = {https://ui.adsabs.harvard.edu/abs/2009A&A...505..205D}
}

@INPROCEEDINGS{white15,
       author = {{White}, T.~R. and {Silva Aguirre}, V. and {Boyajian}, T. and {Creevey}, O. and {Huber}, D. and {von Braun}, K. and {Bedding}, T.~R. and {Elsworth}, Y. and {Hekker}, S. and {Stello}, D. and {Weiss}, A.},
        title = "{Testing Asteroseismic Scaling Relations with Interferometry}",
    booktitle = {European Physical Journal Web of Conferences},
         year = 2015,
       series = {European Physical Journal Web of Conferences},
       volume = {101},
        month = sep,
          eid = {06068},
        pages = {06068},
          doi = {10.1051/epjconf/201510106068},
       adsurl = {https://ui.adsabs.harvard.edu/abs/2015EPJWC.10106068W}
}

@ARTICLE{huber12,
       author = {{Huber}, D. and {Ireland}, M.~J. and {Bedding}, T.~R. and {Brand{\~a}o}, I.~M. and {Piau}, L. and {Maestro}, V. and {White}, T.~R. and {Bruntt}, H. and {Casagrande}, L. and {Molenda-{\.Z}akowicz}, J. and {Silva Aguirre}, V. and {Sousa}, S.~G. and {Barclay}, T. and {Burke}, C.~J. and {Chaplin}, W.~J. and {Christensen-Dalsgaard}, J. and {Cunha}, M.~S. and {De Ridder}, J. and {Farrington}, C.~D. and {Frasca}, A. and {Garc{\'\i}a}, R.~A. and {Gilliland}, R.~L. and {Goldfinger}, P.~J. and {Hekker}, S. and {Kawaler}, S.~D. and {Kjeldsen}, H. and {McAlister}, H.~A. and {Metcalfe}, T.~S. and {Miglio}, A. and {Monteiro}, M.~J.~P.~F.~G. and {Pinsonneault}, M.~H. and {Schaefer}, G.~H. and {Stello}, D. and {Stumpe}, M.~C. and {Sturmann}, J. and {Sturmann}, L. and {ten Brummelaar}, T.~A. and {Thompson}, M.~J. and {Turner}, N. and {Uytterhoeven}, K.},
        title = "{Fundamental Properties of Stars Using Asteroseismology from Kepler and CoRoT and Interferometry from the CHARA Array}",
      journal = {\apj},
         year = 2012,
        month = nov,
       volume = {760},
       number = {1},
          eid = {32},
        pages = {32},
          doi = {10.1088/0004-637X/760/1/32},
archivePrefix = {arXiv},
       eprint = {1210.0012},
 primaryClass = {astro-ph.SR},
       adsurl = {https://ui.adsabs.harvard.edu/abs/2012ApJ...760...32H}
}

@ARTICLE{kane17,
       author = {{Kane}, Stephen R. and {von Braun}, Kaspar and {Henry}, Gregory W. and {Waters}, Miranda A. and {Boyajian}, Tabetha S. and {Mann}, Andrew W.},
        title = "{Characterization of the Wolf 1061 Planetary System}",
      journal = {\apj},
         year = 2017,
        month = feb,
       volume = {835},
       number = {2},
          eid = {200},
        pages = {200},
          doi = {10.3847/1538-4357/835/2/200},
archivePrefix = {arXiv},
       eprint = {1612.09324},
 primaryClass = {astro-ph.EP},
       adsurl = {https://ui.adsabs.harvard.edu/abs/2017ApJ...835..200K}
}

@ARTICLE{boy15,
       author = {{Boyajian}, Tabetha and {von Braun}, Kaspar and {Feiden}, Gregory A. and {Huber}, Daniel and {Basu}, Sarbani and {Demarque}, Pierre and {Fischer}, Debra A. and {Schaefer}, Gail and {Mann}, Andrew W. and {White}, Timothy R. and {Maestro}, Vicente and {Brewer}, John and {Lamell}, C. Brooke and {Spada}, Federico and {L{\'o}pez-Morales}, Mercedes and {Ireland}, Michael and {Farrington}, Chris and {van Belle}, Gerard T. and {Kane}, Stephen R. and {Jones}, Jeremy and {ten Brummelaar}, Theo A. and {Ciardi}, David R. and {McAlister}, Harold A. and {Ridgway}, Stephen and {Goldfinger}, P.~J. and {Turner}, Nils H. and {Sturmann}, Laszlo},
        title = "{Stellar diameters and temperatures - VI. High angular resolution measurements of the transiting exoplanet host stars HD 189733 and HD 209458 and implications for models of cool dwarfs}",
      journal = {\mnras},
         year = 2015,
        month = feb,
       volume = {447},
       number = {1},
        pages = {846-857},
          doi = {10.1093/mnras/stu2502},
archivePrefix = {arXiv},
       eprint = {1411.5638},
 primaryClass = {astro-ph.SR},
       adsurl = {https://ui.adsabs.harvard.edu/abs/2015MNRAS.447..846B}
}

@ARTICLE{caba22,
       author = {{Caballero}, J.~A. and {Gonz{\'a}lez-{\'A}lvarez}, E. and {Brady}, M. and {Trifonov}, T. and {Ellis}, T.~G. and {Dorn}, C. and {Cifuentes}, C. and {Molaverdikhani}, K. and {Bean}, J.~L. and {Boyajian}, T. and {Rodr{\'\i}guez}, E. and {Sanz-Forcada}, J. and {Zapatero Osorio}, M.~R. and {Abia}, C. and {Amado}, P.~J. and {Anugu}, N. and {B{\'e}jar}, V.~J.~S. and {Davies}, C.~L. and {Dreizler}, S. and {Dubois}, F. and {Ennis}, J. and {Espinoza}, N. and {Farrington}, C.~D. and {L{\'o}pez}, A. Garc{\'\i}a and {Gardner}, T. and {Hatzes}, A.~P. and {Henning}, Th. and {Herrero}, E. and {Herrero-Cisneros}, E. and {Kaminski}, A. and {Kasper}, D. and {Klement}, R. and {Kraus}, S. and {Labdon}, A. and {Lanthermann}, C. and {Le Bouquin}, J. -B. and {L{\'o}pez Gonz{\'a}lez}, M.~J. and {Luque}, R. and {Mann}, A.~W. and {Marfil}, E. and {Monnier}, J.~D. and {Montes}, D. and {Morales}, J.~C. and {Pall{\'e}}, E. and {Pedraz}, S. and {Quirrenbach}, A. and {Reffert}, S. and {Reiners}, A. and {Ribas}, I. and {Rodr{\'\i}guez-L{\'o}pez}, C. and {Schaefer}, G. and {Schweitzer}, A. and {Seifahrt}, A. and {Setterholm}, B.~R. and {Shan}, Y. and {Shulyak}, D. and {Solano}, E. and {Sreenivas}, K.~R. and {Stef{\'a}nsson}, G. and {St{\"u}rmer}, J. and {Tabernero}, H.~M. and {Tal-Or}, L. and {ten Brummelaar}, T. and {Vanaverbeke}, S. and {von Braun}, K. and {Youngblood}, A. and {Zechmeister}, M.},
        title = "{A detailed analysis of the Gl 486 planetary system}",
      journal = {\aap},
         year = 2022,
        month = sep,
       volume = {665},
          eid = {A120},
        pages = {A120},
          doi = {10.1051/0004-6361/202243548},
archivePrefix = {arXiv},
       eprint = {2206.09990},
 primaryClass = {astro-ph.EP},
       adsurl = {https://ui.adsabs.harvard.edu/abs/2022A&A...665A.120C}
}

@ARTICLE{ligi12,
       author = {{Ligi}, R. and {Mourard}, D. and {Lagrange}, A.~M. and {Perraut}, K. and {Boyajian}, T. and {B{\'e}rio}, Ph. and {Nardetto}, N. and {Tallon-Bosc}, I. and {McAlister}, H. and {ten Brummelaar}, T. and {Ridgway}, S. and {Sturmann}, J. and {Sturmann}, L. and {Turner}, N. and {Farrington}, C. and {Goldfinger}, P.~J.},
        title = "{A new interferometric study of four exoplanet host stars: {\ensuremath{\theta}} Cygni, 14 Andromedae, {\ensuremath{\upsilon}} Andromedae and 42 Draconis}",
      journal = {\aap},
         year = 2012,
        month = sep,
       volume = {545},
          eid = {A5},
        pages = {A5},
          doi = {10.1051/0004-6361/201219467},
archivePrefix = {arXiv},
       eprint = {1208.3895},
 primaryClass = {astro-ph.SR},
       adsurl = {https://ui.adsabs.harvard.edu/abs/2012A&A...545A...5L}
}

@ARTICLE{vonbraun12,
       author = {{von Braun}, Kaspar and {Boyajian}, Tabetha S. and {Kane}, Stephen R. and {Hebb}, Leslie and {van Belle}, Gerard T. and {Farrington}, Chris and {Ciardi}, David R. and {Knutson}, Heather A. and {ten Brummelaar}, Theo A. and {L{\'o}pez-Morales}, Mercedes and {McAlister}, Harold A. and {Schaefer}, Gail and {Ridgway}, Stephen and {Collier Cameron}, Andrew and {Goldfinger}, P.~J. and {Turner}, Nils H. and {Sturmann}, Laszlo and {Sturmann}, Judit},
        title = "{The GJ 436 System: Directly Determined Astrophysical Parameters of an M Dwarf and Implications for the Transiting Hot Neptune}",
      journal = {\apj},
         year = 2012,
        month = jul,
       volume = {753},
       number = {2},
          eid = {171},
        pages = {171},
          doi = {10.1088/0004-637X/753/2/171},
archivePrefix = {arXiv},
       eprint = {1202.0083},
 primaryClass = {astro-ph.EP},
       adsurl = {https://ui.adsabs.harvard.edu/abs/2012ApJ...753..171V}
}

@ARTICLE{elliott24,
       author = {{Elliott}, Ashley and {Boyajian}, Tabetha and {Ellis}, Tyler and {von Braun}, Kaspar and {Mann}, Andrew W. and {Schaefer}, Gail},
        title = "{Measuring the stellar and planetary parameters of the 51 Eridani system}",
      journal = {\pasa},
         year = 2024,
        month = sep,
       volume = {41},
          eid = {e043},
        pages = {e043},
          doi = {10.1017/pasa.2024.40},
archivePrefix = {arXiv},
       eprint = {2401.01468},
 primaryClass = {astro-ph.EP},
       adsurl = {https://ui.adsabs.harvard.edu/abs/2024PASA...41...43E}
}

@ARTICLE{chelli16,
       author = {{Chelli}, Alain and {Duvert}, Gilles and {Bourg{\`e}s}, Laurent and {Mella}, Guillaume and {Lafrasse}, Sylvain and {Bonneau}, Daniel and {Chesneau}, Olivier},
        title = "{Pseudomagnitudes and differential surface brightness: Application to the apparent diameter of stars}",
      journal = {\aap},
         year = 2016,
        month = may,
       volume = {589},
          eid = {A112},
        pages = {A112},
          doi = {10.1051/0004-6361/201527484},
archivePrefix = {arXiv},
       eprint = {1604.07700},
 primaryClass = {astro-ph.SR},
       adsurl = {https://ui.adsabs.harvard.edu/abs/2016A&A...589A.112C}
}

@INPROCEEDINGS{bourges16,
       author = {{Bourg{\`e}s}, L. and {Duvert}, G.},
        title = "{ASPRO2: get ready for VLTI's instruments GRAVITY and MATISSE}",
    booktitle = {Optical and Infrared Interferometry and Imaging V},
         year = 2016,
       editor = {{Malbet}, Fabien and {Creech-Eakman}, Michelle J. and {Tuthill}, Peter G.},
       series = {Society of Photo-Optical Instrumentation Engineers (SPIE) Conference Series},
       volume = {9907},
        month = aug,
          eid = {990711},
        pages = {990711},
          doi = {10.1117/12.2234426},
       adsurl = {https://ui.adsabs.harvard.edu/abs/2016SPIE.9907E..11B}
}

@ARTICLE{baines25,
       author = {{Baines}, Ellyn K. and {Clark}, James H. and {Kingsley}, Bradley I. and {Schmitt}, Henrique R. and {Stone}, Jordan M.},
        title = "{Vintage NPOI: New and Updated Angular Diameters for 145 Stars}",
      journal = {\aj},
         year = 2025,
        month = jun,
       volume = {169},
       number = {6},
          eid = {293},
        pages = {293},
          doi = {10.3847/1538-3881/adc930},
archivePrefix = {arXiv},
       eprint = {2506.02912},
 primaryClass = {astro-ph.SR},
       adsurl = {https://ui.adsabs.harvard.edu/abs/2025AJ....169..293B}
}

@software{radpy,
  author       = {Elliott, Ashley},
  title        = {{RADPy: Robust Angular Diameters in Python}},
  year         = {2025},
  version      = {0.3.9},
  month        = {July},
  url          = {https://github.com/spaceashley/radpy},
  note         = {GNU General Public License v3. Louisiana State University.},
}

@ARTICLE{stanford20,
       author = {{Stanford-Moore}, S. Adam and {Nielsen}, Eric L. and {De Rosa}, Robert J. and {Macintosh}, Bruce and {Czekala}, Ian},
        title = "{BAFFLES: Bayesian Ages for Field Lower-mass Stars}",
      journal = {\apj},
         year = 2020,
        month = jul,
       volume = {898},
       number = {1},
          eid = {27},
        pages = {27},
          doi = {10.3847/1538-4357/ab9a35},
archivePrefix = {arXiv},
       eprint = {2006.04811},
 primaryClass = {astro-ph.SR},
       adsurl = {https://ui.adsabs.harvard.edu/abs/2020ApJ...898...27S}
}

@ARTICLE{ramirez12,
       author = {{Ram{\'\i}rez}, I. and {Fish}, J.~R. and {Lambert}, D.~L. and {Allende Prieto}, C.},
        title = "{Lithium Abundances in nearby FGK Dwarf and Subgiant Stars: Internal Destruction, Galactic Chemical Evolution, and Exoplanets}",
      journal = {\apj},
         year = 2012,
        month = sep,
       volume = {756},
       number = {1},
          eid = {46},
        pages = {46},
          doi = {10.1088/0004-637X/756/1/46},
archivePrefix = {arXiv},
       eprint = {1207.0499},
 primaryClass = {astro-ph.SR},
       adsurl = {https://ui.adsabs.harvard.edu/abs/2012ApJ...756...46R}
}

@ARTICLE{lorenzo19,
       author = {{Lorenzo-Oliveira}, Diego and {Mel{\'e}ndez}, Jorge and {Yana Galarza}, Jhon and {Ponte}, Geisa and {dos Santos}, Leonardo A. and {Spina}, Lorenzo and {Bedell}, Megan and {Ram{\'\i}rez}, Iv{\'a}n and {Bean}, Jacob L. and {Asplund}, Martin},
        title = "{Constraining the evolution of stellar rotation using solar twins}",
      journal = {\mnras},
         year = 2019,
        month = may,
       volume = {485},
       number = {1},
        pages = {L68-L72},
          doi = {10.1093/mnrasl/slz034},
archivePrefix = {arXiv},
       eprint = {1903.02630},
 primaryClass = {astro-ph.SR},
       adsurl = {https://ui.adsabs.harvard.edu/abs/2019MNRAS.485L..68L}
}

@ARTICLE{shej24,
       author = {{Shejeelammal}, J. and {Mel{\'e}ndez}, Jorge and {Rathsam}, Anne and {Martos}, Giulia},
        title = "{The [Y/Mg] chemical clock in the Galactic disk: The influence of metallicity and the Galactic population in the solar neighbourhood}",
      journal = {\aap},
         year = 2024,
        month = oct,
       volume = {690},
          eid = {A107},
        pages = {A107},
          doi = {10.1051/0004-6361/202449669},
archivePrefix = {arXiv},
       eprint = {2407.07283},
 primaryClass = {astro-ph.GA},
       adsurl = {https://ui.adsabs.harvard.edu/abs/2024A&A...690A.107S}
}

@ARTICLE{gould92,
       author = {{Gould}, Andrew},
        title = "{Extending the MACHO Search to approximately 10 6 M sub sun}",
      journal = {\apj},
         year = 1992,
        month = jun,
       volume = {392},
        pages = {442},
          doi = {10.1086/171443},
       adsurl = {https://ui.adsabs.harvard.edu/abs/1992ApJ...392..442G}
}

@ARTICLE{an02,
       author = {{An}, Jin H. and {Albrow}, M.~D. and {Beaulieu}, J. -P. and {Caldwell}, J.~A.~R. and {DePoy}, D.~L. and {Dominik}, M. and {Gaudi}, B.~S. and {Gould}, A. and {Greenhill}, J. and {Hill}, K. and {Kane}, S. and {Martin}, R. and {Menzies}, J. and {Pogge}, R.~W. and {Pollard}, K.~R. and {Sackett}, P.~D. and {Sahu}, K.~C. and {Vermaak}, P. and {Watson}, R. and {Williams}, A.},
        title = "{First Microlens Mass Measurement: PLANET Photometry of EROS BLG-2000-5}",
      journal = {\apj},
         year = 2002,
        month = jun,
       volume = {572},
       number = {1},
        pages = {521-539},
          doi = {10.1086/340191},
archivePrefix = {arXiv},
       eprint = {astro-ph/0110095},
 primaryClass = {astro-ph},
       adsurl = {https://ui.adsabs.harvard.edu/abs/2002ApJ...572..521A}
}

@ARTICLE{bennett16,
       author = {{Bennett}, D.~P. and {Rhie}, S.~H. and {Udalski}, A. and {Gould}, A. and {Tsapras}, Y. and {Kubas}, D. and {Bond}, I.~A. and {Greenhill}, J. and {Cassan}, A. and {Rattenbury}, N.~J. and {Boyajian}, T.~S. and {Luhn}, J. and {Penny}, M.~T. and {Anderson}, J. and {Abe}, F. and {Bhattacharya}, A. and {Botzler}, C.~S. and {Donachie}, M. and {Freeman}, M. and {Fukui}, A. and {Hirao}, Y. and {Itow}, Y. and {Koshimoto}, N. and {Li}, M.~C.~A. and {Ling}, C.~H. and {Masuda}, K. and {Matsubara}, Y. and {Muraki}, Y. and {Nagakane}, M. and {Ohnishi}, K. and {Oyokawa}, H. and {Perrott}, Y.~C. and {Saito}, To. and {Sharan}, A. and {Sullivan}, D.~J. and {Sumi}, T. and {Suzuki}, D. and {Tristram}, P.~J. and {Yonehara}, A. and {Yock}, P.~C.~M. and {MOA Collaboration} and {Szyma{\'n}ski}, M.~K. and {Soszy{\'n}ski}, I. and {Ulaczyk}, K. and {Wyrzykowski}, {\L}. and {OGLE Collaboration} and {Allen}, W. and {DePoy}, D. and {Gal-Yam}, A. and {Gaudi}, B.~S. and {Han}, C. and {Monard}, I.~A.~G. and {Ofek}, E. and {Pogge}, R.~W. and {{\ensuremath{\mu}}FUN Collaboration} and {Street}, R.~A. and {Bramich}, D.~M. and {Dominik}, M. and {Horne}, K. and {Snodgrass}, C. and {Steele}, I.~A. and {Robonet Collaboration} and {Albrow}, M.~D. and {Bachelet}, E. and {Batista}, V. and {Beaulieu}, J. -P. and {Brillant}, S. and {Caldwell}, J.~A.~R. and {Cole}, A. and {Coutures}, C. and {Dieters}, S. and {Dominis Prester}, D. and {Donatowicz}, J. and {Fouqu{\'e}}, P. and {Hundertmark}, M. and {J{\o}rgensen}, U.~G. and {Kains}, N. and {Kane}, S.~R. and {Marquette}, J. -B. and {Menzies}, J. and {Pollard}, K.~R. and {Ranc}, C. and {Sahu}, K.~C. and {Wambsganss}, J. and {Williams}, A. and {Zub}, M. and {PLANET Collaboration}},
        title = "{The First Circumbinary Planet Found by Microlensing: OGLE-2007-BLG-349L(AB)c}",
      journal = {\aj},
         year = 2016,
        month = nov,
       volume = {152},
       number = {5},
          eid = {125},
        pages = {125},
          doi = {10.3847/0004-6256/152/5/125},
archivePrefix = {arXiv},
       eprint = {1609.06720},
 primaryClass = {astro-ph.EP},
       adsurl = {https://ui.adsabs.harvard.edu/abs/2016AJ....152..125B}
}

@ARTICLE{baines12,
       author = {{Baines}, Ellyn K. and {White}, Russel J. and {Huber}, Daniel and {Jones}, Jeremy and {Boyajian}, Tabetha and {McAlister}, Harold A. and {ten Brummelaar}, Theo A. and {Turner}, Nils H. and {Sturmann}, Judit and {Sturmann}, Laszlo and {Goldfinger}, P.~J. and {Farrington}, Christopher D. and {Riedel}, Adric R. and {Ireland}, Michael and {von Braun}, Kaspar and {Ridgway}, Stephen T.},
        title = "{The CHARA Array Angular Diameter of HR 8799 Favors Planetary Masses for its Imaged Companions}",
      journal = {\apj},
         year = 2012,
        month = dec,
       volume = {761},
       number = {1},
          eid = {57},
        pages = {57},
          doi = {10.1088/0004-637X/761/1/57},
archivePrefix = {arXiv},
       eprint = {1210.0556},
 primaryClass = {astro-ph.SR},
       adsurl = {https://ui.adsabs.harvard.edu/abs/2012ApJ...761...57B}
}

@ARTICLE{johnson14,
       author = {{Johnson}, John Asher and {Huber}, Daniel and {Boyajian}, Tabetha and {Brewer}, John M. and {White}, Timothy R. and {von Braun}, Kaspar and {Maestro}, Vicente and {Stello}, Dennis and {Barclay}, Thomas},
        title = "{The Physical Parameters of the Retired a Star HD 185351}",
      journal = {\apj},
         year = 2014,
        month = oct,
       volume = {794},
       number = {1},
          eid = {15},
        pages = {15},
          doi = {10.1088/0004-637X/794/1/15},
archivePrefix = {arXiv},
       eprint = {1407.2329},
 primaryClass = {astro-ph.SR},
       adsurl = {https://ui.adsabs.harvard.edu/abs/2014ApJ...794...15J}
}

@ARTICLE{code76,
       author = {{Code}, A.~D. and {Davis}, J. and {Bless}, R.~C. and {Brown}, R.~H.},
        title = "{Empirical effective temperatures and bolometric corrections for early-type stars.}",
      journal = {\apj},
         year = 1976,
        month = jan,
       volume = {203},
        pages = {417-434},
          doi = {10.1086/154093},
       adsurl = {https://ui.adsabs.harvard.edu/abs/1976ApJ...203..417C}
}

@ARTICLE{mann15,
       author = {{Mann}, Andrew W. and {Feiden}, Gregory A. and {Gaidos}, Eric and {Boyajian}, Tabetha and {von Braun}, Kaspar},
        title = "{How to Constrain Your M Dwarf: Measuring Effective Temperature, Bolometric Luminosity, Mass, and Radius}",
      journal = {\apj},
         year = 2015,
        month = may,
       volume = {804},
       number = {1},
          eid = {64},
        pages = {64},
          doi = {10.1088/0004-637X/804/1/64},
archivePrefix = {arXiv},
       eprint = {1501.01635},
 primaryClass = {astro-ph.SR},
       adsurl = {https://ui.adsabs.harvard.edu/abs/2015ApJ...804...64M}
}

@ARTICLE{rabus19,
       author = {{Rabus}, Markus and {Lachaume}, R{\'e}gis and {Jord{\'a}n}, Andr{\'e}s and {Brahm}, Rafael and {Boyajian}, Tabetha and {von Braun}, Kaspar and {Espinoza}, N{\'e}stor and {Berger}, Jean-Philippe and {Le Bouquin}, Jean-Baptiste and {Absil}, Olivier},
        title = "{A discontinuity in the T$_{eff}$-radius relation of M-dwarfs}",
      journal = {\mnras},
         year = 2019,
        month = apr,
       volume = {484},
       number = {2},
        pages = {2674-2683},
          doi = {10.1093/mnras/sty3430},
archivePrefix = {arXiv},
       eprint = {1901.08077},
 primaryClass = {astro-ph.SR},
       adsurl = {https://ui.adsabs.harvard.edu/abs/2019MNRAS.484.2674R}
}

@ARTICLE{ligi16,
       author = {{Ligi}, R. and {Creevey}, O. and {Mourard}, D. and {Crida}, A. and {Lagrange}, A. -M. and {Nardetto}, N. and {Perraut}, K. and {Schultheis}, M. and {Tallon-Bosc}, I. and {ten Brummelaar}, T.},
        title = "{Radii, masses, and ages of 18 bright stars using interferometry and new estimations of exoplanetary parameters}",
      journal = {\aap},
         year = 2016,
        month = feb,
       volume = {586},
          eid = {A94},
        pages = {A94},
          doi = {10.1051/0004-6361/201527054},
archivePrefix = {arXiv},
       eprint = {1511.03197},
 primaryClass = {astro-ph.SR},
       adsurl = {https://ui.adsabs.harvard.edu/abs/2016A&A...586A..94L}
}

@ARTICLE{soubiran22,
       author = {{Soubiran}, C. and {Brouillet}, N. and {Casamiquela}, L.},
        title = "{Assessment of [Fe/H] determinations for FGK stars in spectroscopic surveys}",
      journal = {\aap},
         year = 2022,
        month = jul,
       volume = {663},
          eid = {A4},
        pages = {A4},
          doi = {10.1051/0004-6361/202142409},
archivePrefix = {arXiv},
       eprint = {2112.07545},
 primaryClass = {astro-ph.SR},
       adsurl = {https://ui.adsabs.harvard.edu/abs/2022A&A...663A...4S}
}

@ARTICLE{GAIA,
       author = {{Gaia Collaboration}},
        title = "{VizieR Online Data Catalog: Gaia EDR3 (Gaia Collaboration, 2020)}",
      journal = {VizieR Online Data Catalog},
         year = 2020,
        month = nov,
          eid = {I/350},
        pages = {I/350},
       adsurl = {https://ui.adsabs.harvard.edu/abs/2020yCat.1350....0G}
}

@ARTICLE{ttb05,
       author = {{ten Brummelaar}, T.~A. and {McAlister}, H.~A. and {Ridgway}, S.~T. and {Bagnuolo}, W.~G., Jr. and {Turner}, N.~H. and {Sturmann}, L. and {Sturmann}, J. and {Berger}, D.~H. and {Ogden}, C.~E. and {Cadman}, R. and {Hartkopf}, W.~I. and {Hopper}, C.~H. and {Shure}, M.~A.},
        title = "{First Results from the CHARA Array. II. A Description of the Instrument}",
      journal = {\apj},
         year = 2005,
        month = jul,
       volume = {628},
       number = {1},
        pages = {453-465},
          doi = {10.1086/430729},
archivePrefix = {arXiv},
       eprint = {astro-ph/0504082},
 primaryClass = {astro-ph},
       adsurl = {https://ui.adsabs.harvard.edu/abs/2005ApJ...628..453T}
}

@INPROCEEDINGS{Ire08,
       author = {{Ireland}, Michael J. and {M{\'e}rand}, Antoine and {ten Brummelaar}, Theo A. and {Tuthill}, Peter G. and {Schaefer}, Gail H. and {Turner}, Nils H. and {Sturmann}, Judit and {Sturmann}, Laszlo and {McAlister}, Harold A.},
        title = "{Sensitive visible interferometry with PAVO}",
    booktitle = {Optical and Infrared Interferometry},
         year = 2008,
       editor = {{Sch{\"o}ller}, Markus and {Danchi}, William C. and {Delplancke}, Fran{\c{c}}oise},
       series = {Society of Photo-Optical Instrumentation Engineers (SPIE) Conference Series},
       volume = {7013},
        month = jul,
          eid = {701324},
        pages = {701324},
          doi = {10.1117/12.788386},
       adsurl = {https://ui.adsabs.harvard.edu/abs/2008SPIE.7013E..24I}
}

@ARTICLE{kopp11,
   author = {{Kopp}, G. and {Lean}, J.~L.},
    title = "{A new, lower value of total solar irradiance: Evidence and climate significance}",
  journal = {\grl},
     year = 2011,
    month = jan,
   volume = 38,
      eid = {L01706},
    pages = {1706},
      doi = {10.1029/2010GL045777},
   adsurl = {http://adsabs.harvard.edu/abs/2011GeoRL..38.1706K}
}

@ARTICLE{han74,
   author = {{Hanbury Brown}, R.~H. and {Davis}, J. and {Lake}, R.~J.~W. and {Thompson}, R.~J.
	},
    title = "{The effects of limb darkening on measurements of angular size with an intensity interferometer}",
  journal = {\mnras},
     year = 1974,
    month = jun,
   volume = 167,
    pages = {475-484},
   adsurl = {http://adsabs.harvard.edu/abs/1974MNRAS.167..475H}
}

@ARTICLE{boy12b,
   author = {{Boyajian}, T.~S. and {von Braun}, K. and {van Belle}, G. and 
	{McAlister}, H.~A. and {ten Brummelaar}, T.~A. and {Kane}, S.~R. and 
	{Muirhead}, P.~S. and {Jones}, J. and {White}, R. and {Schaefer}, G. and 
	{Ciardi}, D. and {Henry}, T. and {L{\'o}pez-Morales}, M. and 
	{Ridgway}, S. and {Gies}, D. and {Jao}, W.-C. and {Rojas-Ayala}, B. and 
	{Parks}, J.~R. and {Sturmann}, L. and {Sturmann}, J. and {Turner}, N.~H. and 
	{Farrington}, C. and {Goldfinger}, P.~J. and {Berger}, D.~H.
	},
    title = "{Stellar Diameters and Temperatures. II. Main-sequence K- and M-stars}",
  journal = {\apj},
archivePrefix = "arXiv",
   eprint = {1208.2431},
 primaryClass = "astro-ph.SR",
     year = 2012,
    month = oct,
   volume = 757,
      eid = {112},
    pages = {112},
      doi = {10.1088/0004-637X/757/2/112},
   adsurl = {http://adsabs.harvard.edu/abs/2012ApJ...757..112B}
}

@ARTICLE{boy13a,
   author = {{Boyajian}, T.~S. and {von Braun}, K. and {van Belle}, G. and 
	{Farrington}, C. and {Schaefer}, G. and {Jones}, J. and {White}, R. and 
	{McAlister}, H.~A. and {ten Brummelaar}, T.~A. and {Ridgway}, S. and 
	{Gies}, D. and {Sturmann}, L. and {Sturmann}, J. and {Turner}, N.~H. and 
	{Goldfinger}, P.~J. and {Vargas}, N.},
    title = "{Stellar Diameters and Temperatures. III. Main-sequence A, F, G, and K Stars: Additional High-precision Measurements and Empirical Relations}",
  journal = {\apj},
archivePrefix = "arXiv",
   eprint = {1306.2974},
 primaryClass = "astro-ph.SR",
     year = 2013,
    month = jul,
   volume = 771,
      eid = {40},
    pages = {40},
      doi = {10.1088/0004-637X/771/1/40},
   adsurl = {http://adsabs.harvard.edu/abs/2013ApJ...771...40B}
}

@ARTICLE{tayar22,
       author = {{Tayar}, Jamie and {Claytor}, Zachary R. and {Huber}, Daniel and {van Saders}, Jennifer},
        title = "{A Guide to Realistic Uncertainties on the Fundamental Properties of Solar-type Exoplanet Host Stars}",
      journal = {\apj},
         year = 2022,
        month = mar,
       volume = {927},
       number = {1},
          eid = {31},
        pages = {31},
          doi = {10.3847/1538-4357/ac4bbc},
archivePrefix = {arXiv},
       eprint = {2012.07957},
 primaryClass = {astro-ph.EP},
       adsurl = {https://ui.adsabs.harvard.edu/abs/2022ApJ...927...31T}
}

@ARTICLE{claytor20,
       author = {{Claytor}, Zachary R. and {van Saders}, Jennifer L. and {Santos}, {\^A}ngela R.~G. and {Garc{\'\i}a}, Rafael A. and {Mathur}, Savita and {Tayar}, Jamie and {Pinsonneault}, Marc H. and {Shetrone}, Matthew},
        title = "{Chemical Evolution in the Milky Way: Rotation-based Ages for APOGEE-Kepler Cool Dwarf Stars}",
      journal = {\apj},
         year = 2020,
        month = jan,
       volume = {888},
       number = {1},
          eid = {43},
        pages = {43},
          doi = {10.3847/1538-4357/ab5c24},
archivePrefix = {arXiv},
       eprint = {1911.04518},
 primaryClass = {astro-ph.SR},
       adsurl = {https://ui.adsabs.harvard.edu/abs/2020ApJ...888...43C}
}

@ARTICLE{mic21,
       author = {{Michelson}, A.~A. and {Pease}, F.~G.},
        title = "{Measurement of the Diameter of {\ensuremath{\alpha}} Orionis with the Interferometer.}",
      journal = {\apj},
         year = 1921,
        month = may,
       volume = {53},
        pages = {249-259},
          doi = {10.1086/142603},
       adsurl = {https://ui.adsabs.harvard.edu/abs/1921ApJ....53..249M}
}

@ARTICLE{soubiran05,
       author = {{Soubiran}, C. and {Girard}, P.},
        title = "{Abundance trends in kinematical groups of the Milky Way's disk}",
      journal = {\aap},
         year = 2005,
        month = jul,
       volume = {438},
       number = {1},
        pages = {139-151},
          doi = {10.1051/0004-6361:20042390},
archivePrefix = {arXiv},
       eprint = {astro-ph/0503498},
 primaryClass = {astro-ph},
       adsurl = {https://ui.adsabs.harvard.edu/abs/2005A&A...438..139S}
}

@ARTICLE{mam12,
   author = {{Mamajek}, E.~E.},
    title = "{On the Age and Binarity of Fomalhaut}",
  journal = {\apjl},
archivePrefix = "arXiv",
   eprint = {1206.6353},
 primaryClass = "astro-ph.SR",
     year = 2012,
    month = aug,
   volume = 754,
      eid = {L20},
    pages = {L20},
      doi = {10.1088/2041-8205/754/2/L20},
   adsurl = {http://adsabs.harvard.edu/abs/2012ApJ...754L..20M}
}

@INPROCEEDINGS{JMMC,
       author = {{Bourg{\'e}s}, L. and {Lafrasse}, S. and {Mella}, G. and {Chesneau}, O. and {Bouquin}, J.~L. and {Duvert}, G. and {Chelli}, A. and {Delfosse}, X.},
        title = "{The JMMC Stellar Diameters Catalog v2 (JSDC): A New Release Based on SearchCal Improvements}",
    booktitle = {Astronomical Data Analysis Software and Systems XXIII},
         year = 2014,
       editor = {{Manset}, N. and {Forshay}, P.},
       series = {Astronomical Society of the Pacific Conference Series},
       volume = {485},
        month = may,
        pages = {223},
       adsurl = {https://ui.adsabs.harvard.edu/abs/2014ASPC..485..223B}
}

@ARTICLE{TESS,
       author = {{Ricker}, George R. and {Winn}, Joshua N. and {Vanderspek}, Roland and {Latham}, David W. and {Bakos}, G{\'a}sp{\'a}r {\'A}. and {Bean}, Jacob L. and {Berta-Thompson}, Zachory K. and {Brown}, Timothy M. and {Buchhave}, Lars and {Butler}, Nathaniel R. and {Butler}, R. Paul and {Chaplin}, William J. and {Charbonneau}, David and {Christensen-Dalsgaard}, J{\o}rgen and {Clampin}, Mark and {Deming}, Drake and {Doty}, John and {De Lee}, Nathan and {Dressing}, Courtney and {Dunham}, Edward W. and {Endl}, Michael and {Fressin}, Francois and {Ge}, Jian and {Henning}, Thomas and {Holman}, Matthew J. and {Howard}, Andrew W. and {Ida}, Shigeru and {Jenkins}, Jon M. and {Jernigan}, Garrett and {Johnson}, John Asher and {Kaltenegger}, Lisa and {Kawai}, Nobuyuki and {Kjeldsen}, Hans and {Laughlin}, Gregory and {Levine}, Alan M. and {Lin}, Douglas and {Lissauer}, Jack J. and {MacQueen}, Phillip and {Marcy}, Geoffrey and {McCullough}, Peter R. and {Morton}, Timothy D. and {Narita}, Norio and {Paegert}, Martin and {Palle}, Enric and {Pepe}, Francesco and {Pepper}, Joshua and {Quirrenbach}, Andreas and {Rinehart}, Stephen A. and {Sasselov}, Dimitar and {Sato}, Bun'ei and {Seager}, Sara and {Sozzetti}, Alessandro and {Stassun}, Keivan G. and {Sullivan}, Peter and {Szentgyorgyi}, Andrew and {Torres}, Guillermo and {Udry}, Stephane and {Villasenor}, Joel},
        title = "{Transiting Exoplanet Survey Satellite (TESS)}",
      journal = {Journal of Astronomical Telescopes, Instruments, and Systems},
         year = 2015,
        month = jan,
       volume = {1},
          eid = {014003},
        pages = {014003},
          doi = {10.1117/1.JATIS.1.1.014003},
       adsurl = {https://ui.adsabs.harvard.edu/abs/2015JATIS...1a4003R}
}

@misc{10.26093/cds/vizier,
  doi = {10.26093/CDS/VIZIER},
  url = {https://vizier.cds.unistra.fr},
  author = {Ochsenbein,  Francois},
  title = {The VizieR database of astronomical catalogues},
  publisher = {CDS,  Centre de DonnÃ©es astronomiques de Strasbourg},
  year = {1996},
  copyright = {Refer to CDS usage}
}

@ARTICLE{vizier2000,
       author = {{Ochsenbein}, F. and {Bauer}, P. and {Marcout}, J.},
        title = "{The VizieR database of astronomical catalogues}",
      journal = {\aaps},
         year = 2000,
        month = apr,
       volume = {143},
        pages = {23-32},
          doi = {10.1051/aas:2000169},
archivePrefix = {arXiv},
       eprint = {astro-ph/0002122},
 primaryClass = {astro-ph},
       adsurl = {https://ui.adsabs.harvard.edu/abs/2000A&AS..143...23O}
}

@ARTICLE{cla11,
       author = {{Claret}, A. and {Bloemen}, S.},
        title = "{Gravity and limb-darkening coefficients for the Kepler, CoRoT, Spitzer, uvby, UBVRIJHK, and Sloan photometric systems}",
      journal = {\aap},
         year = 2011,
        month = may,
       volume = {529},
          eid = {A75},
        pages = {A75},
          doi = {10.1051/0004-6361/201116451},
       adsurl = {https://ui.adsabs.harvard.edu/abs/2011A&A...529A..75C}
}

@ARTICLE{Lindegren21,
       author = {{Lindegren}, L. and {Bastian}, U. and {Biermann}, M. and {Bombrun}, A. and {de Torres}, A. and {Gerlach}, E. and {Geyer}, R. and {Hern{\'a}ndez}, J. and {Hilger}, T. and {Hobbs}, D. and {Klioner}, S.~A. and {Lammers}, U. and {McMillan}, P.~J. and {Ramos-Lerate}, M. and {Steidelm{\"u}ller}, H. and {Stephenson}, C.~A. and {van Leeuwen}, F.},
        title = "{Gaia Early Data Release 3. Parallax bias versus magnitude, colour, and position}",
      journal = {\aap},
         year = 2021,
        month = may,
       volume = {649},
          eid = {A4},
        pages = {A4},
          doi = {10.1051/0004-6361/202039653},
archivePrefix = {arXiv},
       eprint = {2012.01742},
 primaryClass = {astro-ph.IM},
       adsurl = {https://ui.adsabs.harvard.edu/abs/2021A&A...649A...4L}
}

@ARTICLE{Ellis2021,
       author = {{Ellis}, Tyler G. and {Boyajian}, Tabetha and {von Braun}, Kaspar and {Ligi}, Roxanne and {Mourard}, Denis and {Dragomir}, Diana and {Schaefer}, Gail H. and {Farrington}, Christopher D.},
        title = "{Directly Determined Properties of HD 97658 from Interferometric Observations}",
      journal = {\aj},
         year = 2021,
        month = sep,
       volume = {162},
       number = {3},
          eid = {118},
        pages = {118},
          doi = {10.3847/1538-3881/ac141a},
archivePrefix = {arXiv},
       eprint = {2107.06254},
 primaryClass = {astro-ph.SR},
       adsurl = {https://ui.adsabs.harvard.edu/abs/2021AJ....162..118E}
}

@ARTICLE{White2018,
       author = {{White}, T.~R. and {Huber}, D. and {Mann}, A.~W. and {Casagrande}, L. and {Grunblatt}, S.~K. and {Justesen}, A.~B. and {Silva Aguirre}, V. and {Bedding}, T.~R. and {Ireland}, M.~J. and {Schaefer}, G.~H. and {Tuthill}, P.~G.},
        title = "{Interferometric diameters of five evolved intermediate-mass planet-hosting stars measured with PAVO at the CHARA Array}",
      journal = {\mnras},
         year = 2018,
        month = jul,
       volume = {477},
       number = {4},
        pages = {4403-4413},
          doi = {10.1093/mnras/sty898},
archivePrefix = {arXiv},
       eprint = {1804.05976},
 primaryClass = {astro-ph.SR},
       adsurl = {https://ui.adsabs.harvard.edu/abs/2018MNRAS.477.4403W}
}

@ARTICLE{Adams2018,
	author = {{Adams}, Arthur D. and {Boyajian}, Tabetha S. and {von Braun}, Kaspar},
	title = "{Predicting stellar angular diameters from V, I$_{C}$, H and K photometry}",
	journal = {\mnras},
	year = 2018,
	month = jan,
	volume = {473},
	number = {3},
	pages = {3608-3614},
	doi = {10.1093/mnras/stx2367},
	archivePrefix = {arXiv},
	eprint = {1709.03902},
	primaryClass = {astro-ph.SR},
	adsurl = {https://ui.adsabs.harvard.edu/abs/2018MNRAS.473.3608A}
}


\clearpage
\begin{appendix}


\section{Observing Log}
{\small
\begin{longtable}{lllll}
\caption{Observing log of interferometric observations.}\label{tab:obs_log}\\
\toprule
\textbf{Science star} & \textbf{UT Date} & \textbf{Calibrator(s)} & \textbf{Baseline} & \textbf{Nobs} \\
\midrule
\endfirsthead

\toprule
\textbf{Science star} & \textbf{UT Date} & \textbf{Calibrator(s)} & \textbf{Baseline} & \textbf{Nobs} \\
\midrule
\endhead

\bottomrule
\endfoot

HD 1461 & 2012/11/12 & HD 1064, HD 224926 & S2W2 & 2  \\
 & 2014/09/07 & HD 315, HD 1064 & E1W1 & 1  \\
 & 2015/10/11 & HD 1064, HD 6530 & E1W1 & 2  \\
HD 8574 & 2015/10/11 & HD 5382, HD 9298 & E1W1 & 2  \\
 & 2018/08/30 & HD 10390 & E2S2 & 1  \\
HD 8673 & 2014/08/16 & HD 10205, HD 10390 & E2S2 & 4  \\
 & 2014/08/17 & HD 9531, HD 10390 & E2W1 & 9  \\
 & 2014/09/07 & HD 9298, HD 10390 & E1W1 & 2  \\
 & 2015/10/11 & HD 5382, HD 9298 & E1W1 & 3  \\
 & 2016/08/02 & HD 5382, HD 7019, HD 10390 & E1W2 & 7  \\
HD 10697 & 2017/10/29 & HD 10982, HD 11285 & S2W1 & 4  \\
 & 2018/08/30 & HD 10982, HD 11285 & E2S2 & 4  \\
HD 16141 & 2012/10/04 & HD 14840, HD 15004 & E1S1 & 6  \\
 & 2013/09/07 & HD 14417, HD 15633 & E2S2 & 4  \\
 & 2015/10/12 & HD 15633, HD 18633 & E2S2 & 1  \\
 & 2016/11/11 & HD 16582, HD 17943, HD 22243 & E2W1 & 3  \\
HD 20367 & 2011/08/13 & HD 18928, HD 21402 & S2W2 & 2  \\
 & 2011/08/14 & HD 18769, HD 21402 & S2W2 & 4  \\
 & 2011/08/15 & HD 18769, HD 18928 & E2S2 & 4  \\
 & 2012/10/04 & HD 19600 & E1S1 & 1  \\
 & 2013/09/05 & HD 19600, HD 21402 & E2W2 & 5  \\
 & 2013/09/06 & HD 19600, HD 21402 & S2W2 & 4  \\
HD 22879 & 2014/08/17 & HD 25340, HD 27563 & E2W1 & 3  \\
 & 2016/11/13 & HD 21790, HD 22920 & E1W1 & 3  \\
HD 30562 & 2013/09/05 & HD 30321 & E2W2 & 2  \\
 & 2013/09/06 & HD 27563, HD 30321 & S2W2 & 4  \\
 & 2015/10/12 & HD 27563, HD 28375 & E2S2 & 5  \\
 & 2016/11/09 & HD 27563 & E2W2 & 3  \\
HD 42807 & 2012/11/13 & HD 39985, HD 43386, HD 43683 & S2W2 & 2  \\
 & 2017/10/29 & HD 38899, HD 43247 & S2W1 & 2  \\
HD 58551 & 2016/11/09 & HD 64648 & E2W2 & 3  \\
 & 2016/11/11 & HD 62510 & E2W1 & 3  \\
 & 2016/11/13 & HD 62510 & E1W1 & 4  \\
HD 76932 & 2016/11/09 & HD 72968, HD 79931 & E2W2 & 2  \\
 & 2016/11/11 & HD 74190, HD 74706 & E2W1 & 4  \\
HD 108510 & 2017/06/04 & HD 109309, HD 106384 & E2W1 & 4  \\
 & 2018/05/04 & HD 109309, HD 109704 & E2W1 & 2  \\
 & 2018/05/05 & HD 109309 & E1W1 & 1  \\
HD 127334 & 2015/05/05 & HD 125642, HD 133029 & E2W2 & 3  \\
 & 2017/06/03 & HD 133962 & E2W1 & 1  \\
 & 2017/06/04 & HD 125349, HD 133962 & E2W1 & 3  \\
 & 2018/05/04 & HD 129002, HD 125349, HD 133962 & E1W2 & 4  \\
 & 2018/05/05 & HD 129002, HD 125349, HD 133962 & E2W1 & 4  \\
HD 144579 & 2018/05/05 & HD 133962, HD 144206 & E2W1 & 4  \\
 & 2018/05/07 & HD 144206, HD 147394 & E2W1 & 5  \\
HD 154345 & 2011/08/13 & HD 154713 & S2W2 & 1  \\
 & 2011/08/14 & HD 154713, HD 158414 & S2W2 & 3  \\
 & 2011/08/15 & HD 154713 & E2S2 & 2  \\
HD 164922 & 2012/05/13 & HD 161019, HD 164900 & E1W1 &  4 \\
 & 2012/05/14 & HD 164900, HD 165373 & E1W1 &  4 \\
HD 181096 & 2014/08/16 & HD 177003, HD 185872 & E2S2 & 5  \\
 & 2014/08/17 & HD 177003, HD 185872 & E2W1 & 2  \\
 & 2014/08/18 & HD 177003, HD 185872 & E2W2 & 2  \\
HD 182488 & 2011/08/13 & HD 184381, HD 186440  & S2W2 & 6  \\
 & 2011/08/14 & HD 186357, HD 186440 & S2W2 & 4  \\
 & 2016/08/01 & HD 179527, HD 183986 & E2W2 & 4  \\
 & 2016/08/03 & HD 179527 & E2W2 & 4  \\
HD 184499 & 2016/08/03 & HD 179527 & E2W2 & 4  \\
 & 2016/08/05 & HD 183986 & E2W1 & 3  \\
 & 2017/06/03 & HD 182255, HD186440 & E2W1 & 3  \\
 & 2017/06/04 & HD 186440 & E2W1 & 1  \\
 & 2017/06/05 & HD 182255, HD186440 & E2W1 & 1  \\
 & 2017/06/06 & HD 182255, HD186440 & S1W1 & 4  \\
HD 185269 & 2012/08/24 & HD 185268, HD 186440 & E1W1 & 3  \\
 & 2013/09/07 & HD 185268, HD 186440 & E2S2 & 3  \\
 & 2014/09/06 & HD 186440, HD 189944 & E1S1 & 1  \\
 & 2016/08/01 & HD 182255, HD 186440 & E2W2 & 2  \\
 & 2017/06/03 & HD 182255, HD 186440 & E2W1 & 1  \\
 & 2017/06/04 & HD 182255, HD 186440 & E2W1 & 1  \\
 & 2017/06/05 & HD 182255, HD 186440 & E2W1 & 1  \\
 & 2017/06/06 & HD 182255, HD 186440 & S1W1 & 1  \\
HD 190228 & 2013/09/07 & HD 189944, HD 186998 & E2S2 & 5  \\
 & 2017/06/06 & HD 189944, HD 191243 & S1W1 & 1  \\
 & 2018/08/30 & HD 189944, HD 187640 & E2S2 & 2  \\
HD 193664 & 2011/08/13 & HD 197950 & S2W2 & 2  \\
 & 2011/08/14 & HD 189900, HD197950 & S2W2 & 4  \\
 & 2011/08/15 & HD 189900, HD187764 & E2S2 & 3  \\
HD 195019 & 2015/08/17 & HD 193707, HD 195943 & E2S1 & 1  \\
 & 2016/08/01 & HD 198090, HD 196724 & E2W2 & 2  \\
 & 2016/08/02 & HD 196724 & E1W2 & 1  \\
 & 2016/08/03 & HD 196724 & E2W2 & 1  \\
 & 2016/11/11 & HD 195810, HD 196724 & E2W1 & 3  \\
HD 196885 & 2013/09/05 & HD 192895, HD 196182 & S2W2 & 3  \\
 & 2013/09/06 & HD 192895, HD 196182 & E2W2 & 5  \\
 & 2015/08/14 & HD 196426, HD 201616 & E2W1 & 2  \\
 & 2015/08/15 & HD 192715, HD 195943, HD 196426 & E2W1 & 6  \\
 & 2015/08/17 & HD 193707, HD 195943 & E2S1 & 3  \\
HD 202568 & 2013/09/05 & HD 202862 & E2W2 & 5  \\
 & 2013/09/06 & HD 202862 & S2W2 & 5  \\
HD 210277 & 2014/08/17 & HD 210424, HD 212717 & E2W1 & 2  \\
 & 2015/08/17 & HD 205244, HD 210424 & E2S1 & 5  \\
 & 2016/06/30 & HD 210424, HD 213326 & S2W2 & 3  \\
HD 222155 & 2016/08/05 & HD 1438, HD 220885 & E2W1 & 4  \\
 & 2016/11/11 & HD 1279, HD 1438 & E2W1 & 1  \\
 & 2016/11/13 & HD 220885, HD 222173 & E1W1 & 3  \\
 \end{longtable}
} 



\clearpage
\section{Calibrators}

{\footnotesize
\begin{longtable}{lc|lc}
\caption{Calibrators used in this work. The estimated angular diameter,
$\theta_{\mathrm{est}}$ is the $R$-band uniform-disk diameter from \citet{JMMC}.
See \S\ref{s:observations} for additional information.}\label{tab:cals}\\
\toprule
\textbf{Calibrator} & \textbf{$\theta_{\mathrm{est}}$} &
\textbf{Calibrator} & \textbf{$\theta_{\mathrm{est}}$} \\
\midrule
\endfirsthead

\toprule
\textbf{Calibrator} & \textbf{$\theta_{\mathrm{est}}$} &
\textbf{Calibrator} & \textbf{$\theta_{\mathrm{est}}$} \\
\midrule
\endhead

\bottomrule
\endfoot

HD 315   & 0.150$\pm$0.004 & HD 111270 & 0.307$\pm$0.009 \\
HD 1064  & 0.187$\pm$0.006 & HD 115735 & 0.268$\pm$0.009 \\
HD 1279  & 0.183$\pm$0.005 & HD 125349 & 0.209$\pm$0.006 \\
HD 1438  & 0.158$\pm$0.005 & HD 125642 & 0.197$\pm$0.006 \\
HD 5382  & 0.246$\pm$0.007 & HD 129002 & 0.287$\pm$0.009 \\
HD 6530  & 0.243$\pm$0.007 & HD 133029 & 0.144$\pm$0.005 \\
HD 7019  & 0.174$\pm$0.005 & HD 133962 & 0.235$\pm$0.007 \\
HD 9298  & 0.132$\pm$0.004 & HD 144206 & 0.314$\pm$0.009 \\
HD 9531  & 0.188$\pm$0.005 & HD 147394 & 0.378$\pm$0.032 \\
HD 10205 & 0.241$\pm$0.008 & HD 154713 & 0.198$\pm$0.005 \\
HD 10390 & 0.210$\pm$0.006 & HD 158414 & 0.289$\pm$0.008 \\
HD 10982 & 0.201$\pm$0.006 & HD 161019 & 0.182$\pm$0.005 \\
HD 11285 & 0.250$\pm$0.006 & HD 164900 & 0.120$\pm$0.004 \\
HD 14417 & 0.181$\pm$0.005 & HD 165373 & 0.296$\pm$0.008 \\
HD 14840 & 0.203$\pm$0.006 & HD 177003 & 0.145$\pm$0.005 \\
HD 15004 & 0.172$\pm$0.005 & HD 179527 & 0.185$\pm$0.005 \\
HD 15633 & 0.256$\pm$0.008 & HD 182255 & 0.212$\pm$0.006 \\
HD 16582 & 0.225$\pm$0.021 & HD 183986 & 0.173$\pm$0.005 \\
HD 17943 & 0.234$\pm$0.007 & HD 184381 & 0.296$\pm$0.008 \\
HD 18633 & 0.216$\pm$0.006 & HD 185268 & 0.119$\pm$0.004 \\
HD 18769 & 0.255$\pm$0.007 & HD 185872 & 0.246$\pm$0.007 \\
HD 18928 & 0.288$\pm$0.008 & HD 186357 & 0.273$\pm$0.007 \\
HD 19600 & 0.170$\pm$0.005 & HD 186440 & 0.196$\pm$0.006 \\
HD 21038 & 0.169$\pm$0.005 & HD 186998 & 0.239$\pm$0.007 \\
HD 21402 & 0.256$\pm$0.007 & HD 187764 & 0.293$\pm$0.009 \\
HD 21790 & 0.287$\pm$0.023 & HD 187640 & 0.119$\pm$0.004 \\
HD 22243 & 0.185$\pm$0.005 & HD 189900 & 0.208$\pm$0.006 \\
HD 22920 & 0.200$\pm$0.006 & HD 189944 & 0.137$\pm$0.004 \\
HD 25152 & 0.174$\pm$0.005 & HD 191243 & 0.204$\pm$0.006 \\
HD 25340 & 0.185$\pm$0.006 & HD 192715 & 0.214$\pm$0.006 \\
HD 27563 & 0.166$\pm$0.006 & HD 192895 & 0.264$\pm$0.007 \\
HD 28375 & 0.175$\pm$0.006 & HD 193707 & 0.155$\pm$0.004 \\
HD 30321 & 0.179$\pm$0.005 & HD 195810 & 0.318$\pm$0.030 \\
HD 38899 & 0.298$\pm$0.030 & HD 195943 & 0.293$\pm$0.013 \\
HD 39985 & 0.190$\pm$0.006 & HD 196182 & 0.142$\pm$0.005 \\
HD 43247 & 0.271$\pm$0.008 & HD 196426 & 0.152$\pm$0.004 \\
HD 43386 & 0.548$\pm$0.057 & HD 196724 & 0.308$\pm$0.027 \\
HD 43683 & 0.222$\pm$0.006 & HD 197950 & 0.349$\pm$0.011 \\
HD 62510 & 0.166$\pm$0.008 & HD 198090 & 0.082$\pm$0.003 \\
HD 64648 & 0.260$\pm$0.007 & HD 201616 & 0.211$\pm$0.006 \\
HD 72968 & 0.228$\pm$0.007 & HD 202862 & 0.149$\pm$0.004 \\
HD 74190 & 0.206$\pm$0.007 & HD 205244 & 0.218$\pm$0.007 \\
HD 74706 & 0.265$\pm$0.008 & HD 210424 & 0.180$\pm$0.006 \\
HD 79931 & 0.172$\pm$0.006 & HD 212717 & 0.229$\pm$0.006 \\
HD 106384 & 0.240$\pm$0.007 & HD 213326 & 0.088$\pm$0.002 \\
HD 109387 & 0.456$\pm$0.042 & HD 220885 & 0.215$\pm$0.006 \\
HD 109309 & 0.225$\pm$0.006 & HD 222173 & 0.329$\pm$0.033 \\
HD 109704 & 0.237$\pm$0.007 & HD 224926 & 0.222$\pm$0.007 \\
HD 110462 & 0.221$\pm$0.007 &                        \\
\end{longtable}
} 
{\small
\begin{longtable}{l}
\caption{Bad Calibrators}\label{tab:bad_cals}\\
\toprule
\textbf{Star} \\
\midrule
\endfirsthead

\toprule
\textbf{Star} \\
\midrule
\endhead

\bottomrule
\endfoot

HD 6116 \\
HD 29335 \\
HD 33224 \\
HD 61219 \\
HD 182568 \\
HD 205939  \\
HD 224093 \\

\end{longtable}
}

\clearpage

\section{Angular Diameter Plots}
\mbox{}\par\vspace{0.5\baselineskip} 

\begin{figure*}
    
\centering

\begin{tabular}{cc}
  \includegraphics[width=0.45\textwidth]{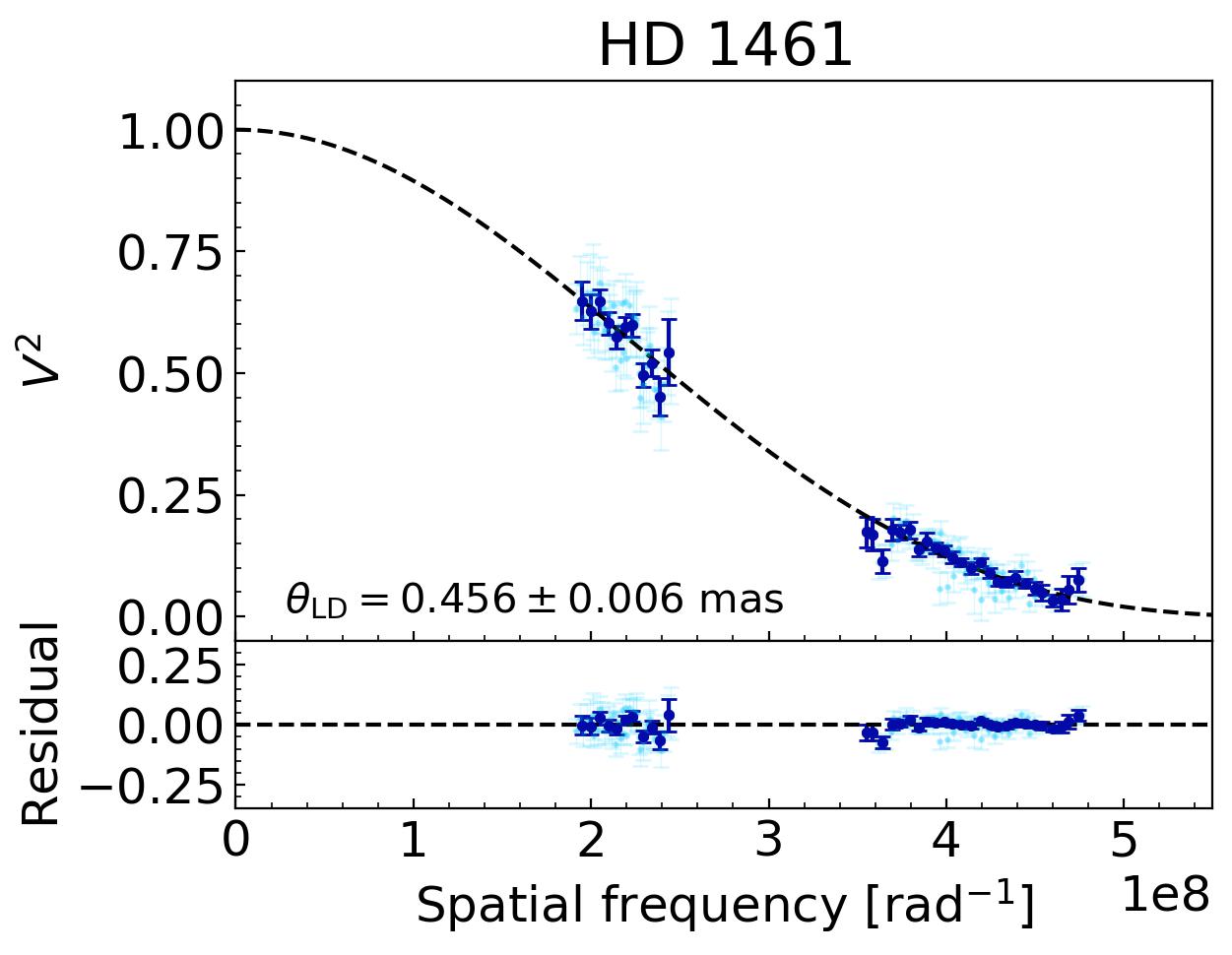} &
  \includegraphics[width=0.45\textwidth]{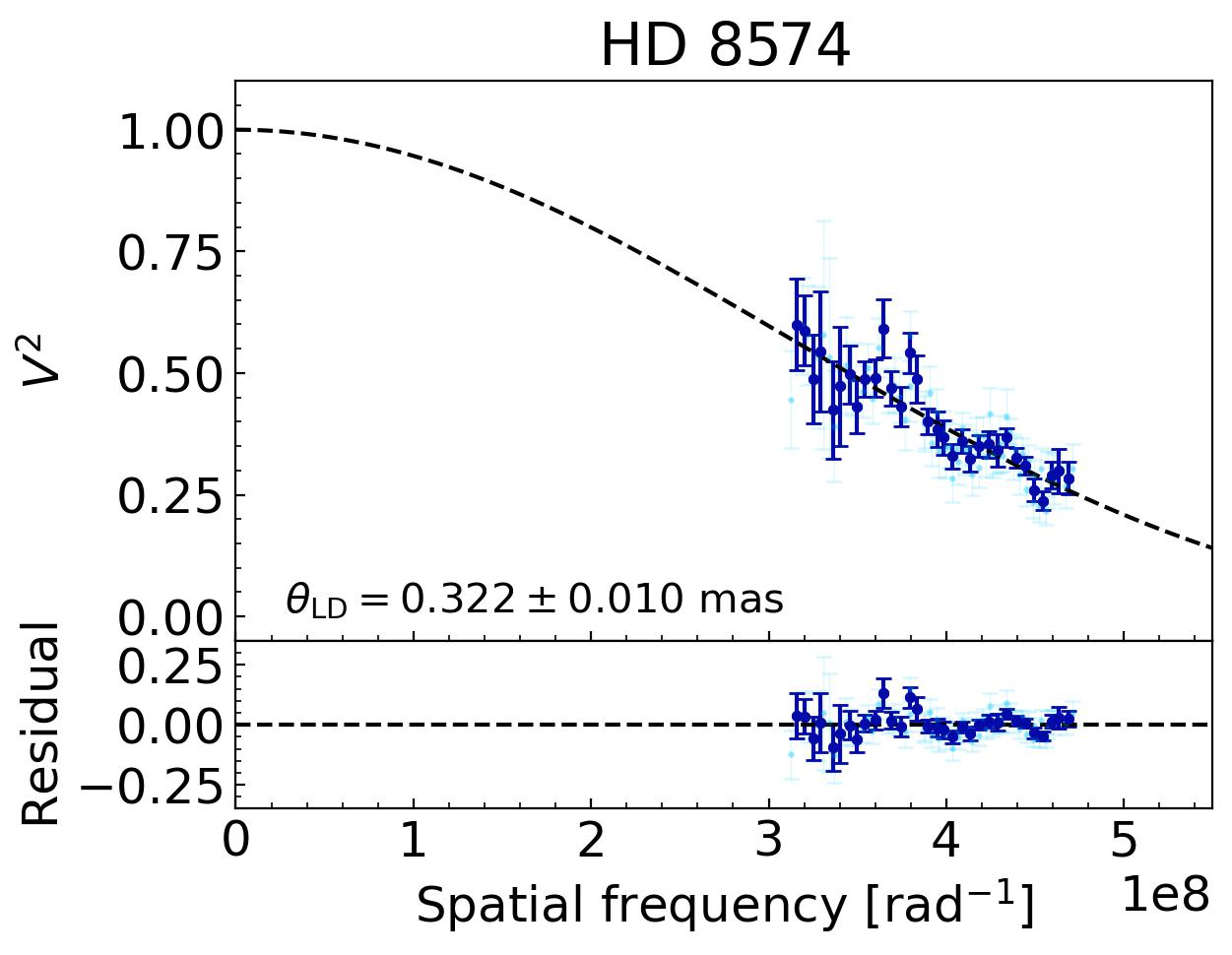} \\

  \includegraphics[width=0.45\textwidth]{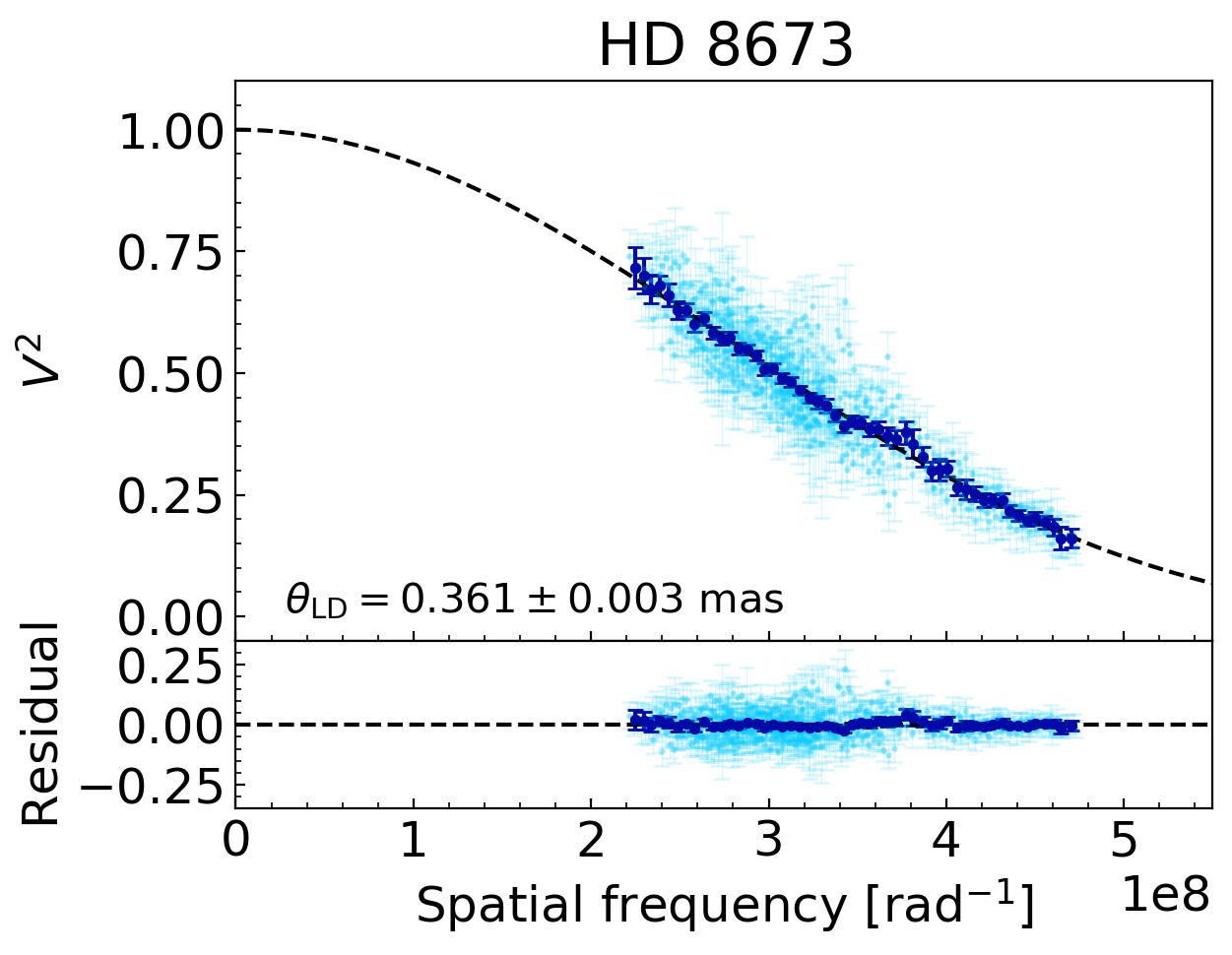} &
  \includegraphics[width=0.45\textwidth]{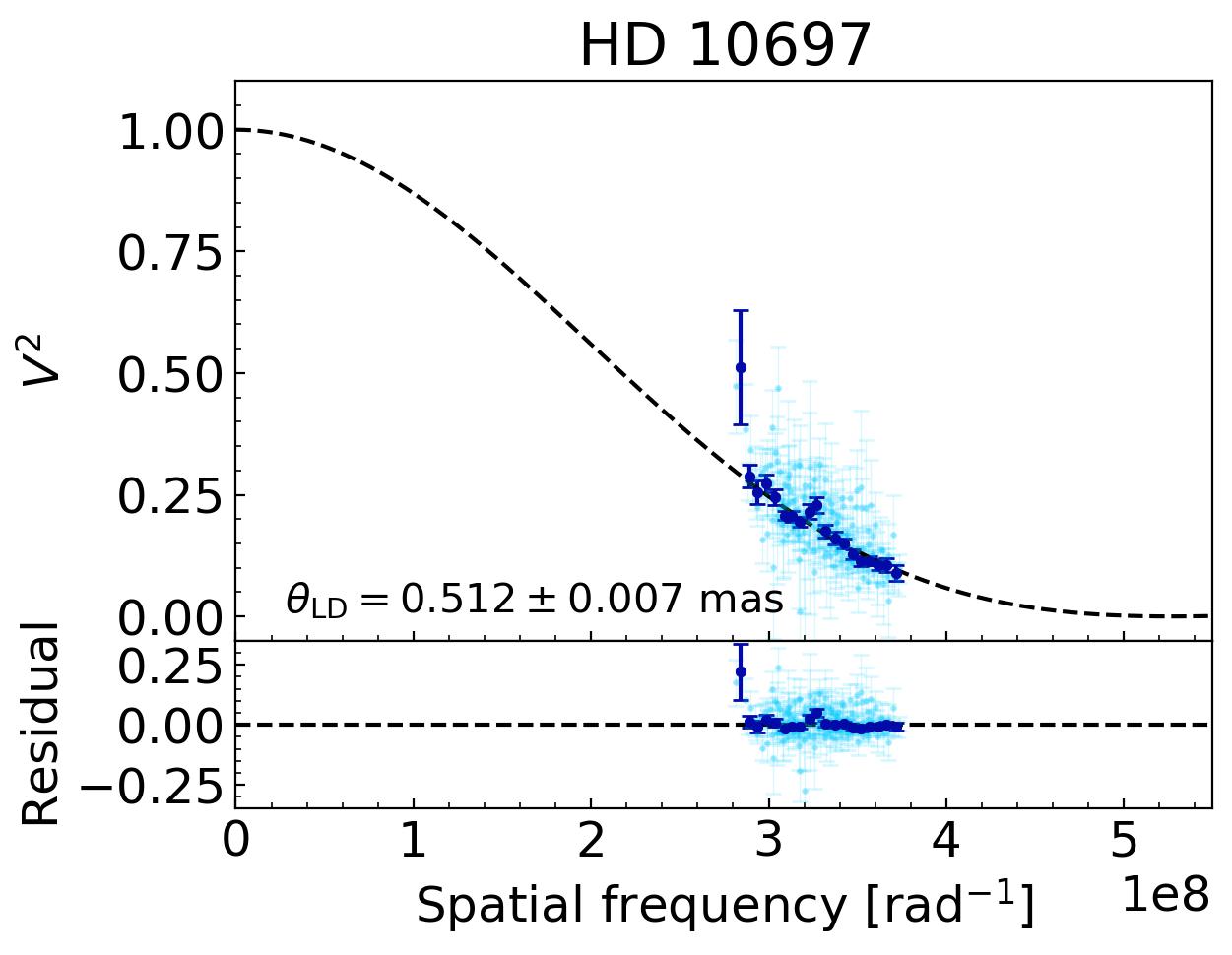} \\

  \includegraphics[width=0.45\textwidth]{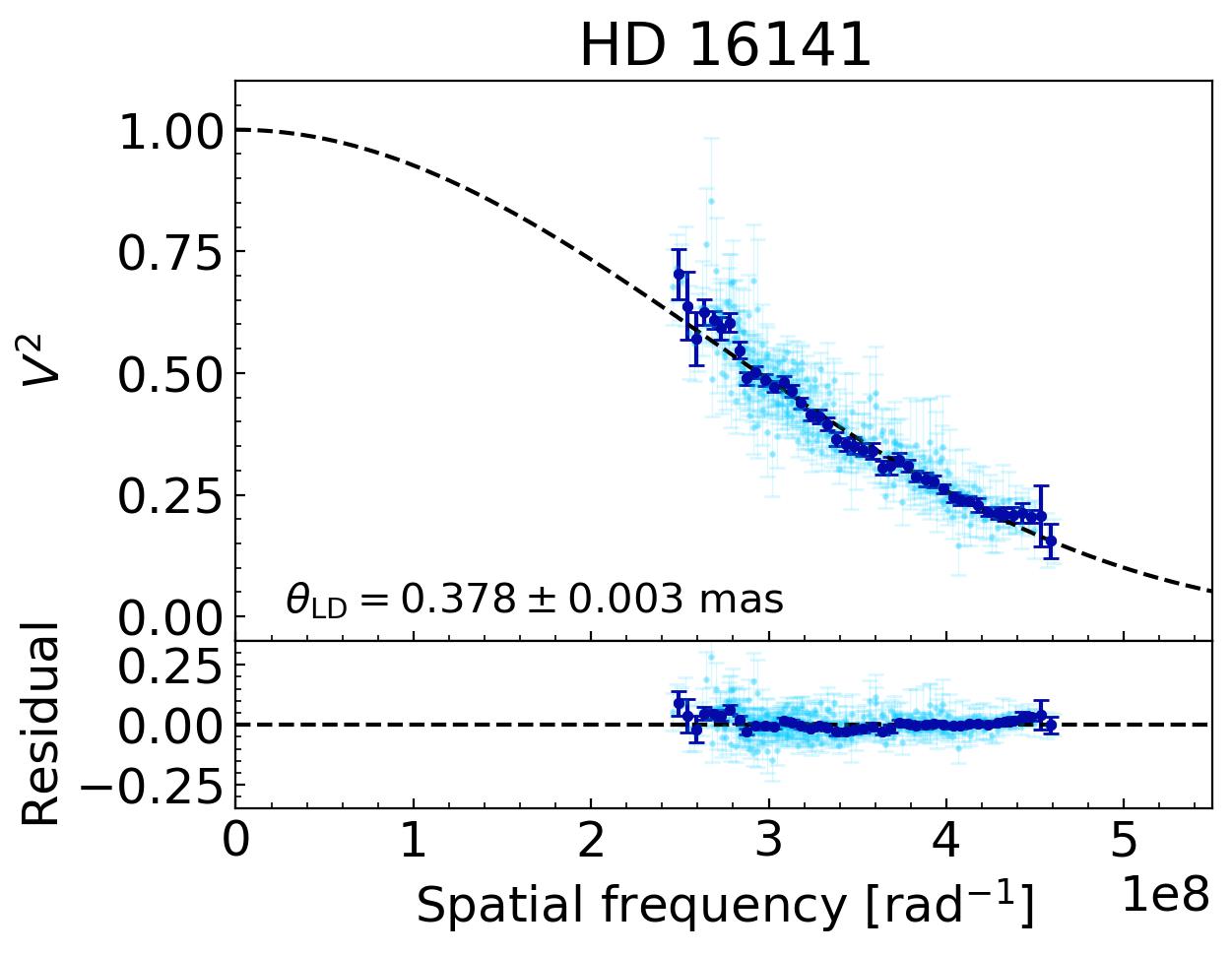} &
  \includegraphics[width=0.45\textwidth]{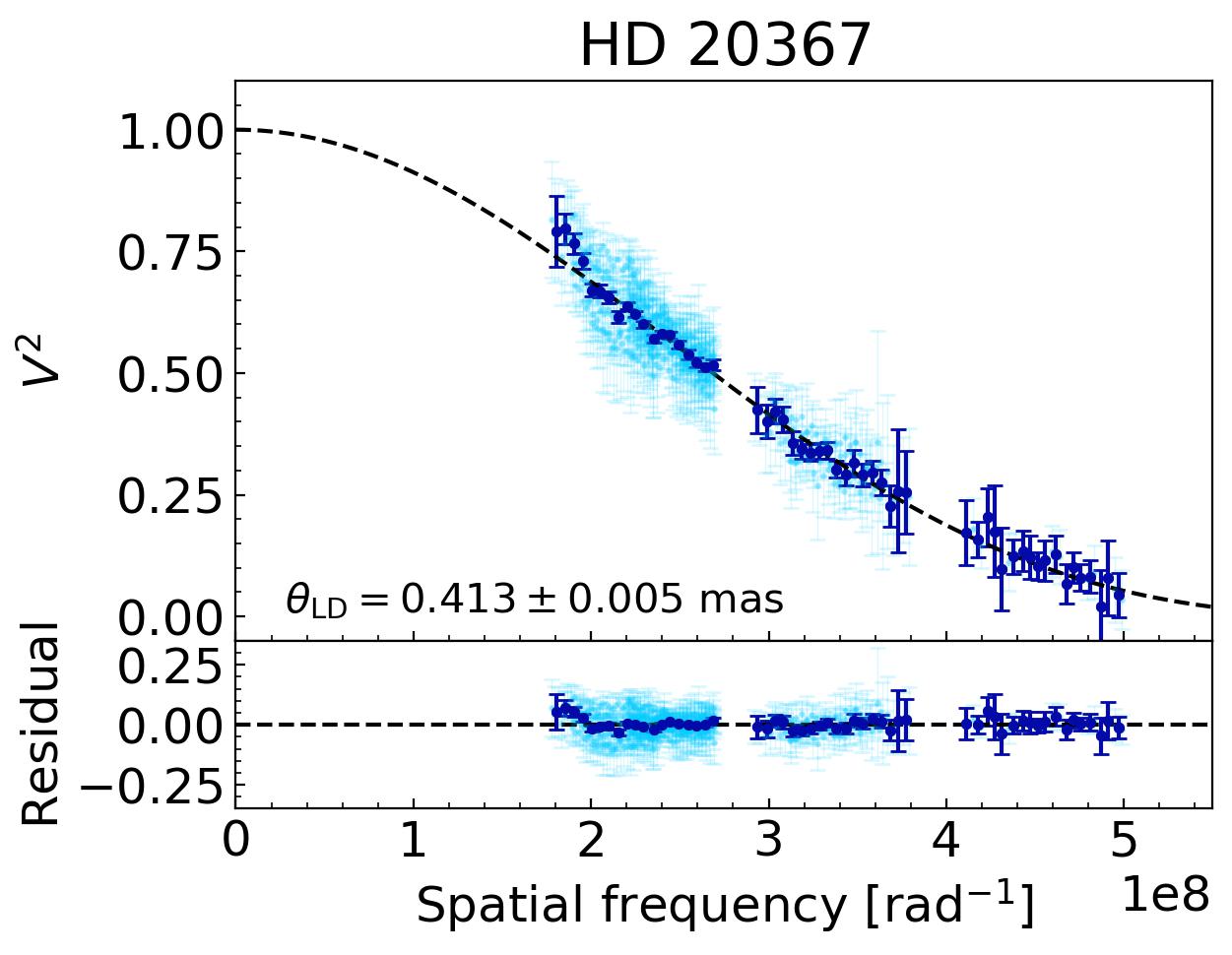} \\

\end{tabular}

\caption{Each panel plots the calibrated interferometric squared visibility $V^2$ as a function of spatial frequency (baseline over wavelength). Light blue marks indicate actual measurements, while dark blue marks are binned data.  The black dashed curve represents the angular diameter fit for each star, which is also printed at the bottom left of each panel.  Residuals to the fit are shown in the lower portion of each plot.}
\label{fig:diameter_plots1}
\end{figure*}
\clearpage

\begin{figure*}
\centering

\begin{tabular}{cc}

  \includegraphics[width=0.45\textwidth]{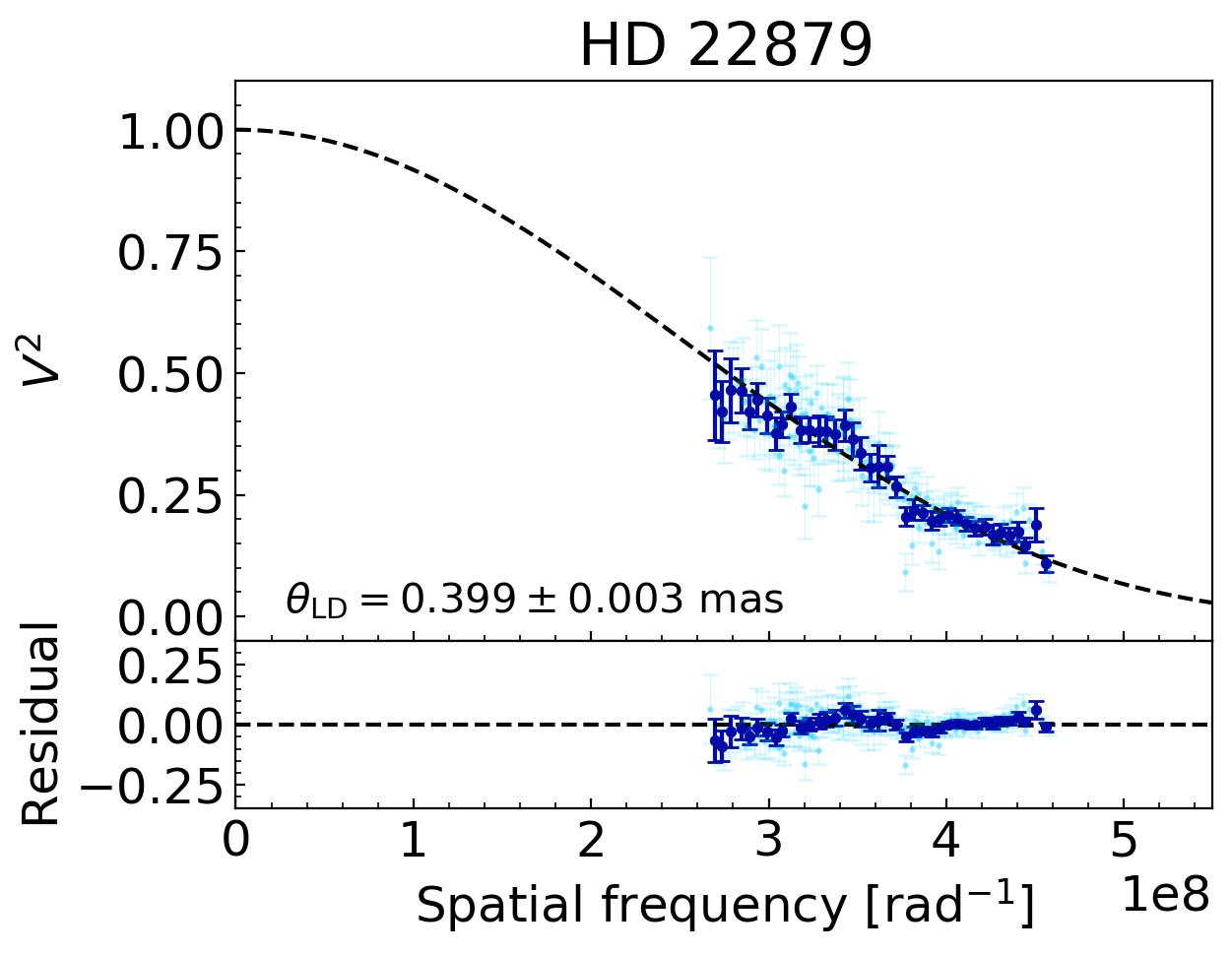} &
  \includegraphics[width=0.45\textwidth]{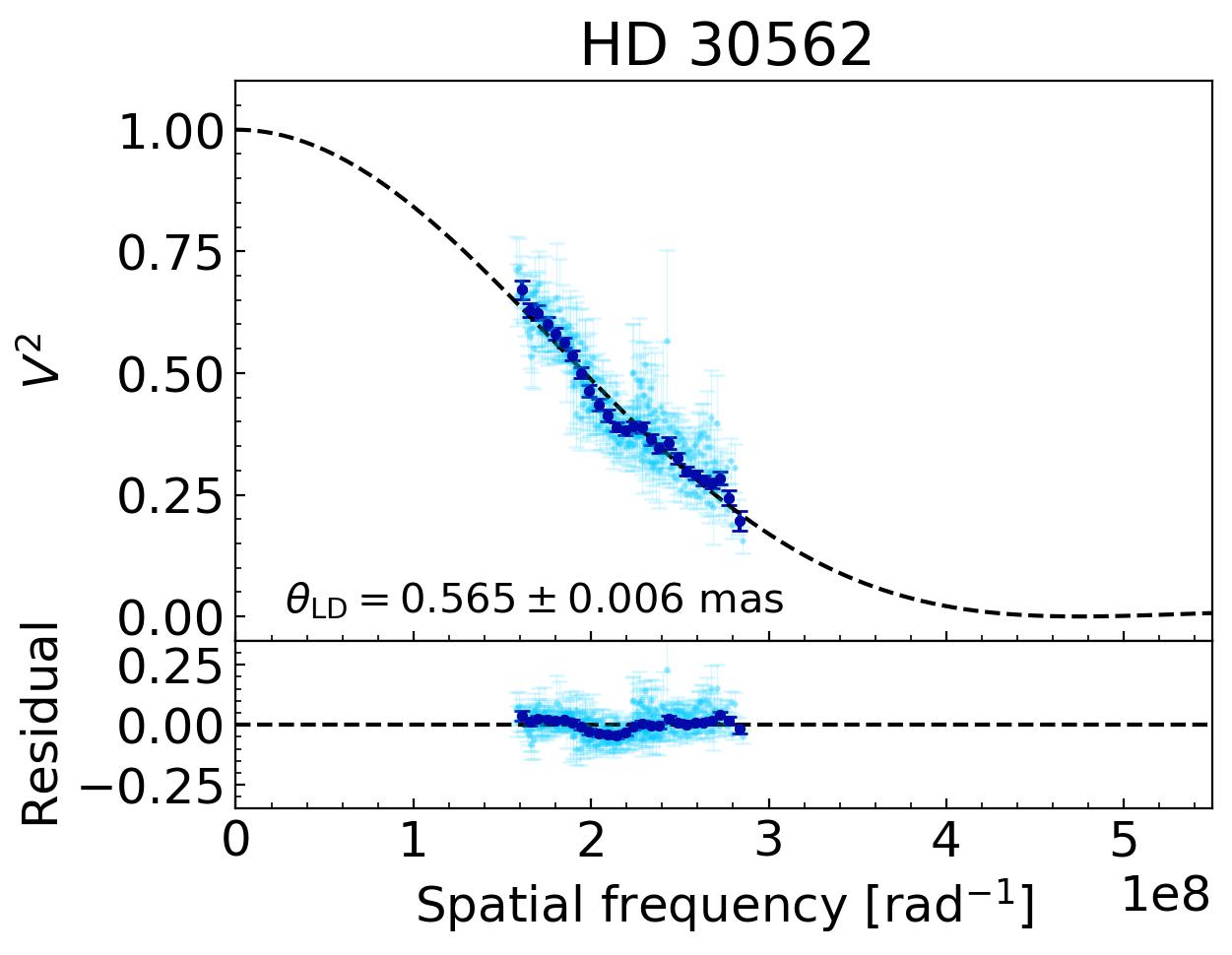} \\
  
  \includegraphics[width=0.45\textwidth]{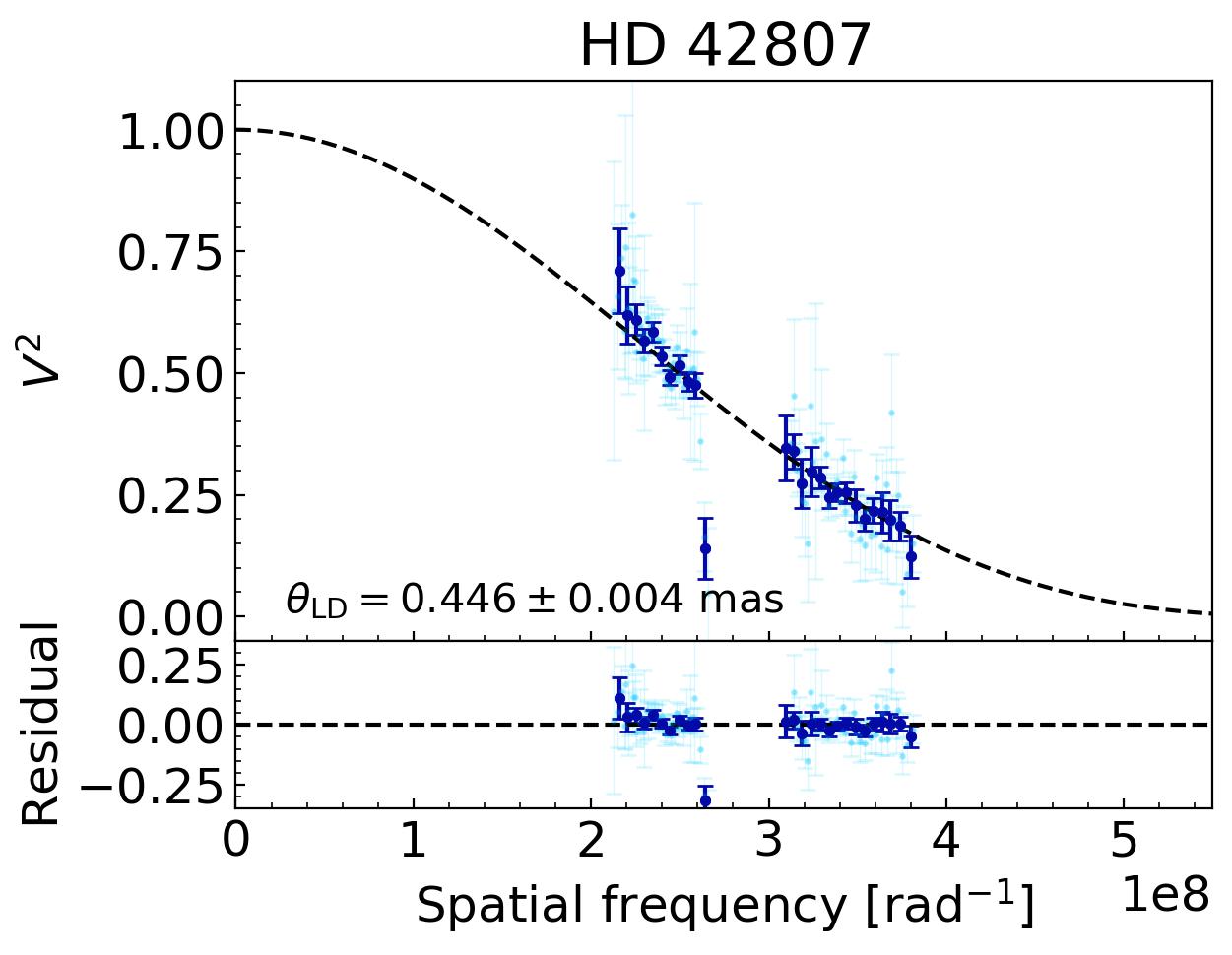} &
  \includegraphics[width=0.45\textwidth]{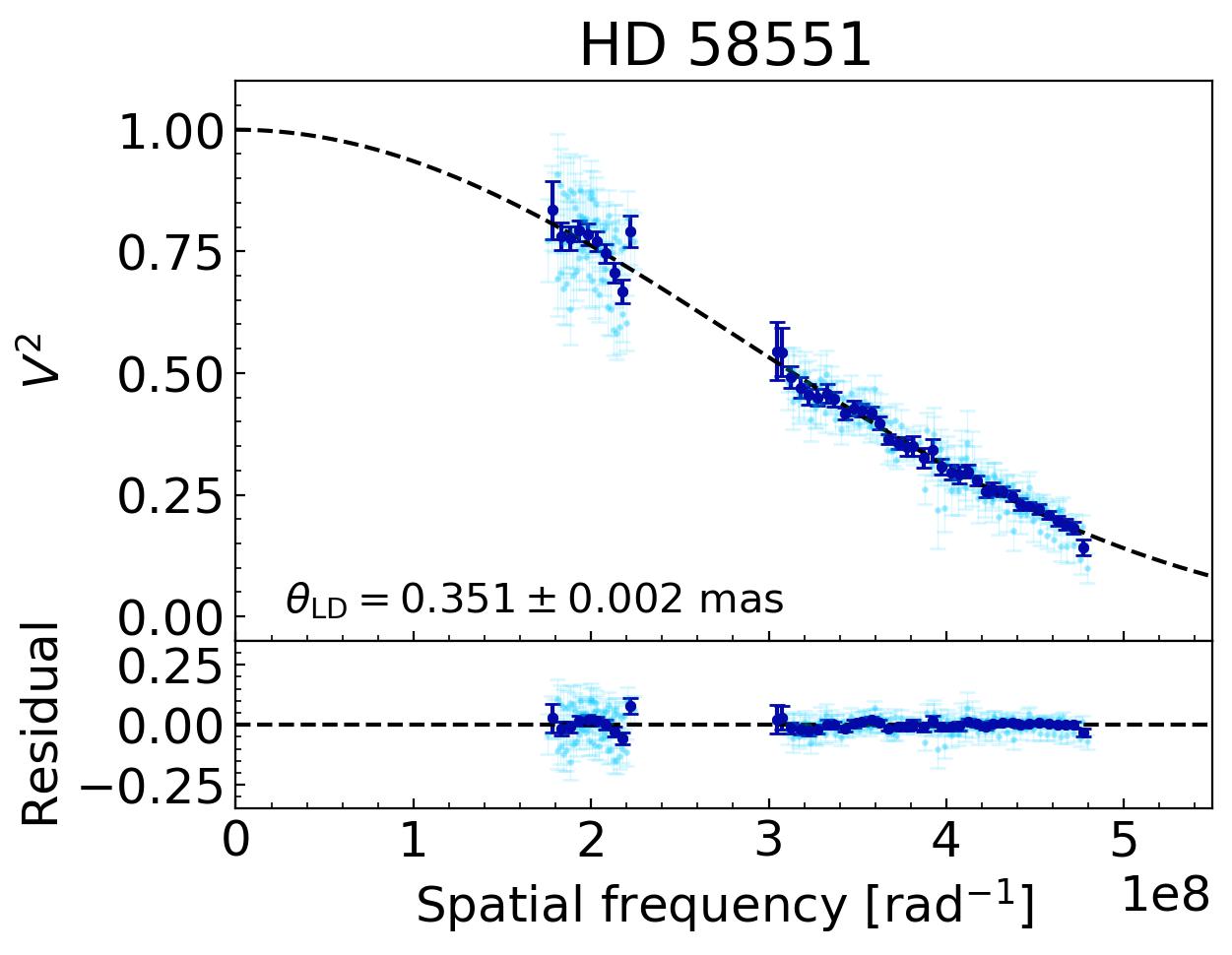} \\

  \includegraphics[width=0.45\textwidth]{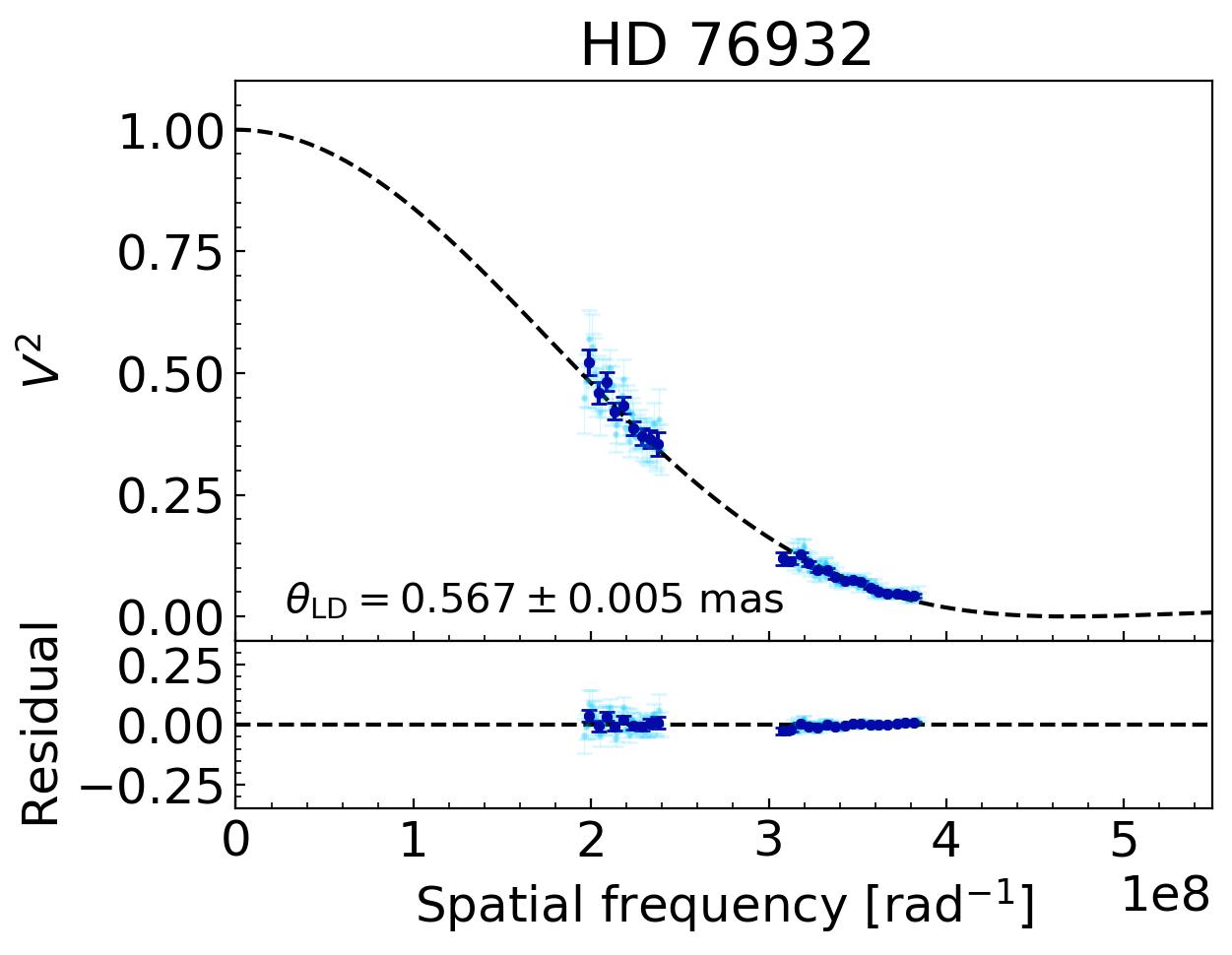} &
  \includegraphics[width=0.45\textwidth]{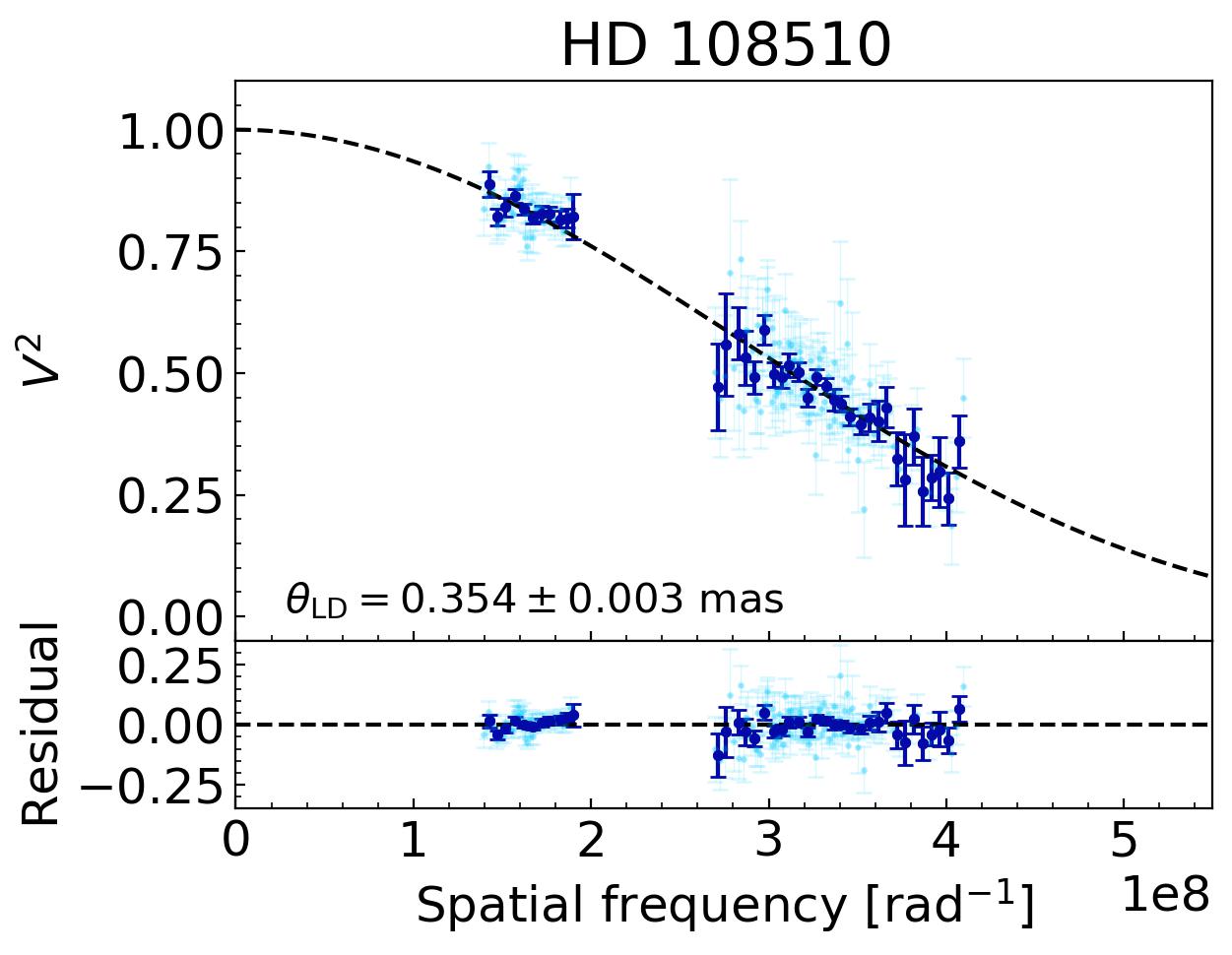} \\

\end{tabular}

\caption{Each panel plots the calibrated interferometric squared visibility $V^2$ as a function of spatial frequency (baseline over wavelength). Light blue marks indicate actual measurements, while dark blue marks are binned data.  The black dashed curve represents the angular diameter fit for each star, which is also printed at the bottom left of each panel.  Residuals to the fit are shown in the lower portion of each plot.}
\label{fig:diameter_plots2}
\end{figure*}
\clearpage

\begin{figure*}
\centering

\begin{tabular}{cc}

  \includegraphics[width=0.45\textwidth]{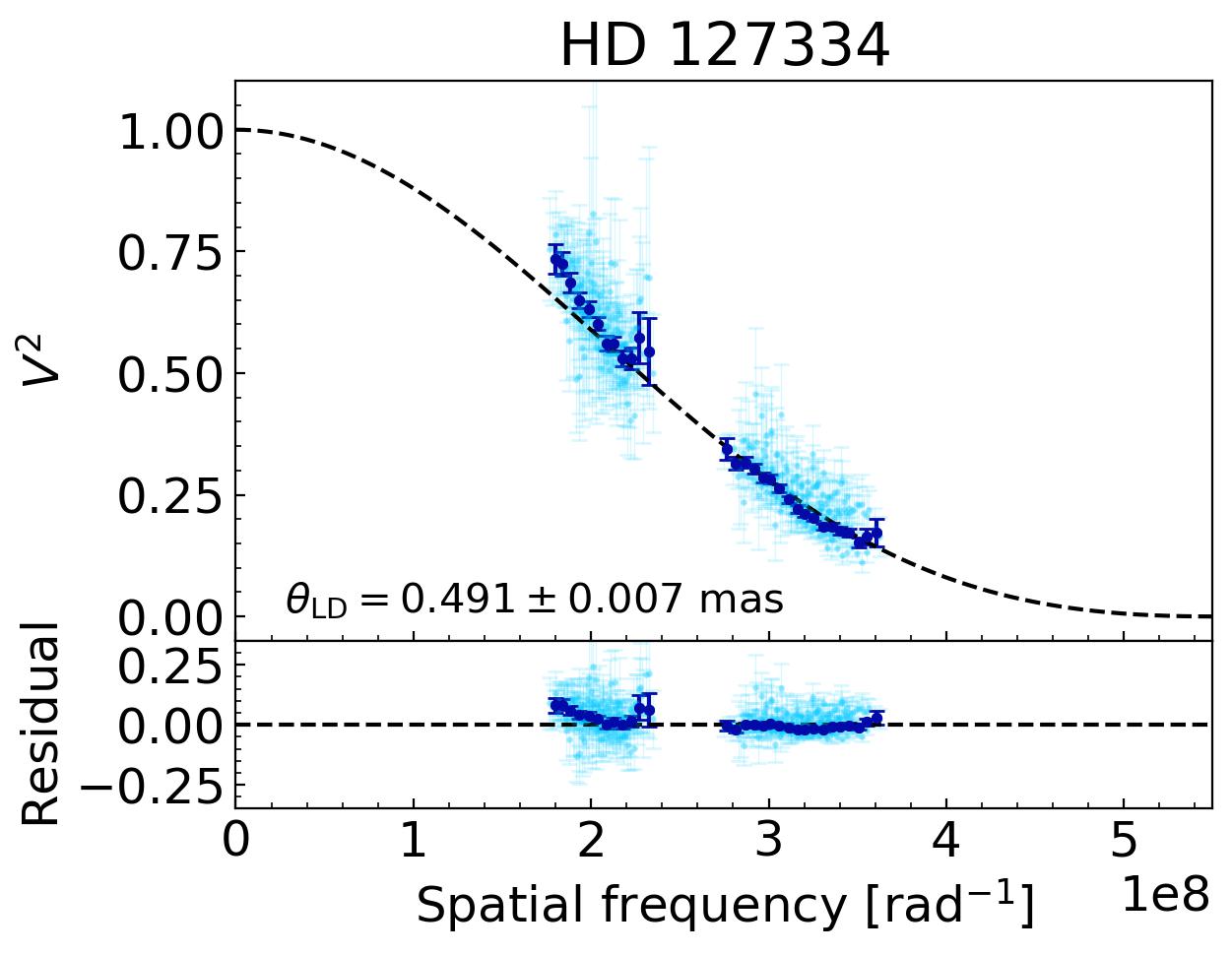} &
  \includegraphics[width=0.45\textwidth]{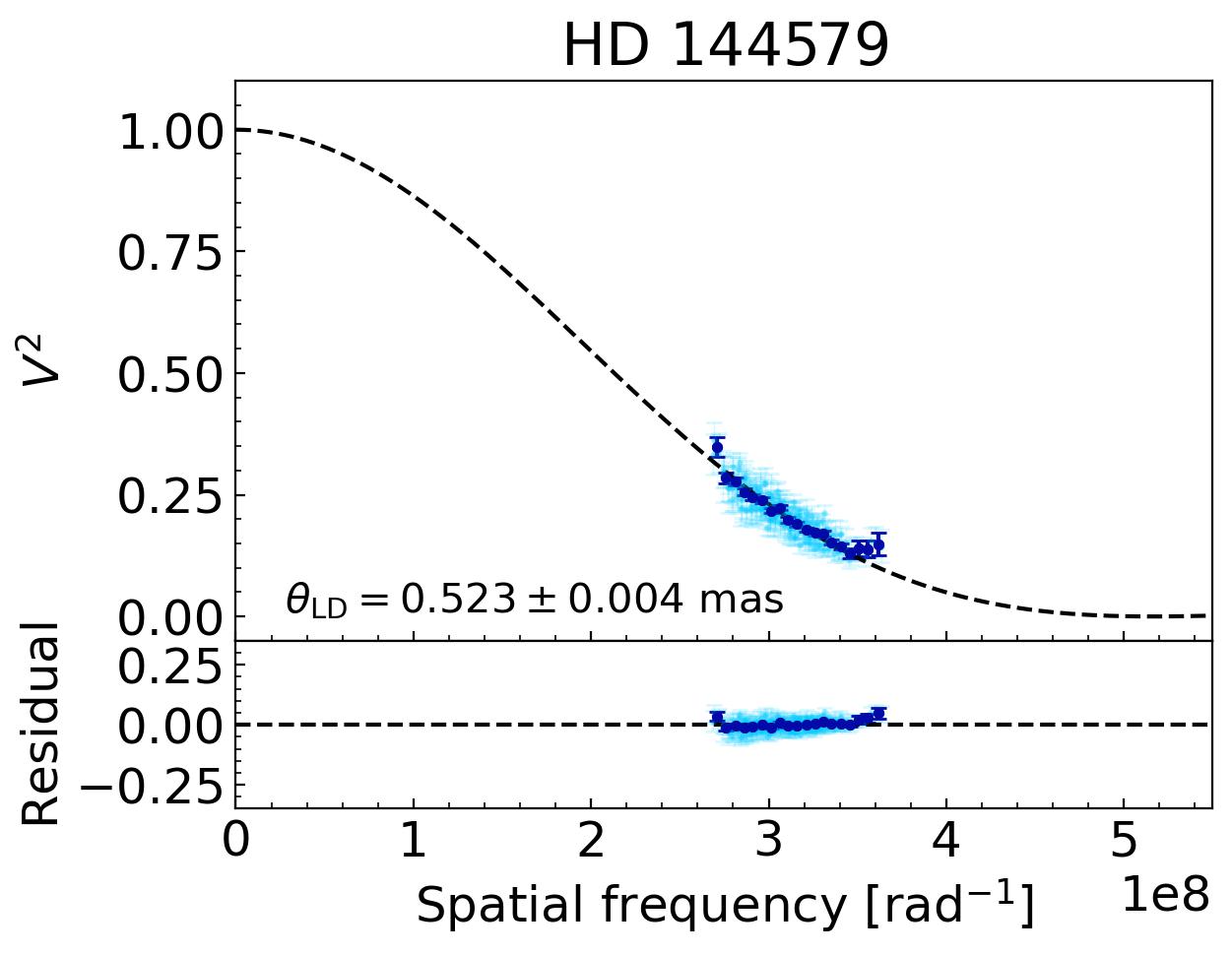} \\

  \includegraphics[width=0.45\textwidth]{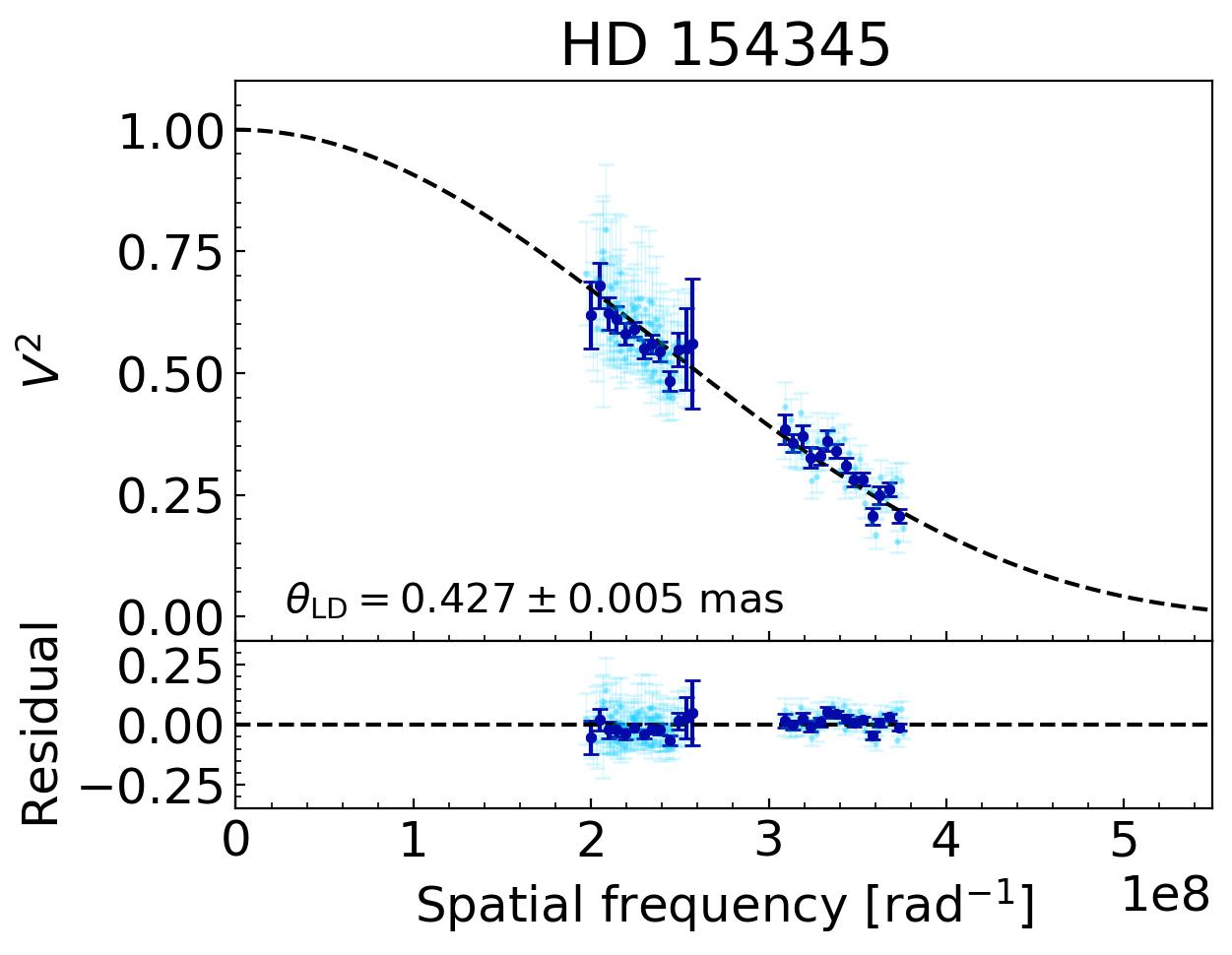} &
  \includegraphics[width=0.45\textwidth]{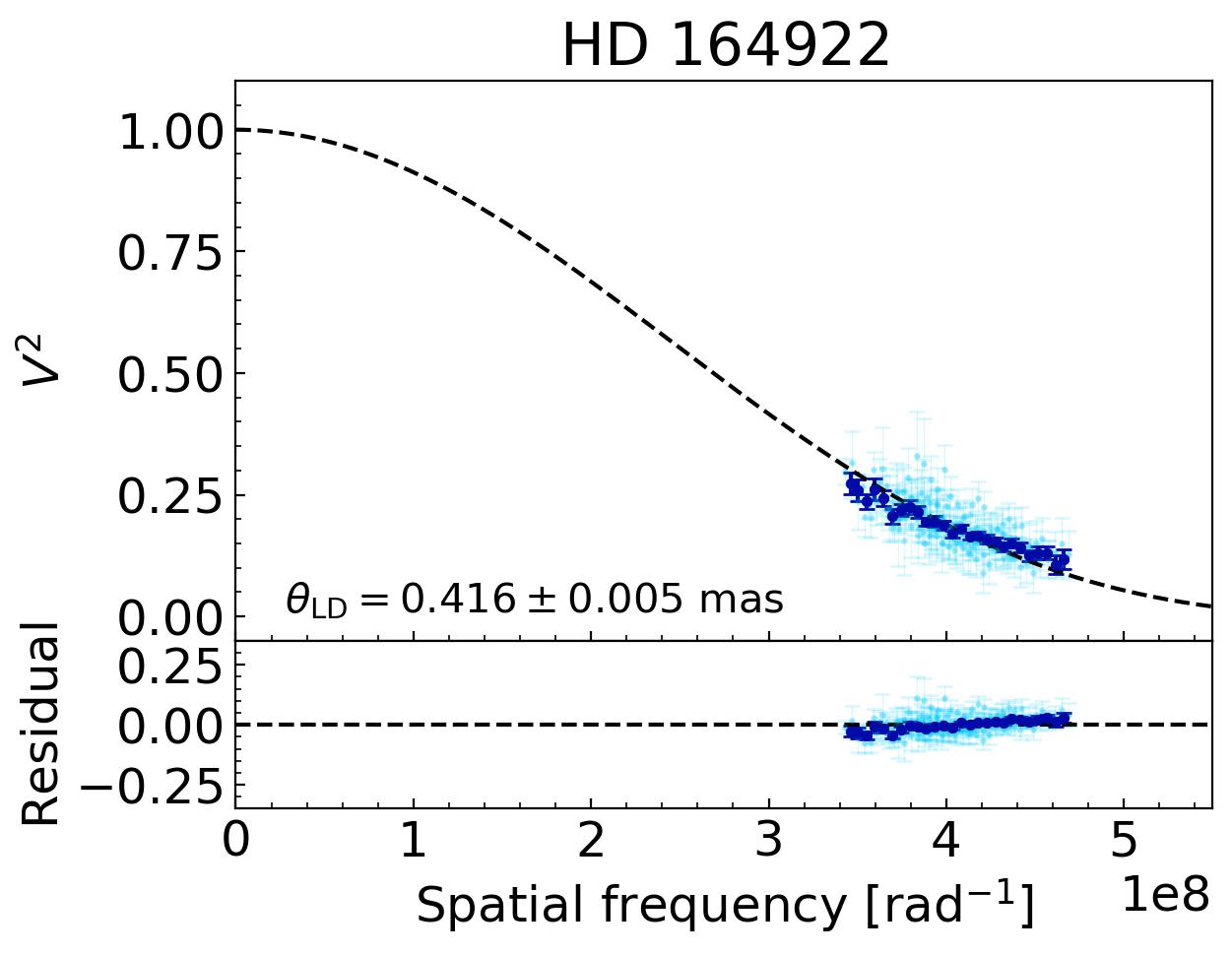} \\

  \includegraphics[width=0.45\textwidth]{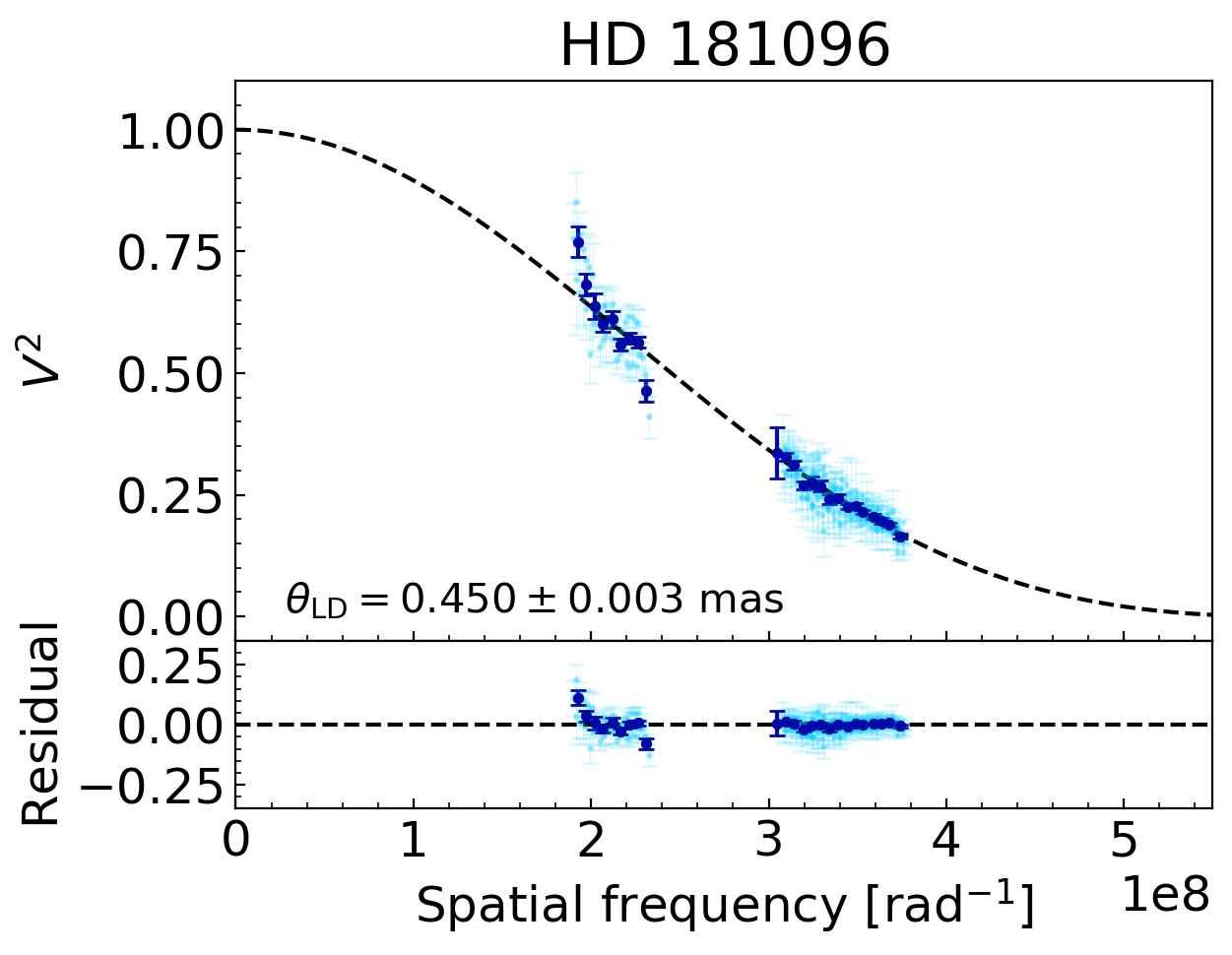} &
  \includegraphics[width=0.45\textwidth]{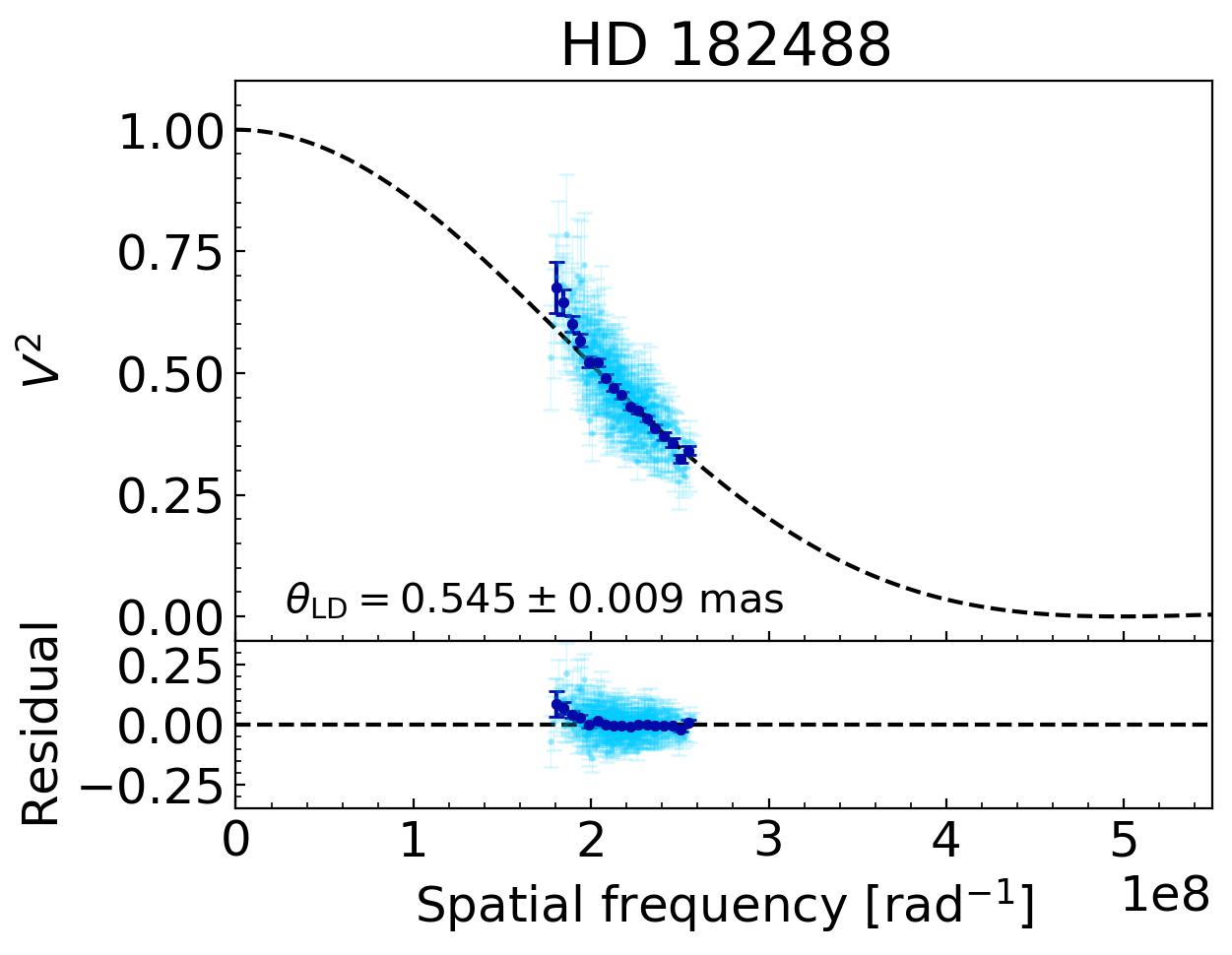} \\

\end{tabular}

\caption{Each panel plots the calibrated interferometric squared visibility $V^2$ as a function of spatial frequency (baseline over wavelength). Light blue marks  indicate actual measurements, while dark blue marks are binned data.  The black dashed curve represents the angular diameter fit for each star, which is also printed at the bottom left of each panel.  Residuals to the fit are shown in the lower portion of each plot.}
\label{fig:diameter_plots3}
\end{figure*}
\clearpage

\begin{figure*}
\centering

\begin{tabular}{cc}

  \includegraphics[width=0.45\textwidth]{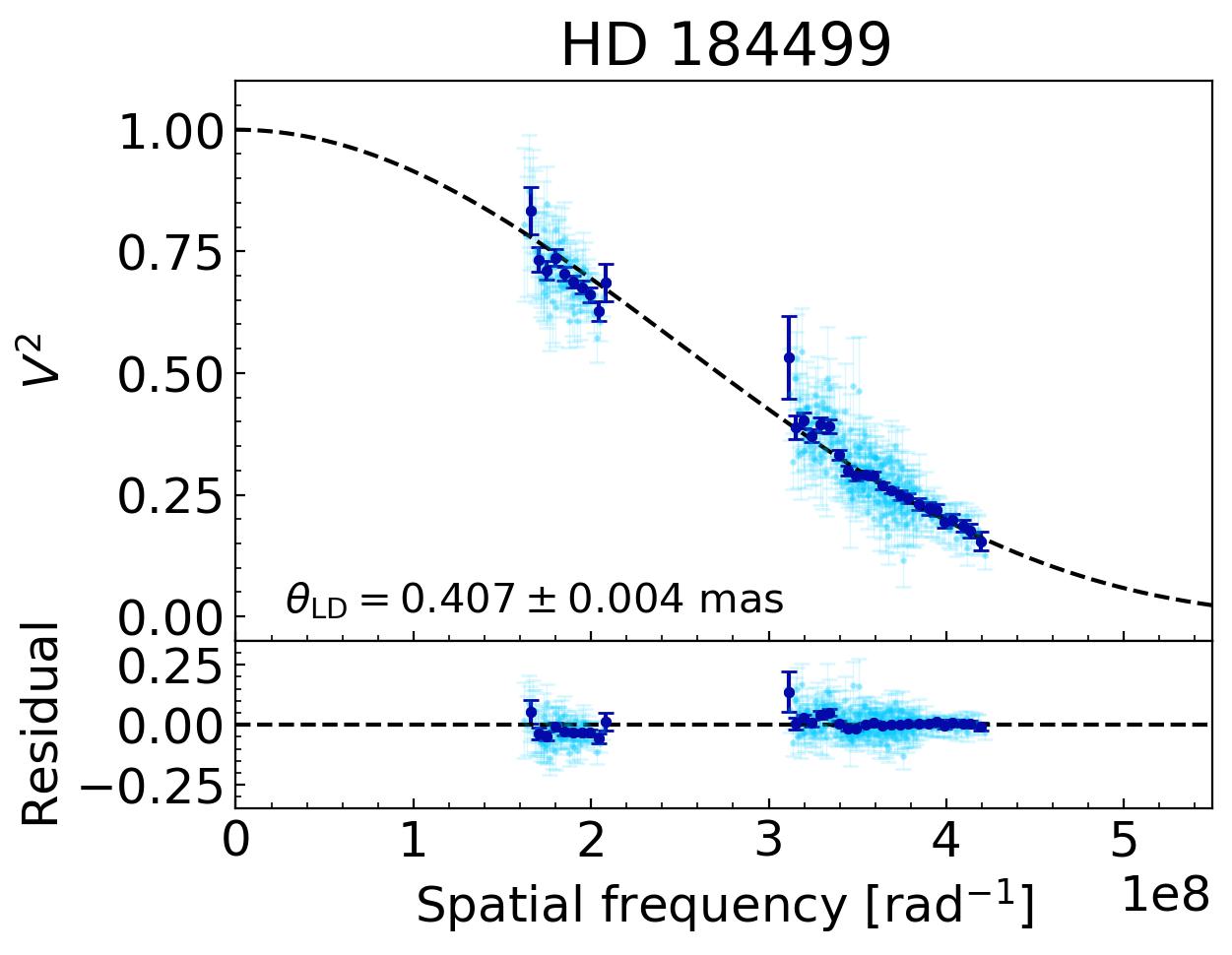} &
  \includegraphics[width=0.45\textwidth]{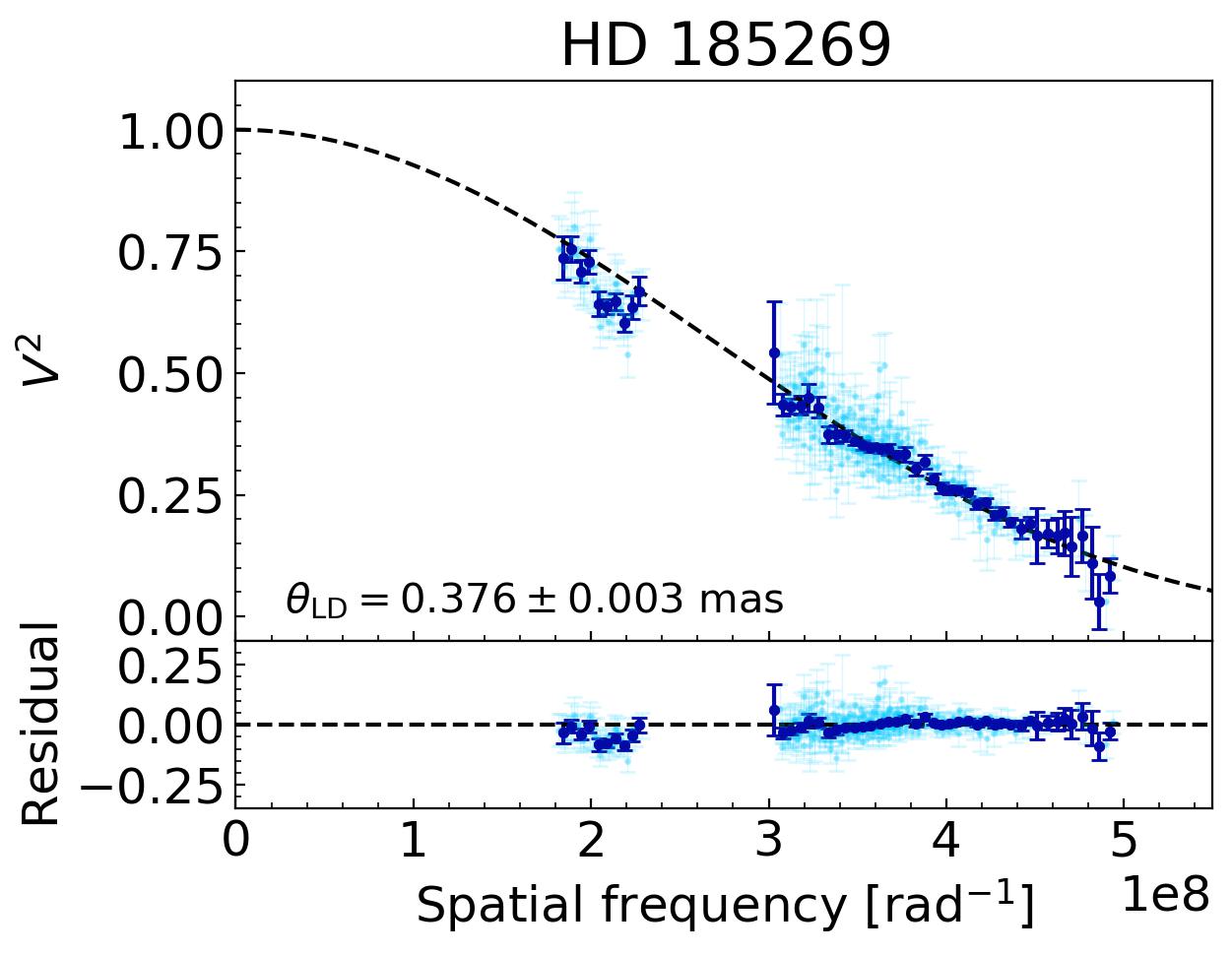} \\

  \includegraphics[width=0.45\textwidth]{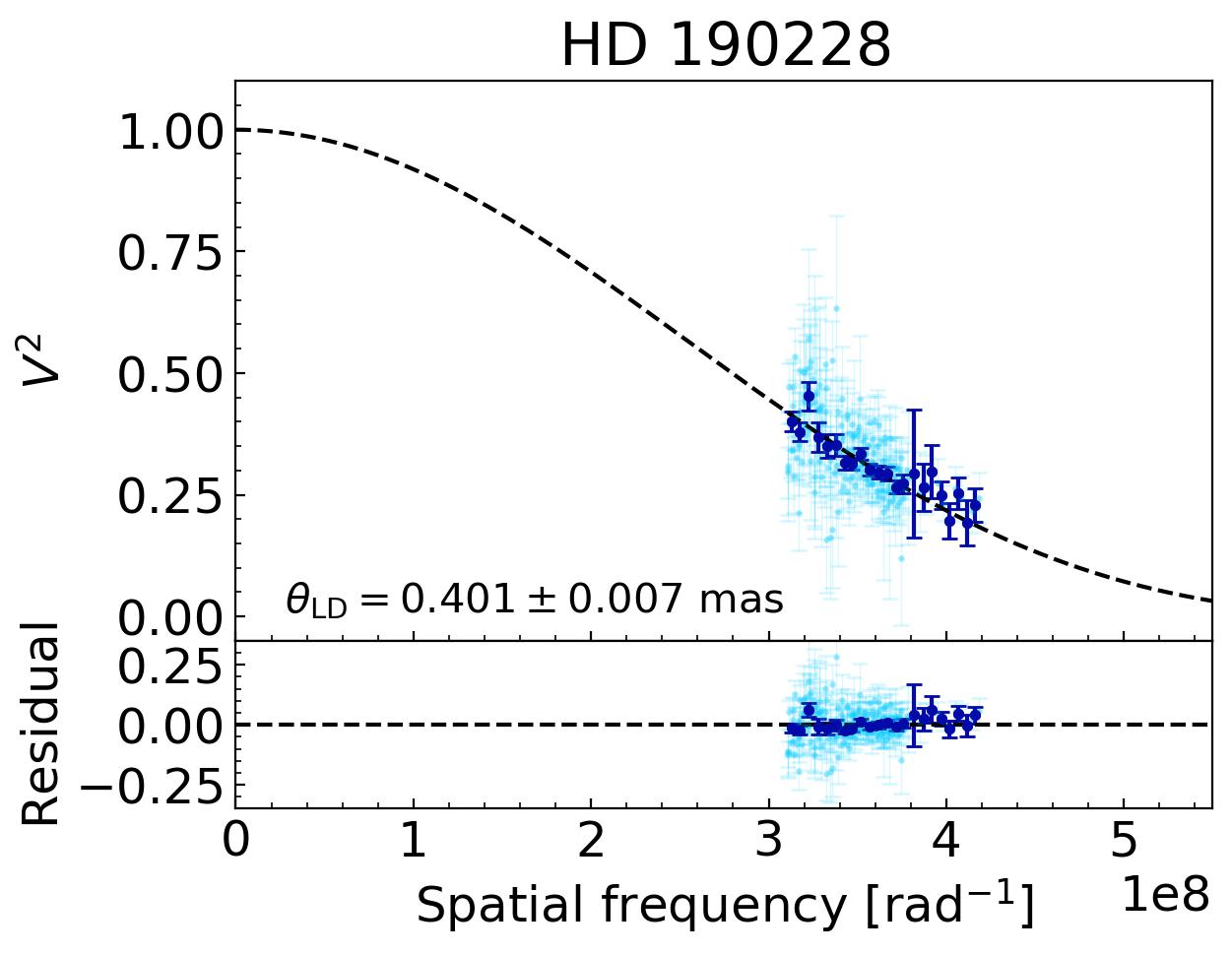} &
  \includegraphics[width=0.45\textwidth]{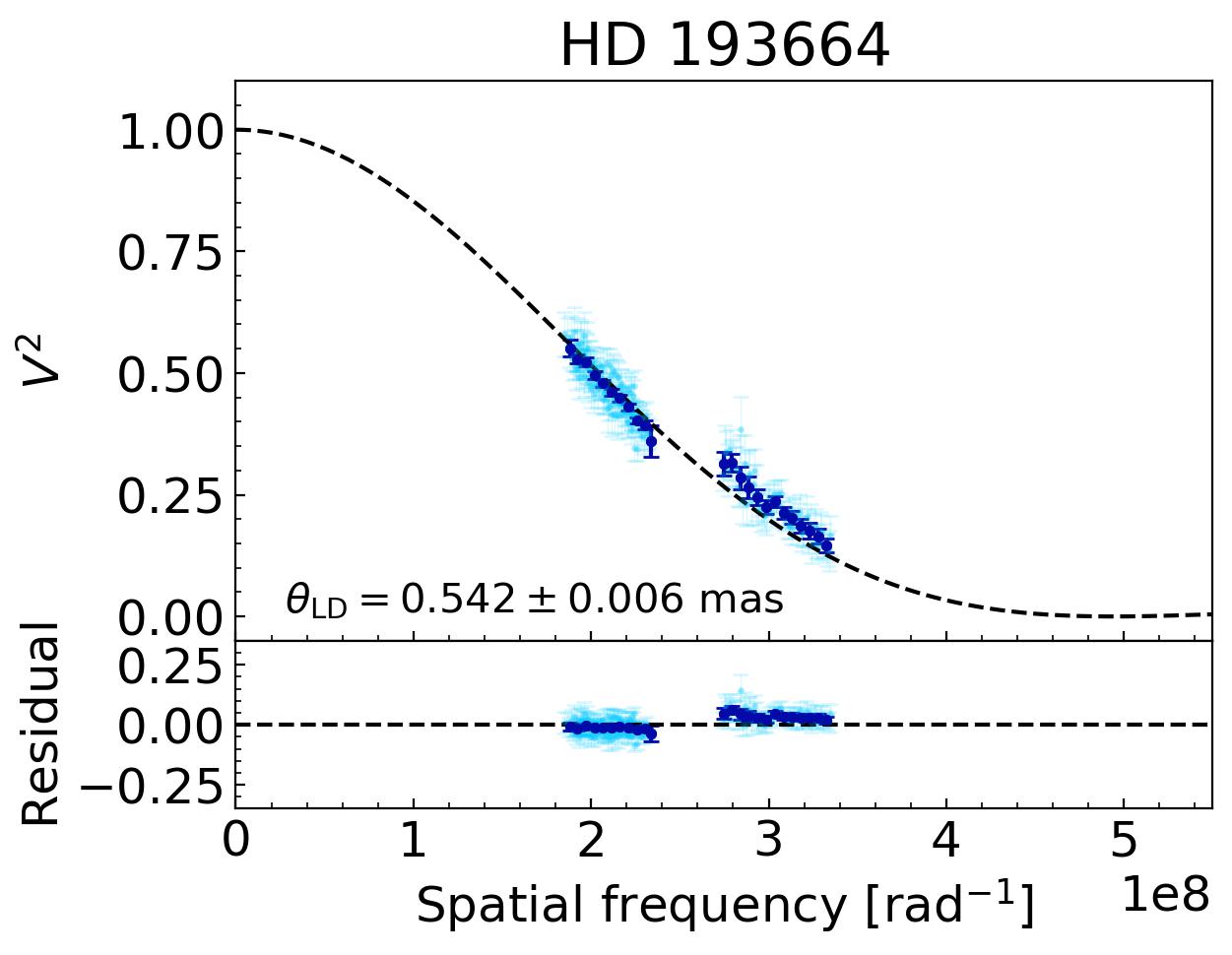} \\

  \includegraphics[width=0.45\textwidth]{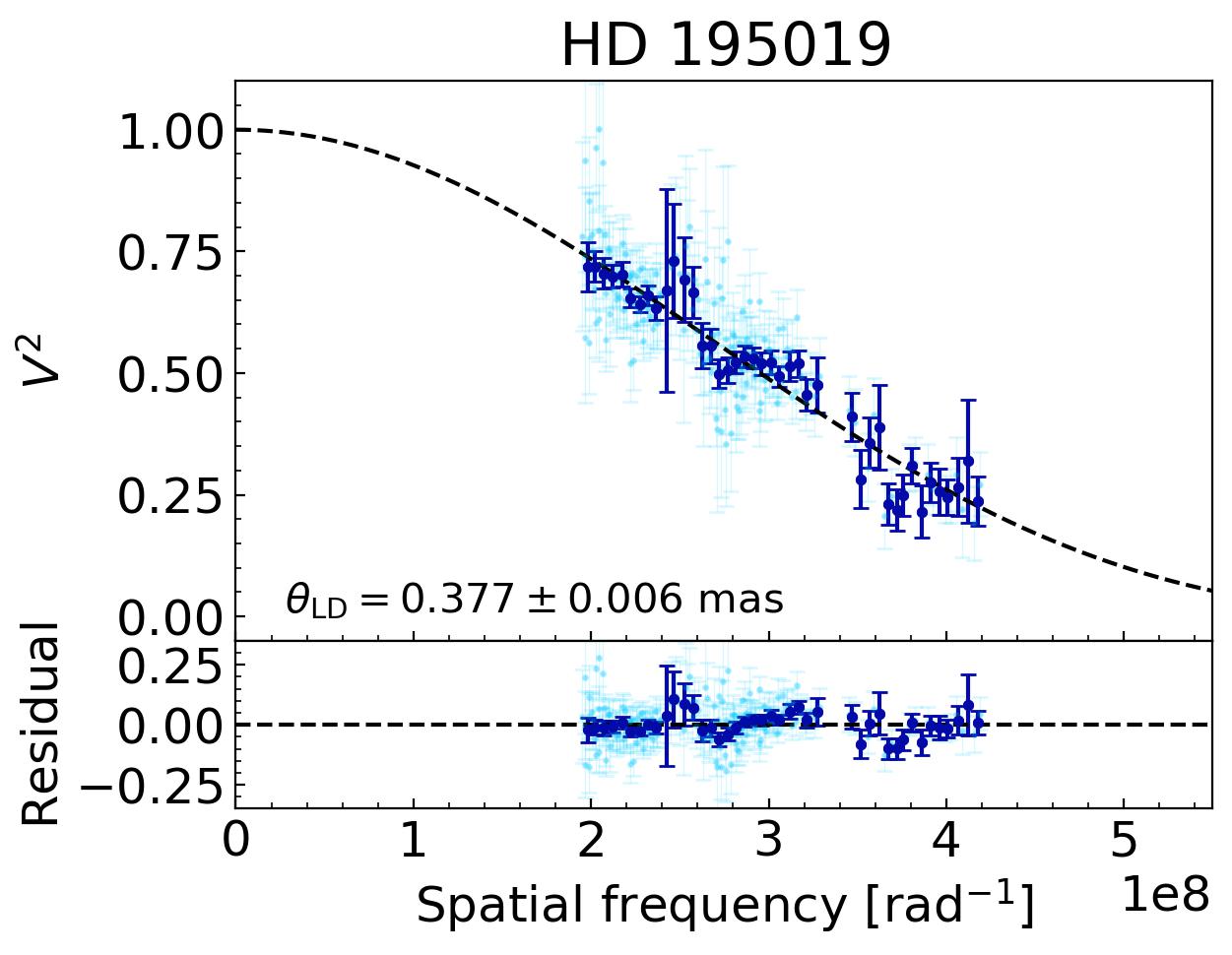} &
  \includegraphics[width=0.45\textwidth]{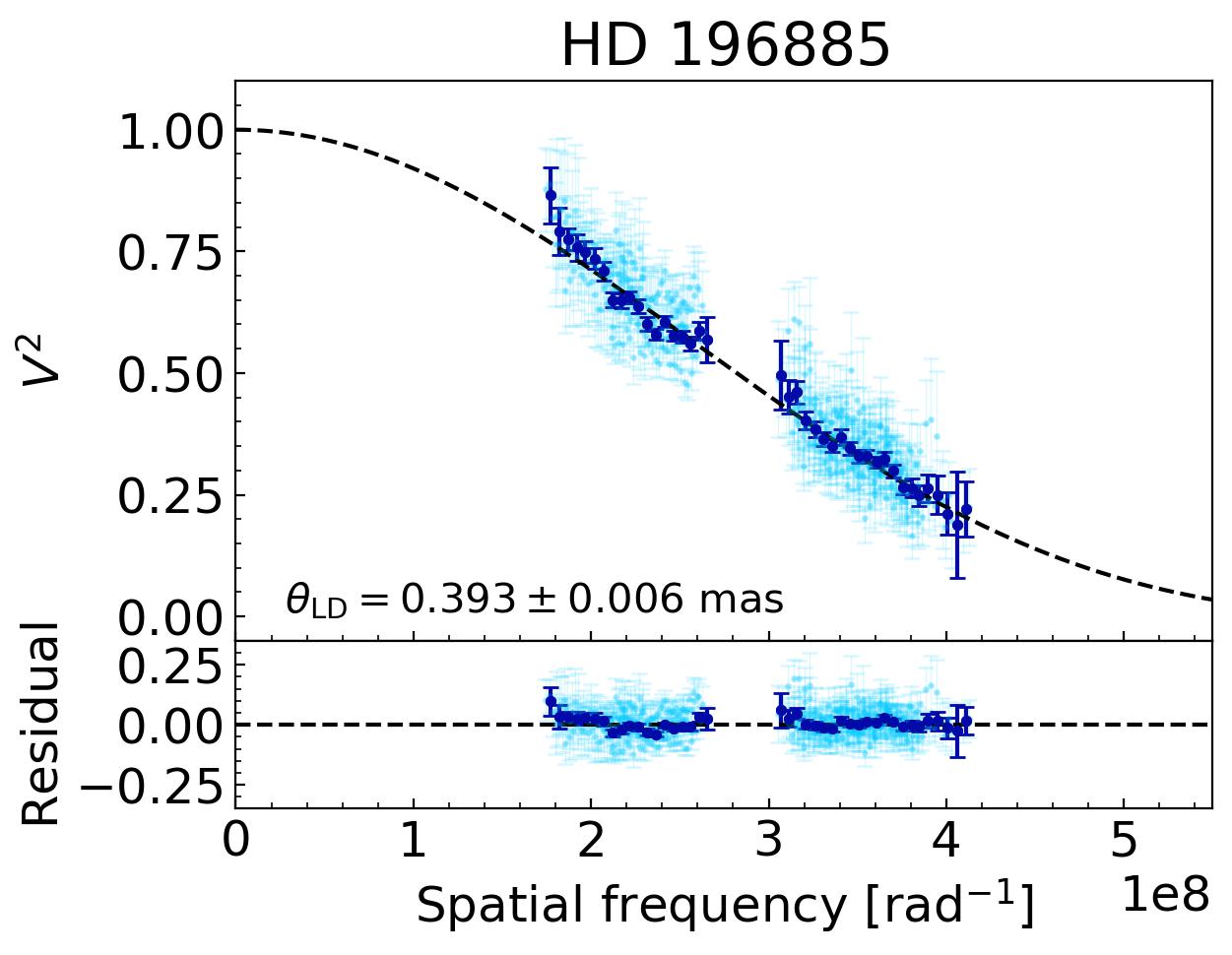} \\

\end{tabular}

\caption{Each panel plots the calibrated interferometric squared visibility $V^2$ as a function of spatial frequency (baseline over wavelength). Light blue marks  indicate actual measurements, while dark blue marks are binned data.  The black dashed curve represents the angular diameter fit for each star, which is also printed at the bottom left of each panel.  Residuals to the fit are shown in the lower portion of each plot.}
\label{fig:diameter_plots4}
\end{figure*}
\clearpage

\begin{figure*}
\centering

\begin{tabular}{cc}
  
  \includegraphics[width=0.45\textwidth]{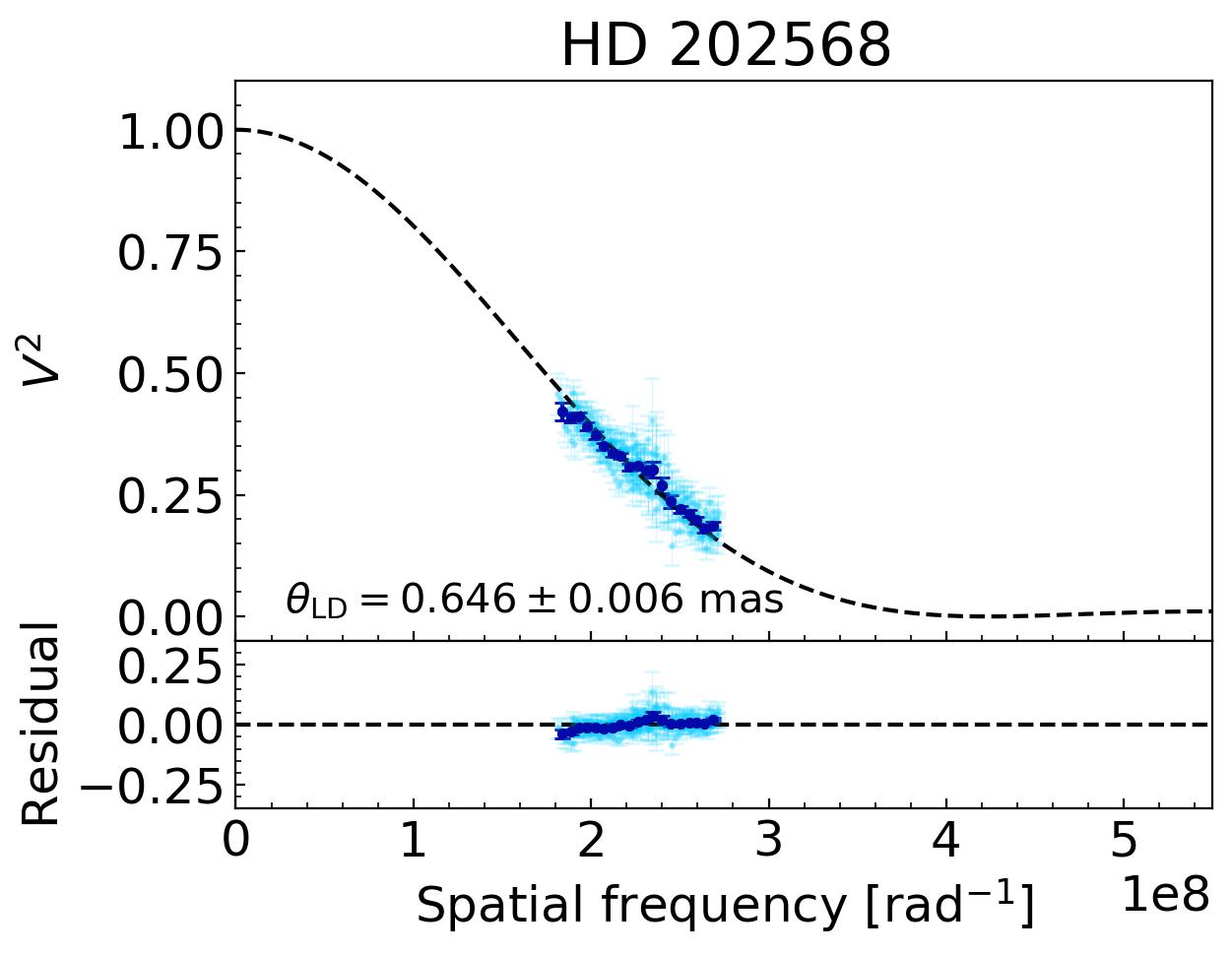} &
  \includegraphics[width=0.45\textwidth]{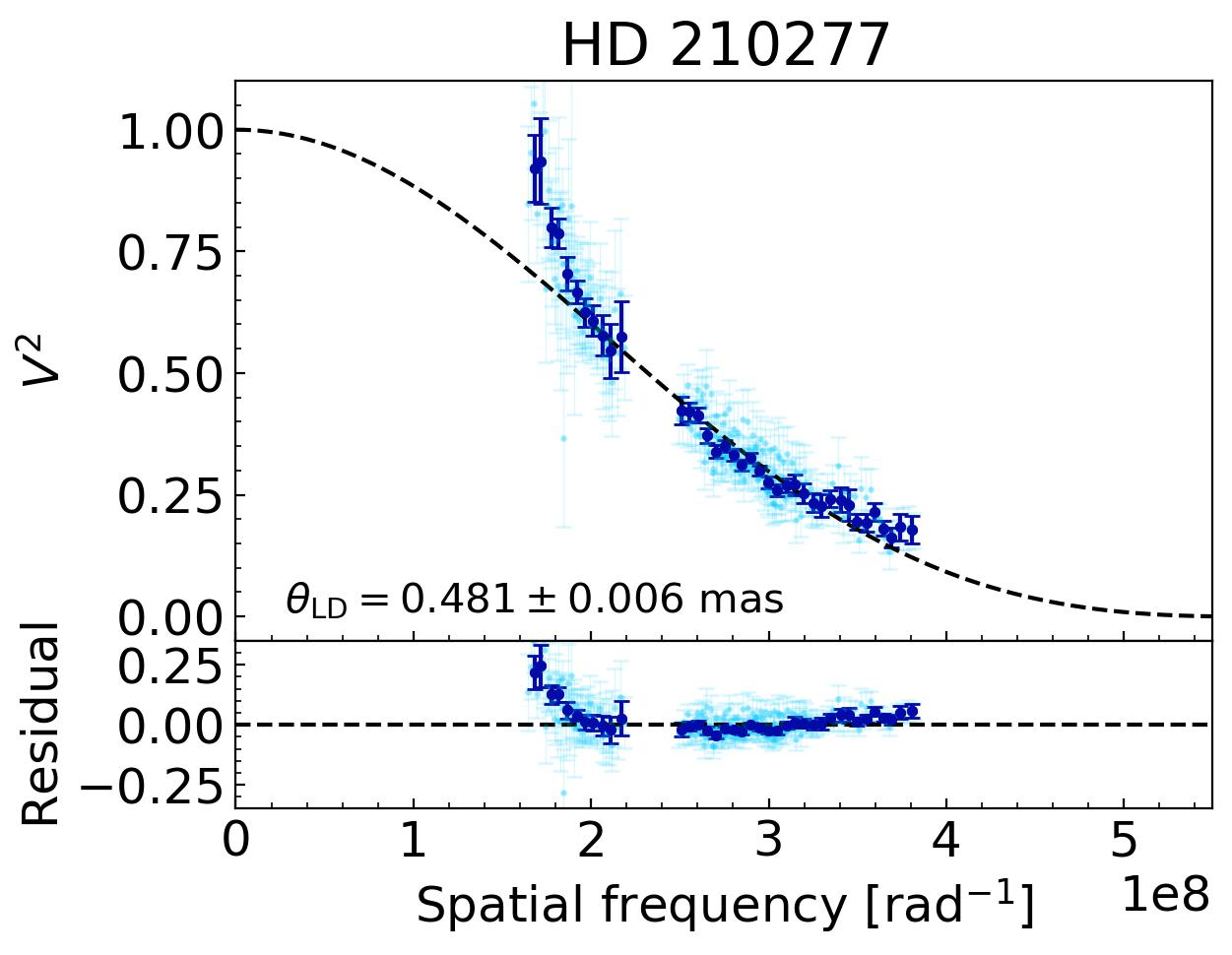} \\

  \includegraphics[width=0.45\textwidth]{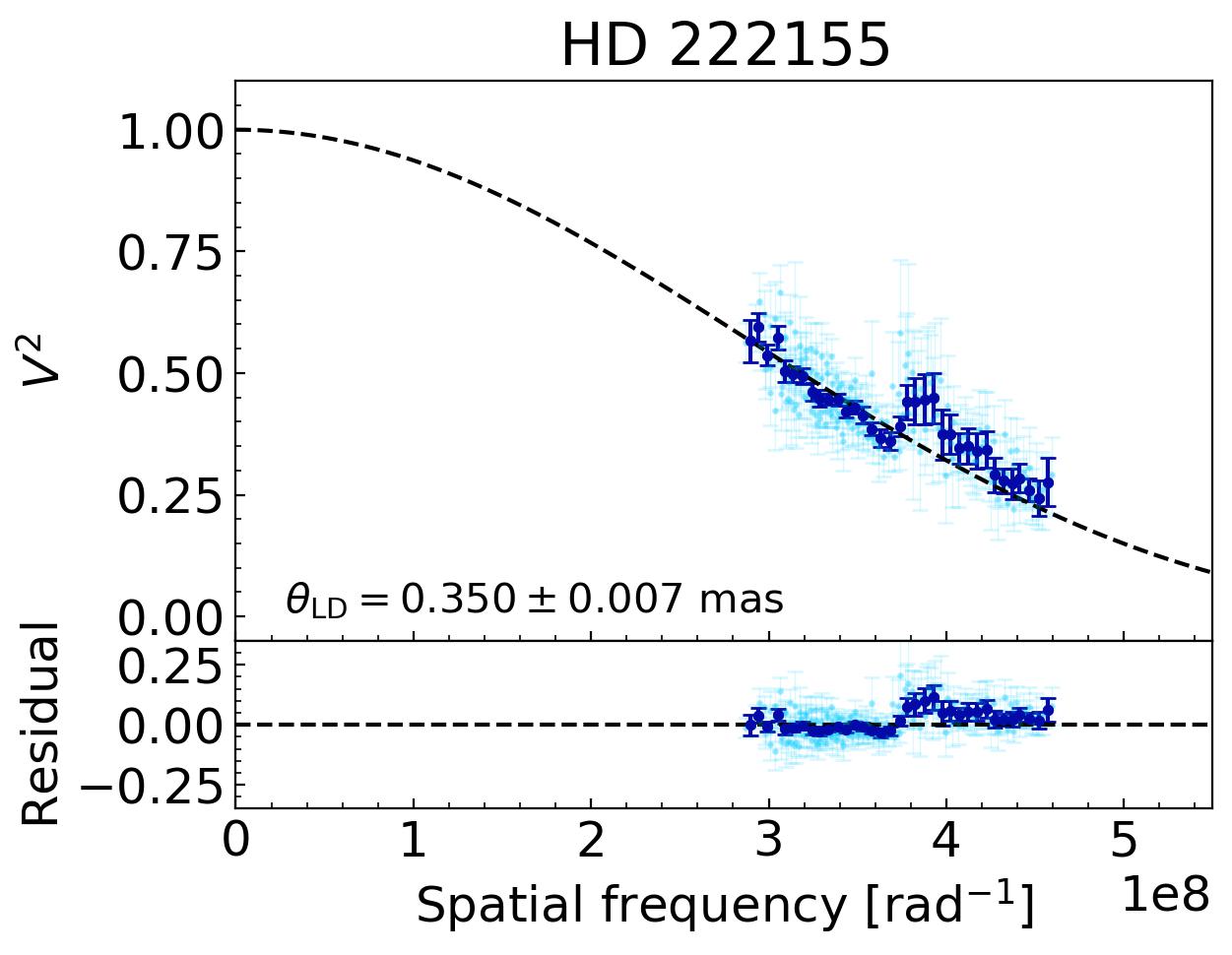} &
  \\

\end{tabular}

\caption{Each panel plots the calibrated interferometric squared visibility $V^2$ as a function of spatial frequency (baseline over wavelength). Light blue marks  indicate actual measurements, while dark blue marks are binned data.  The black dashed curve represents the angular diameter fit for each star, which is also printed at the bottom left of each panel.  Residuals to the fit are shown in the lower portion of each plot.}
\label{fig:diameter_plots5}
\end{figure*}
\clearpage


\end{appendix}


\end{document}